\documentclass[a4paper,12pt]{article}
\usepackage{comment}
\usepackage{authblk}
\usepackage[pdftex]{graphicx}
\usepackage{overpic}
\usepackage{cite}

\title{Development of the Micro Pixel Chamber \\ with resistive electrodes}
\author{Fumiya Yamane\thanks{Corresponding author}}
\author{Atsuhiko Ochi\thanks{E-mail: ochi@kobe-u.ac.jp}}
\author{Kohei Matayoshi}
\author{Keisuke Ogawa}
\author{Yusuke Ishitobi}
\affil{\it Kobe University, 1-1 Rokkodai, Nada, Kobe, Hyogo, Japan}

\date{\empty}

\begin{document}
\maketitle

\begin{abstract}
We developed a novel design of a Micro Pixel Chamber ($\mu$-PIC) with resistive electrodes for a charged-particle-tracking detector in high-rate applications. Diamond-Like Carbon (DLC) thin film is used for the cathodes. The resistivity can be controlled flexibly ($\mathrm{10^{5-7} k\Omega/sq.}$) at high uniformity. The fabrication-process was greatly improved and the resistive $\mu$-PIC could be operated at 10$\times$10 $\mathrm{cm^2}$. Resistors for the HV bias and capacitors for the AC coupling were completely removed by applying PCB and carbon-sputtering techniques, and the resistive $\mu$-PIC became a very compact detector. The performances of our new resistive $\mu$-PIC were measured in various ways. Consequently, it was possible to attain high gas gains ($\mathrm{>10^{4}}$), high detection efficiency, and position resolution exceeding 100 $\mu$m. The spark current was suppressed, and the new resistive $\mu$-PIC was operated stably under fast-neutrons irradiation. These features offer solutions for a charged-particle-tracking detector in future high-rate applications.
\end{abstract}


\section{Introduction}
In high-energy physics experiments, high statistics are needed for precise measurements, and the particle rate has increased in recent years. Therefore detectors are required to withstand high-rate radiation environments. For example, LHC luminosity will probably increase by 2025 from 2$\times$$\mathrm{10^{34} cm^{-2}s^{-1}}$ to 5-7$\times$$\mathrm{10^{34} cm^{-2}s^{-1}}$ \cite{lhc}. Such wire chambers as MWPC and drift chambers are conventionally used for particle tracking with large area coverage. However, since they cannot withstand a high-counting rate ($\mathrm{>10^7 cps/cm^2}$) in the High-Luminosity LHC (HL-LHC), they will be replaced by a new type of devices called Micro Pattern Gaseous Detectors (MPGDs) \cite{mpgd}, where micro-electrodes are used instead of wires, and a high-rate capability over $\mathrm{10^8 cps/cm^2}$ can be achieved. A large detection area can also be achieved at a reasonable cost. Micromegas \cite{megas} will be used for the endcap muon spectrometer of ATLAS \cite{atlastdr}, and GEM \cite{gem,gem2} will be used for that of CMS \cite{cmstdr}. Plans also exist to extend the acceptance for muons to higher $\mid\eta\mid$ regions up to 4.0 ($\eta$: pseudo rapidity) in both experiments. Here a high-rate capability up to 10 MHz/$\mathrm{cm^2}$ and high granularity of a few $\mathrm{mm^2}$ are required, and MPGDs are suggested for those detectors \cite{mtag, mtagtdr}. Thus, MPGDs are expected to play an important role in future high-rate experiments.

The problem with MPGDs is the sparks between electrodes. In gaseous detectors, sparks are caused when the avalanche size exceeds $\mathrm{\sim10^{8}}$ electrons, which is called the Raether limit \cite{raether}. In the case of MPGDs, since electrodes are placed at a very short distance of several tens of microns, the space charge density increases, which reduces the Raether limit to $\mathrm{\sim10^{6-7}}$ \cite{raether2}. The ATLAS Micromegas and the CMS GEM will be operated in the background of fast neutrons that yield heavily ionizing particles in the detector. In this radiation environment, frequent sparks are unavoidable. Fine electrodes are easily damaged by sparks, and electrical breakdown is caused between electrodes. As for the GEM, the spark rate can be reduced using multiple amplification stages. On the other hand, spark-protection structures are needed for such single amplification stage detectors as Micromegas. To overcome this problem, resistive electrodes were introduced in 2010 \cite{spark}. The spark's current is reduced by resistive electrodes and Micromegas with resistive strips could be operated stably under fast-neutrons irradiation \cite{sparkmm}. Over the past few years, MPGDs with resistive electrodes have been developed for high-rate applications. As one of them, the Micro Pixel Chamber ($\mu$-PIC) with resistive electrodes have been developed at Kobe University.

\begin{figure}[htbp]
\begin{center}
\includegraphics[width=8cm,clip]{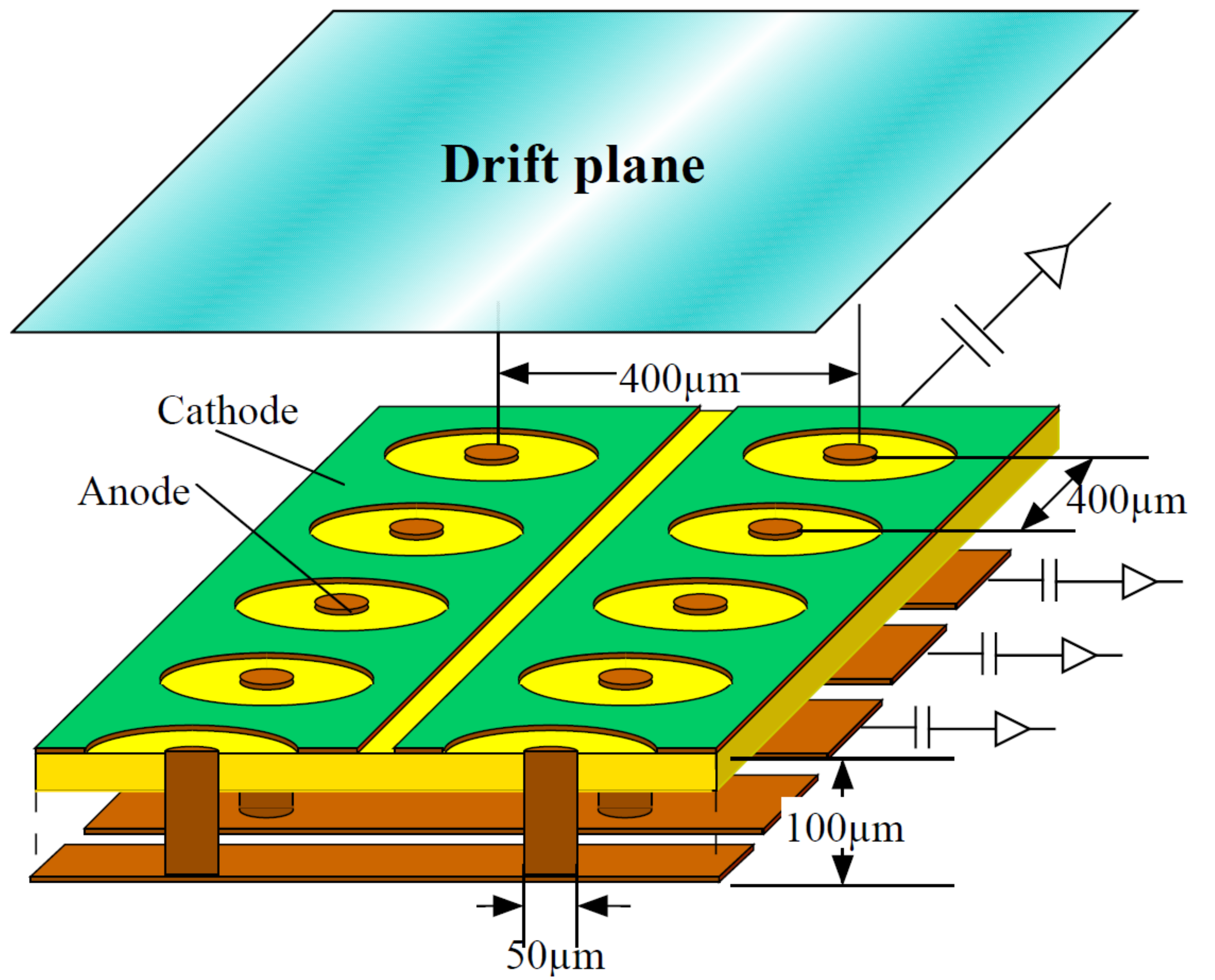}
\end{center}
\caption{Schematic view of $\mu$-PIC \cite{upic}}
\label{upic}
\end{figure}

Fig. \ref{upic} shows a schematic view of a $\mu$-PIC, which is a two-dimensional imaging detector fabricated by printed circuit board (PCB) technique with photolithography. Anode and cathode strips are positioned perpendicularly at a 400$\mu$m pitch, with fine pixels composed of the anode and the surrounding cathode also arranged at this pitch. Gas multiplication occurs in each pixel. One advantage of $\mu$-PIC's operation is that it does not need any floating structures, such as foils and meshes; all of the electrodes are formed rigidly on the substrate, simplifying the assembly procedure. A $\mu$-PIC can be operated by just putting the drift plane above its readout board. Also, it can be extended to a large area by arranging multiple $\mu$-PIC substrates with a very small dead area. Other basic properties of $\mu$-PIC have been reported \cite{upic, upic2}.

\begin{figure}[h]
\centering
\includegraphics[width=8cm,clip]{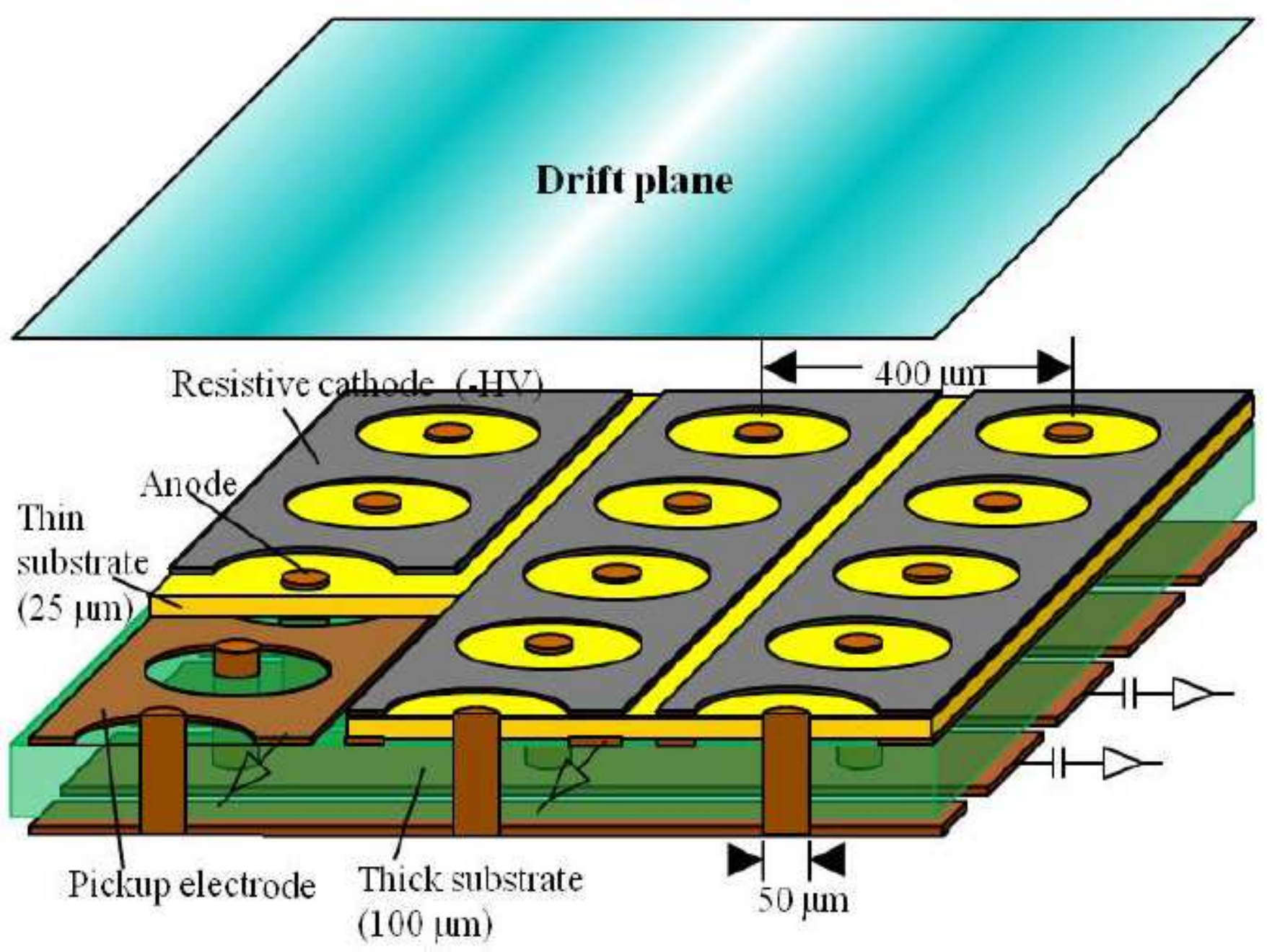}
\caption{Schematic view of resistive $\mu$-PIC \cite{reupic}: Insulating layer is sandwiched between resistive cathode and pickup electrode.}
\label{resistiveupic}
\end{figure}

Fig. \ref{resistiveupic} shows a schematic view of a $\mu$-PIC with resistive electrodes \cite{reupic}. Cathode strips are made of carbon-loaded polyimide paste with $\mathrm{10^{5-7} \Omega/sq}$ resistivity. Resistive cathodes and pickup electrodes are separated by an insulating layer. For the cathode readout, induced signals on the pickup electrodes are used instead of resistive cathode electrodes. Our previous studies argued that the spark current was strongly suppressed and resistive $\mu$-PIC was stably operated under fast-neutrons irradiation. However, there remained problems in the detector design, then it could not be applied for practical use. Therefore, the detector design has been greatly improved.

In this paper, we propose a new design of the resistive $\mu$-PIC to overcome those problems. First, the detector's design and its fabrication-process are described in Section 2, and then the detector's basic performance is reported. In Section 3, operation tests are conducted using X-rays, and in Section 4, performance studies using charged particles are presented. In Section 5, the detector's performance under fast-neutrons irradiation is reported. 

\section{Novel design of new resistive $\mu$-PIC}
\subsection{Innovative ideas for new $\mu$-PIC to overcome problems of former design}
Three approaches have been performed to overcome the problems of the former design. This section delineates its problems and solutions.

\subsubsection{Promising resistive electrodes made of DLC}
Generally, higher tolerance against sparks can be obtained with higher electode resistivity. However, higher resistivity causes serious problems for high-rate applications. The average current of the anodes and cathodes increases during high-rate irradiation. Therefore, the gas gain decreases due to the voltage drop through the resistive electrodes. Hence, the resistivity must be tuned properly depending on the requirements of the high-rate capability and tolerance to sparks. 
A carbon-loaded paste was used for the resistive electrodes of the former $\mu$-PIC \cite{reupic}. The resistivity, which appears by the connections between the carbon black particles in the paste, is determined by the density and the various parameters of the carbon particles. Therefore, the resistivity is affected by the conditions of the carbon particles and their production situations. Hence, achieving precise control and high resistivity uniformity is difficult. Unfortunately, no other materials have proper resistivity around 1 M$\Omega$/sq., which is the ideal resistivity for spark suppression.  

In 2012 at Kobe University, a novel resistive material using a carbon-sputtering technique was developed for MPGD electrodes \cite{dlc}. A thin ($\sim$100nm), uniform film was formed on the substrate using a graphite sputtering target. This film consists of amorphous carbon of an $\mathrm{sp^2}$ and $\mathrm{sp^3}$ hybrid called Diamond Like Carbon (DLC). Production was done at Be-Sputter Co., Ltd. using a large sputtering chamber, which could accommodate a substrate up to 4.5m $\times$ 1m in size. The DLC film firmly adheres to the polyimide substrate and possesses great tolerance to chemicals. Combined with photolithography, fine electrodes can be formed with precision over 10$\mu$m using the liftoff method. The resistivity, which is easily and widely controlled by varying the thickness and the doping nitrogen into the DLC film, can be controlled from 50 k$\Omega$/sq. to 3 G$\Omega$/sq. 

\begin{figure}[htbp]
\begin{center}
\includegraphics[width=10cm,clip]{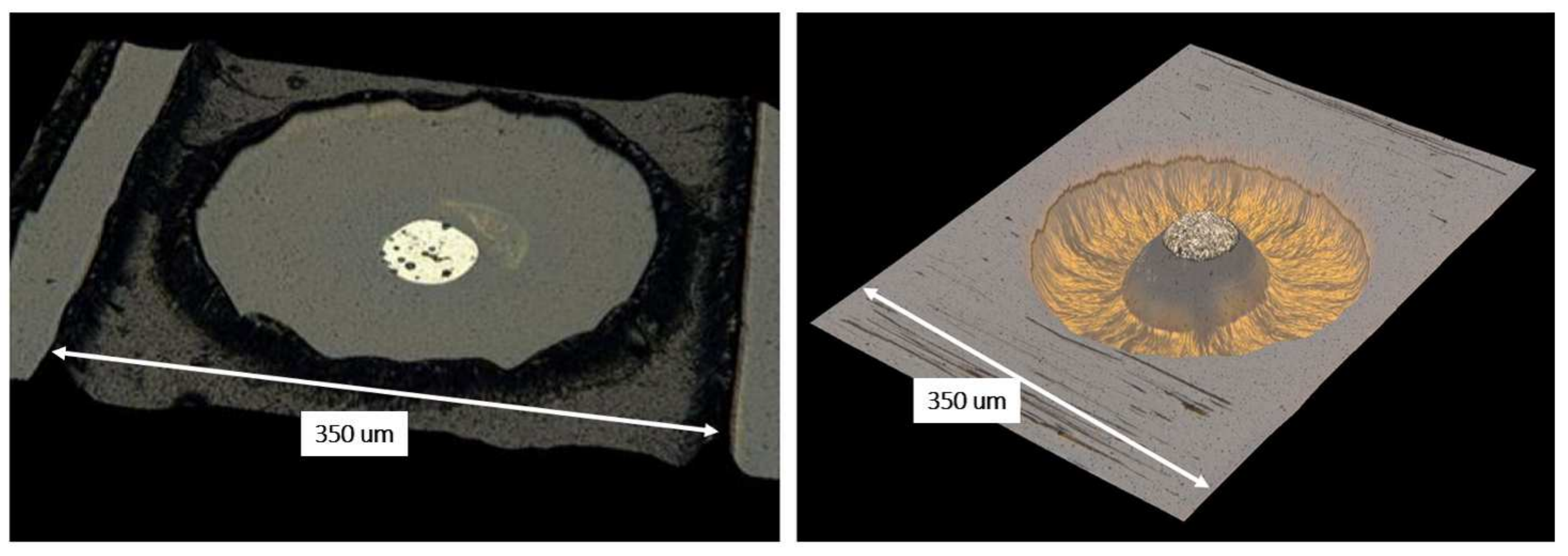}
\end{center}
\caption{Microscopic images of $\mu$-PIC pixel: Left: carbon-loaded paste, right: DLC film for resistive cathode. }
\label{pixel}
\end{figure}

This novel DLC thin film was applied to the resistive $\mu$-PIC. Fig. \ref{pixel} shows microscopic images of a $\mu$-PIC pixel. Carbon-loaded paste is used in the picture on the left and DLC thin film is used on the right. The carbon-loaded paste is about 10$\mu$m thick. The resistivity is affected by the conditions of the carbon particles in the paste. The resistivity variation in the former chambers was a factor of five in a 10 $\times$10 $\mathrm{cm^2}$ detection area. On the other hand, the thickness of the DLC thin film was determined by the sputtering time of the film growth, which is on the order of 10-100nm, which is much thinner than other detector components (10-100$\mu$m). The resistivity variation was within 20\% in a 10 $\times$ 10 $\mathrm{cm^2}$ detection area.

\begin{figure}[htbp]
\begin{center}
\includegraphics[width=12cm,clip]{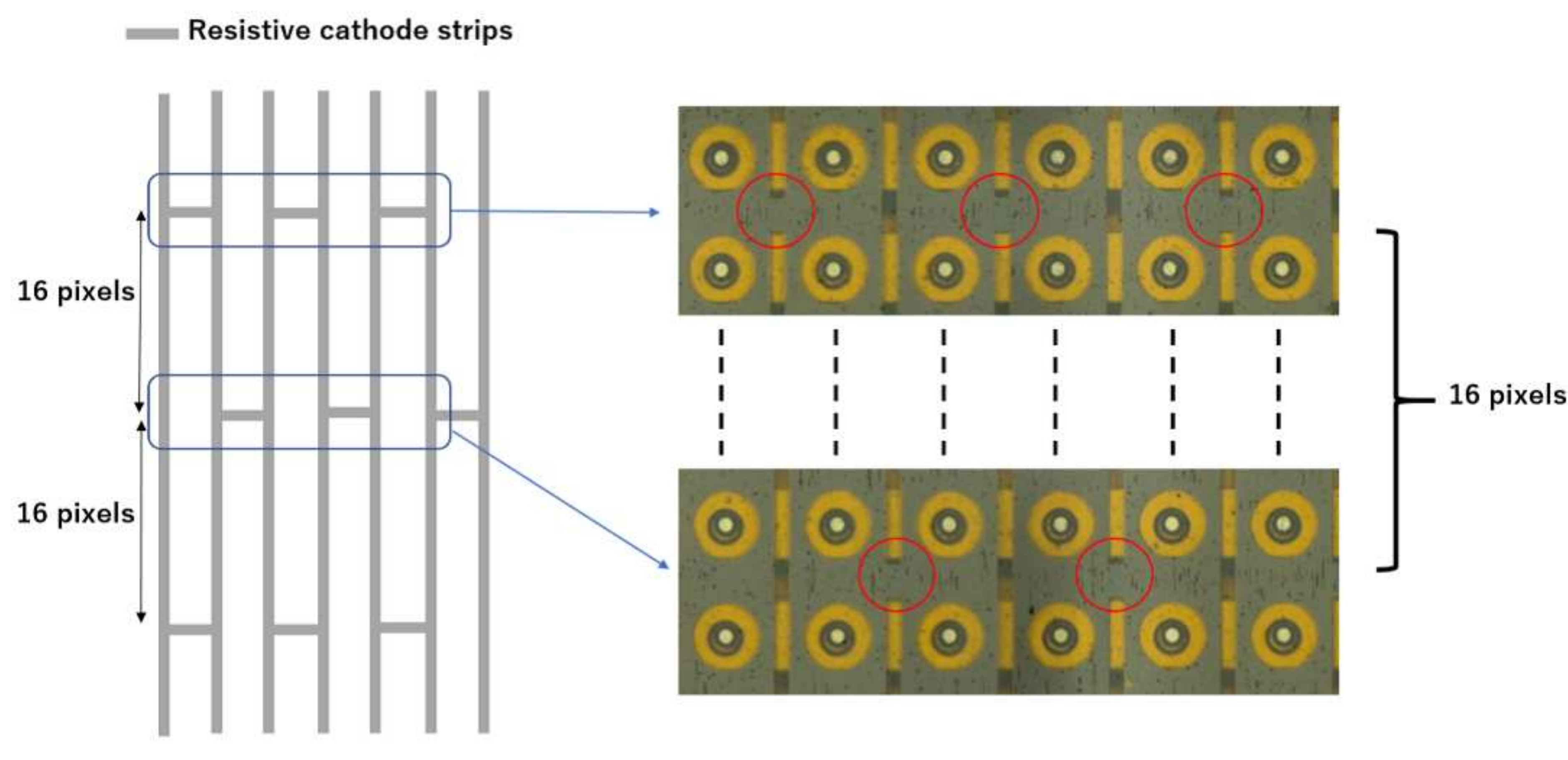}
\end{center}
\caption{Schematic view of the pattern of resistive strips and their photographs: Adjacent resistive cathode strips were connected every 16 pixels. The pattern of resistive strips has ladder-like shape.}
\label{ladder}
\end{figure}

The cathode voltage is supplied from both ends of the resistive cathode strips. In high-rate radiation environments, the voltage supply probably cannot catch up with the voltage drop at the center of the strips. Therefore, adjacent resistive strips were connected every 16 pixels. Fig. \ref{ladder} shows a schematic view of the pattern of the resistive cathode strips and their photographs. The pattern of resistive strips has a ladder-like shape. The readout is not affected by this pattern because it is isolated from the resistive strips.

\subsubsection{Accurate alignment of electrodes}
\begin{figure}[htbp]
\begin{center}
\includegraphics[width=12cm]{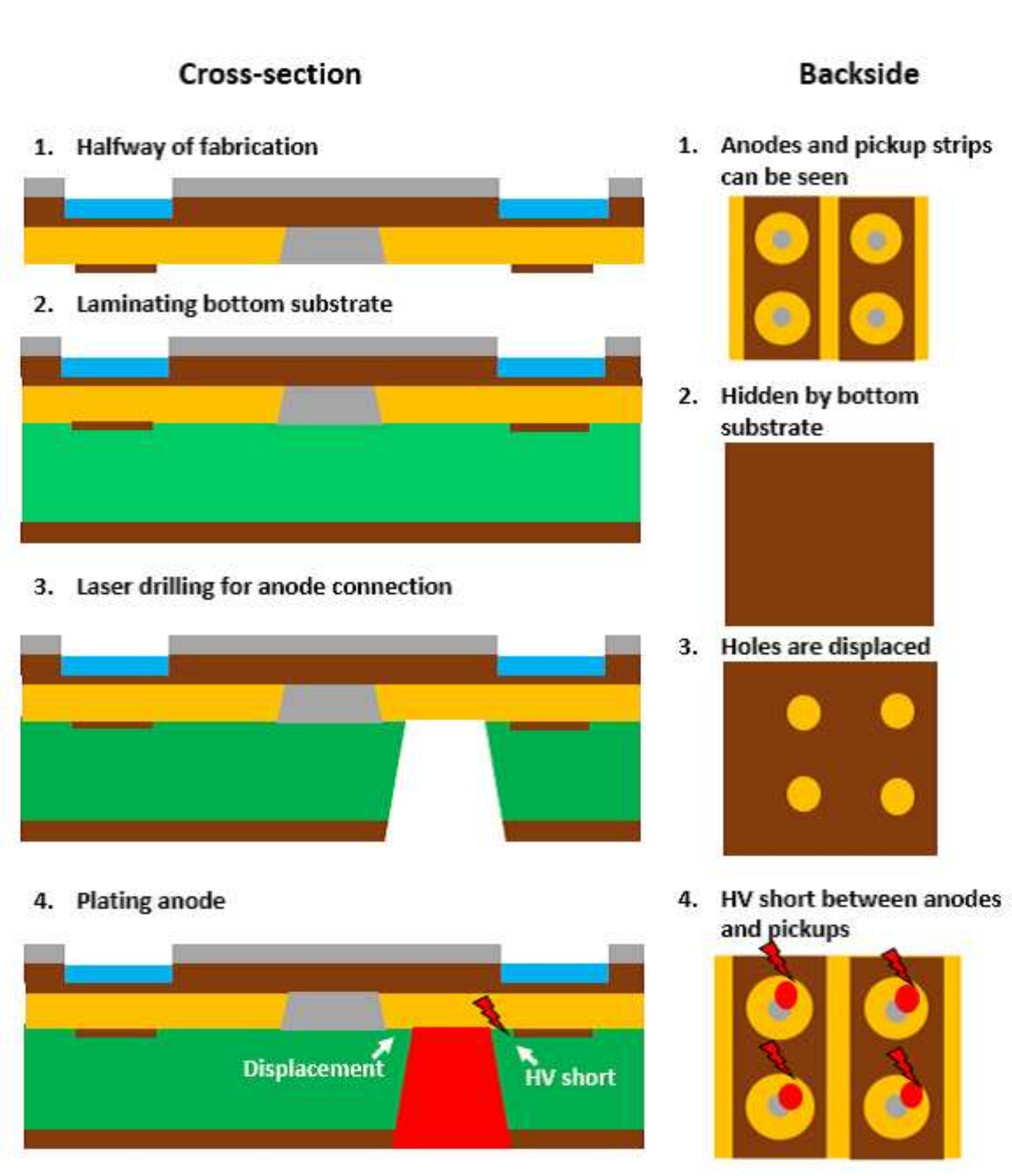}
\end{center}
\caption{Former fabrication process that caused mis-alignment in anodes: Backside views are also shown from cross-section. In laser-drilling process (2, 3), top anodes were hidden by bottom substrate. Therefore, some anodes were displaced near pickups. Such a displacement caused voltage shorts between anode and pickup.}
\label{misalignment}
\end{figure}

The resistive $\mu$-PIC consists of electrodes on double layers. This process of making electrodes is more complicated than the conventional $\mu$-PIC. Accurate alignment between the two layers as well as between the anode and cathode electrodes is critical during fabrication. However, no such technique has been established. Here, we describe our solution to this problem. Fig. \ref{misalignment} shows a part of the fabrication process of our previous chambers. (1) First, pickup strips are formed on the backside of the top substrate, where anodes were filled with nickel. (2) Next, the bottom substrate is laminated. It has a copper layer that is later  formed into anode strips. (3) Holes are made by laser drilling, which are filled with metal by the through-hole plating process, and (4) then anode strips are connected to the top anodes. However, some holes were displaced and mis-connected to the nearby cathode pickups, caused by the difficulties in the manual alignment between the top and bottom substrates. Because holes were drilled from the backside without being able to see the anode pattern on the front side, correctly matching the positions was difficult. Such a displacement caused a short circuit between the anode and the pickup. Only a part of the detection area could be operated in our first trial.

\begin{figure}[htbp]
\begin{center}
\includegraphics[width=12cm,clip]{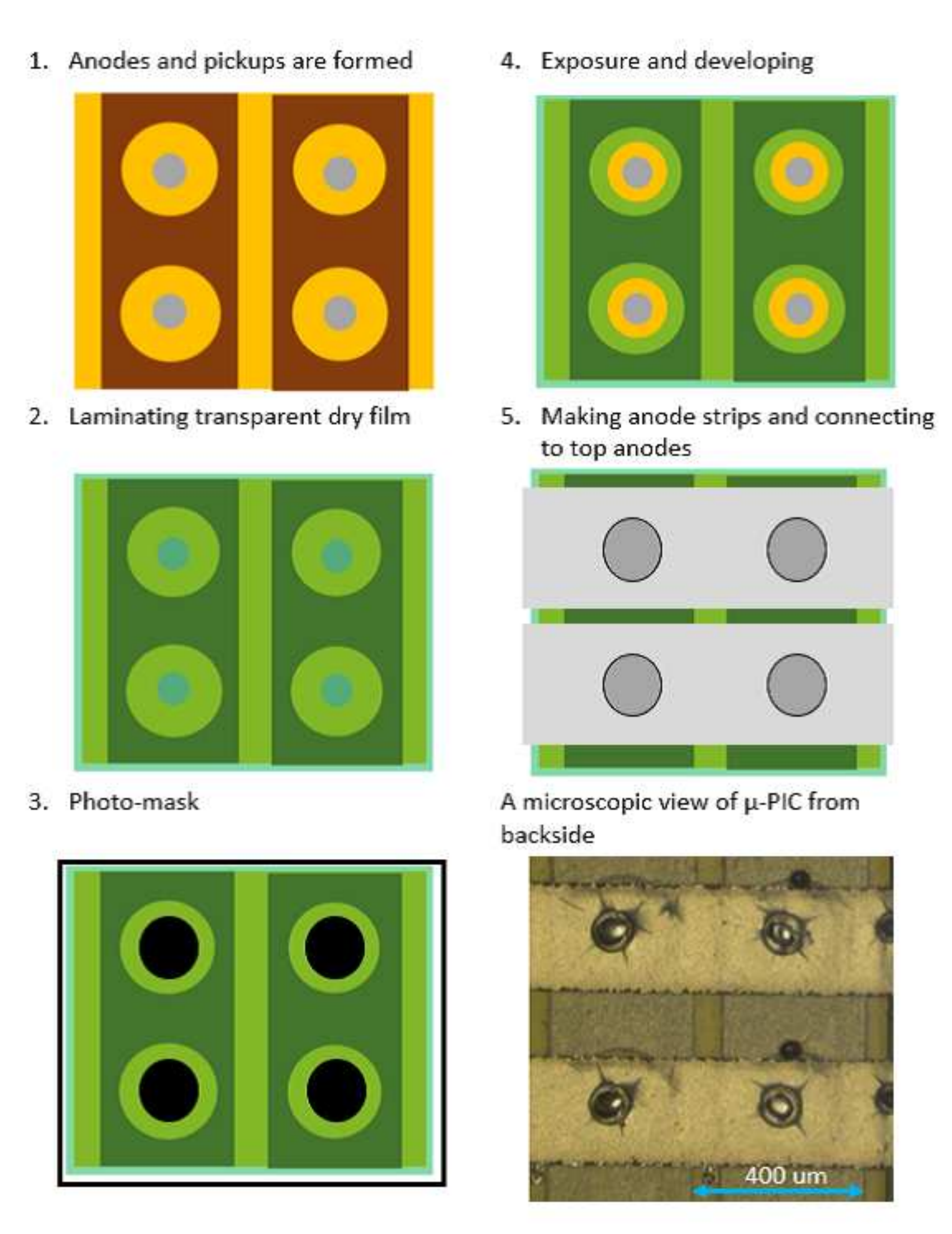}
\end{center}
\caption{New process for making anodes: All were well placed by transparent dry film and photolithography. }
\label{transparent}
\end{figure}

\begin{figure}[htbp]
\begin{center}
\includegraphics[width=12cm,clip]{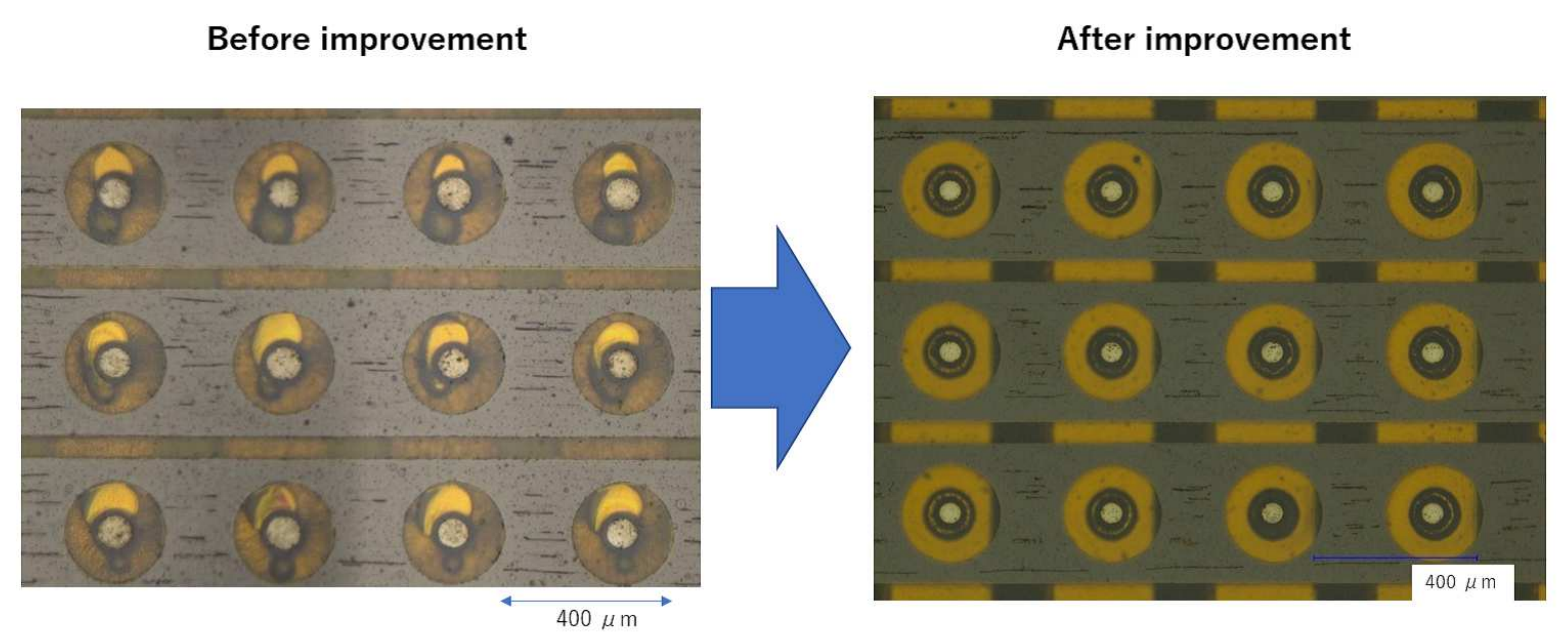}
\end{center}
\caption{Microscopic images of former $\mu$-PIC (left) and new $\mu$-PIC (right): For former, top anodes (white patterns) and bottom anodes (yellow patterns) are mis-aligned, but they are well aligned for new $\mu$-PIC.}
\label{alignment}
\end{figure}

To solve this problem, bottom substrate's material was replaced by a dry film, and the hole-making processes were changed to photolithography. Fig. \ref{transparent} shows the new process. (2) Since the dry film is transparent, the top anodes are seen from the backside. (3-5) Holes were correctly aligned by photolithography and the top anodes and the anode strips were connected properly in each pixel. Fig. \ref{alignment} shows the microscopic images of both the former and the new $\mu$-PIC. For the former, the top anodes (white patterns) and bottom anodes (yellow patterns) are mis-aligned, but they are well aligned for the new $\mu$-PIC. Due to this improvement, no failures were found in any of the pixels in the 10 $\times$ 10$\mathrm{cm^2}$ detection area.
 
\subsubsection{Novel approaches for making a compact detector}
\begin{figure}[htbp]
\begin{center}
\includegraphics[width=10cm,clip]{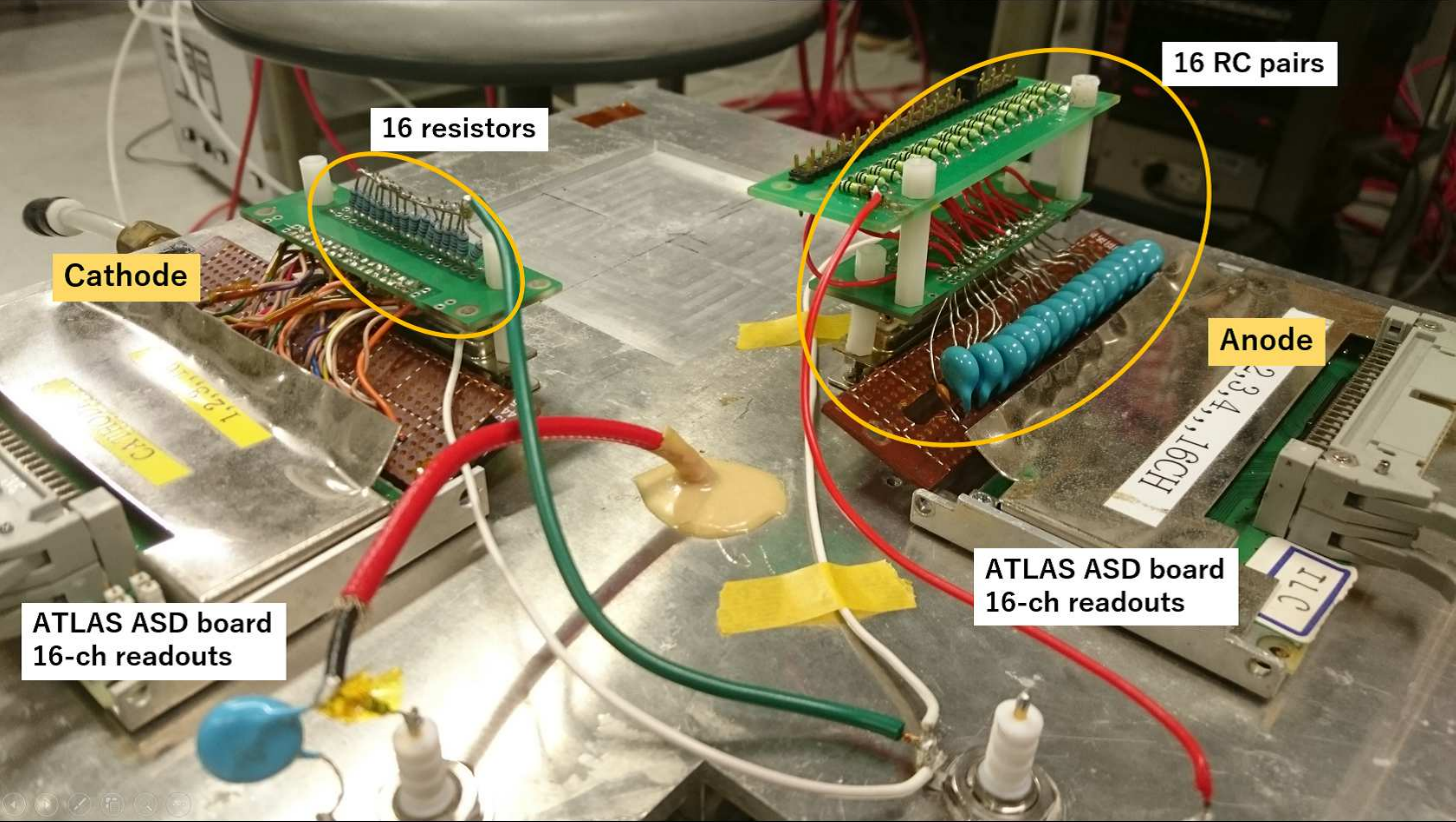}
\end{center}
\caption{RC circuits and readout boards mounted on former design: 16 pairs of RC circuits were mounted on anode. Capacitors were not needed for cathode because it obtains induced charges. }
\label{oldupic}
\end{figure}

If high voltage is applied to an electrode, a resistor and a capacitor pair is mandatory for a signal readout. The resistor is needed for the HV bias, and the capacitor is needed for the AC coupling. Fig. \ref{oldupic} shows the gas package of the former chamber, where resistors are set for the HV bias and capacitors for the AC coupling. In the former design, 16 strips were combined into a 1-channel readout for simplification, and only 16 RC pairs were needed for the anode. However, a large space is needed for setting these parts. Mounting them on a board is the general way for individually reading the signals from all the strips. On the other hand, $\mu$-PIC might become a very compact detector. When a negative HV is supplied to the cathode and the anode voltage is set to 0 V, these parts can be removed. However, this mode is unstable due to the high electric field near the edge of the pickup strips \cite{reupic}. It is preferable to supply HV to the anode for stable operation.

\begin{figure}[htbp]
\begin{center}
\includegraphics[width=10cm,clip]{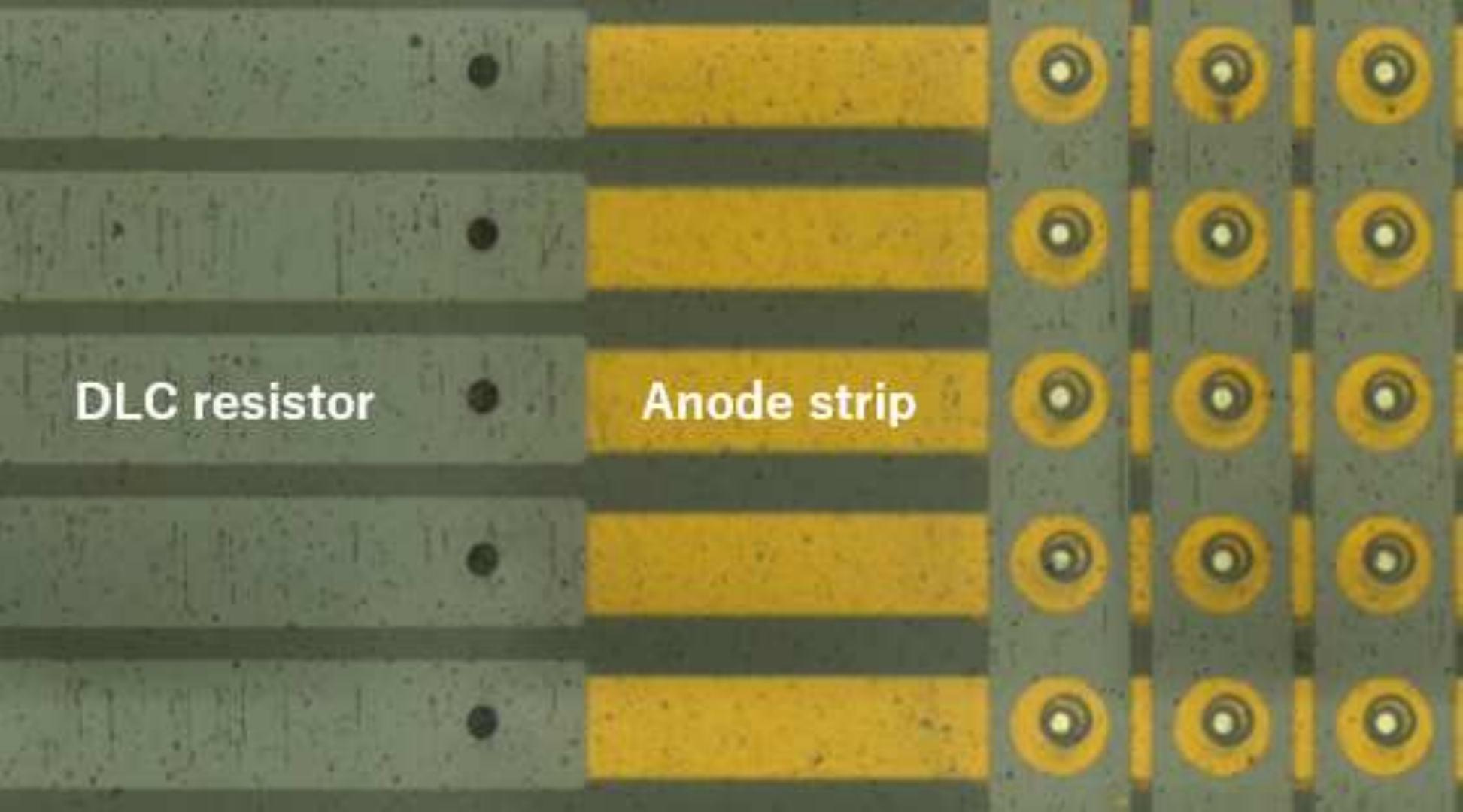}
\end{center}
\caption{Microscopic image of $\mu$-PIC and DLC resistors: Yellow lines are anode strips connected to each resistor.}
\label{anoderesistor}
\end{figure}

\begin{figure}[htbp]
\begin{center}
\includegraphics[width=11cm,clip]{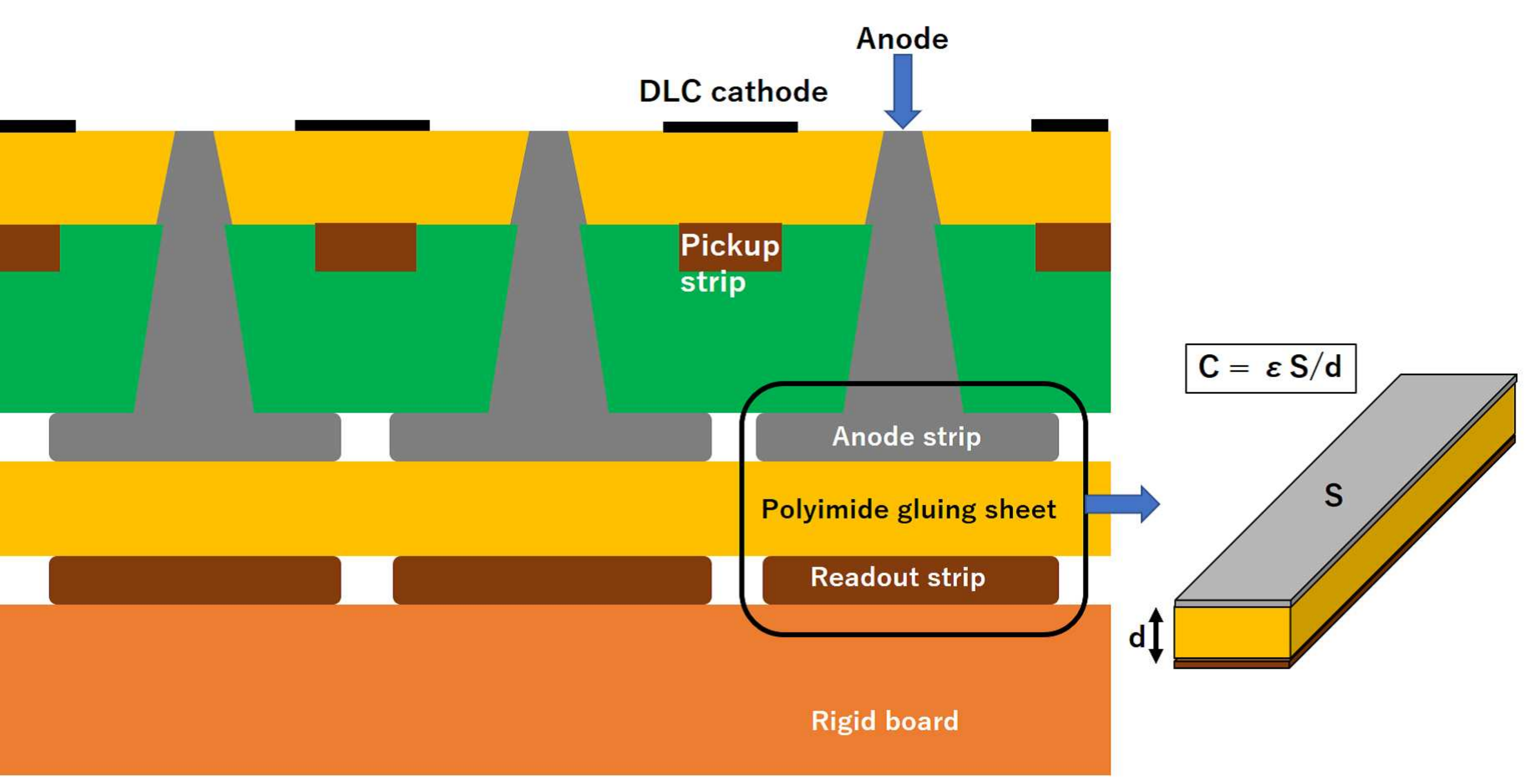}
\end{center}
\caption{Schematic cross section of new prototype on rigid board: Polyimide gluing sheet and an anode strip and a readout strip form a capacitor. Due to this structure, all capacitors for anode readout were removed from our $\mu$-PIC.}
\label{capacitance}
\end{figure}

\begin{figure}[htbp]
\begin{center}
\includegraphics[width=13cm,clip]{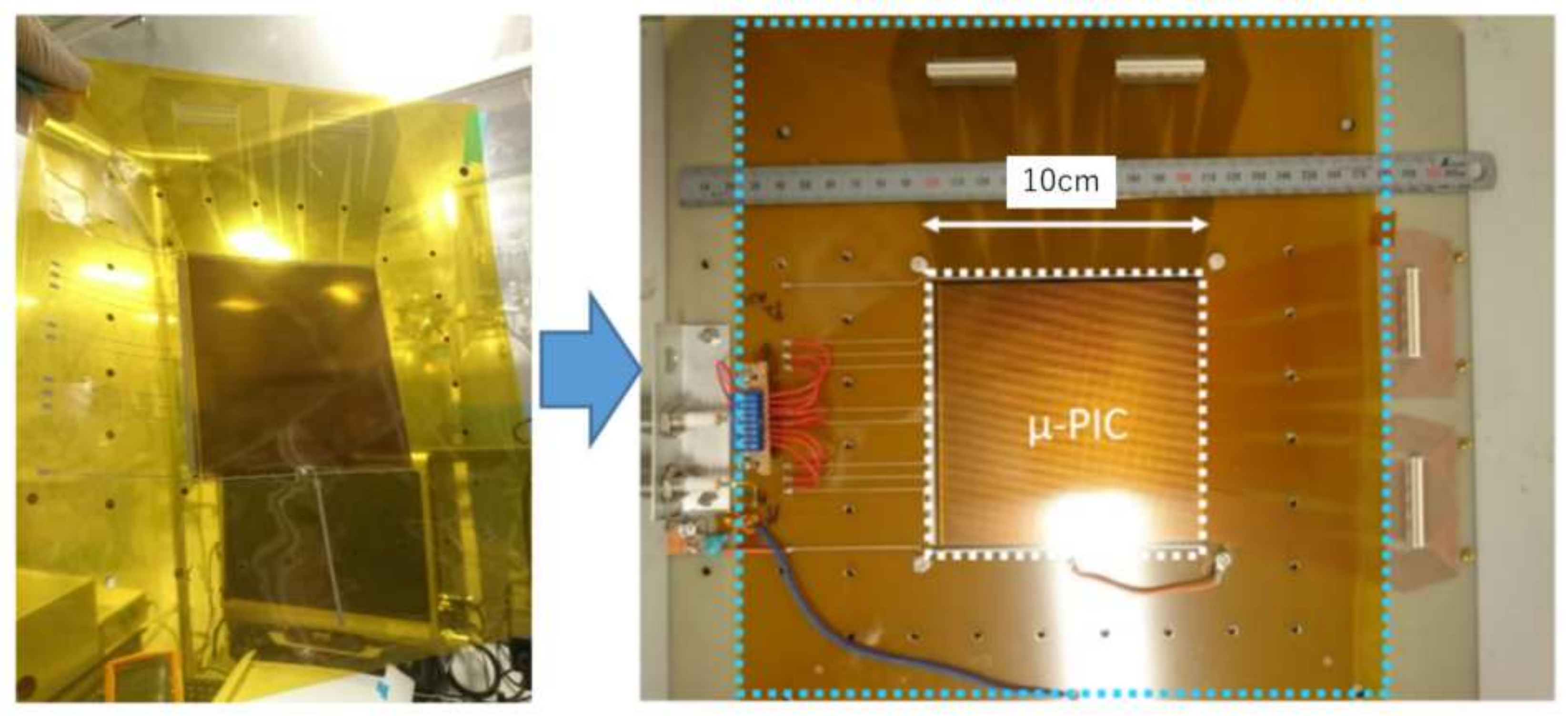}
\end{center}
\caption{Left: $\mu$-PIC substrate before being glued to rigid PCB board. Right: $\mu$-PIC on rigid board. All of the equivalent circuits for readout are in this board. Signals can be directly obtained from on-board connectors.}
\label{glue}
\end{figure}

The best solution is provided by the DLC and PCB techniques. First, HV bias resistors for the anodes were formed on the detector's surface by carbon sputtering. The microscopic image of the DLC resistors for HV bias is shown in Fig. \ref{anoderesistor}. Each DLC resistor is connected to the anode strip via a through hole (Fig. \ref{upiccrosssection}). Second, Fig. \ref{capacitance} shows a cross-section of the $\mu$-PIC. The substrate was glued to a rigid board that has readout strips for anodes. Readout strips are placed parallel to the anode strips of the $\mu$-PIC. Capacitors are formed by a polyimide gluing sheet and anode and readout strips, and the capacitance is about a 22 pF/strip, which is 300 $\mu$m-wide with 10 cm long strips. Although this value is lower than that we want, it is adequate for practical use. Thus, no additional RC circuits are needed for our $\mu$-PIC. Fig. \ref{glue} shows the $\mu$-PIC substrate that is glued on a rigid board. Signals can be obtained directly from the on-board connectors.

\subsection{Fabrication-process}
\begin{figure}[htbp]
\begin{center}
\includegraphics[width=14cm,clip]{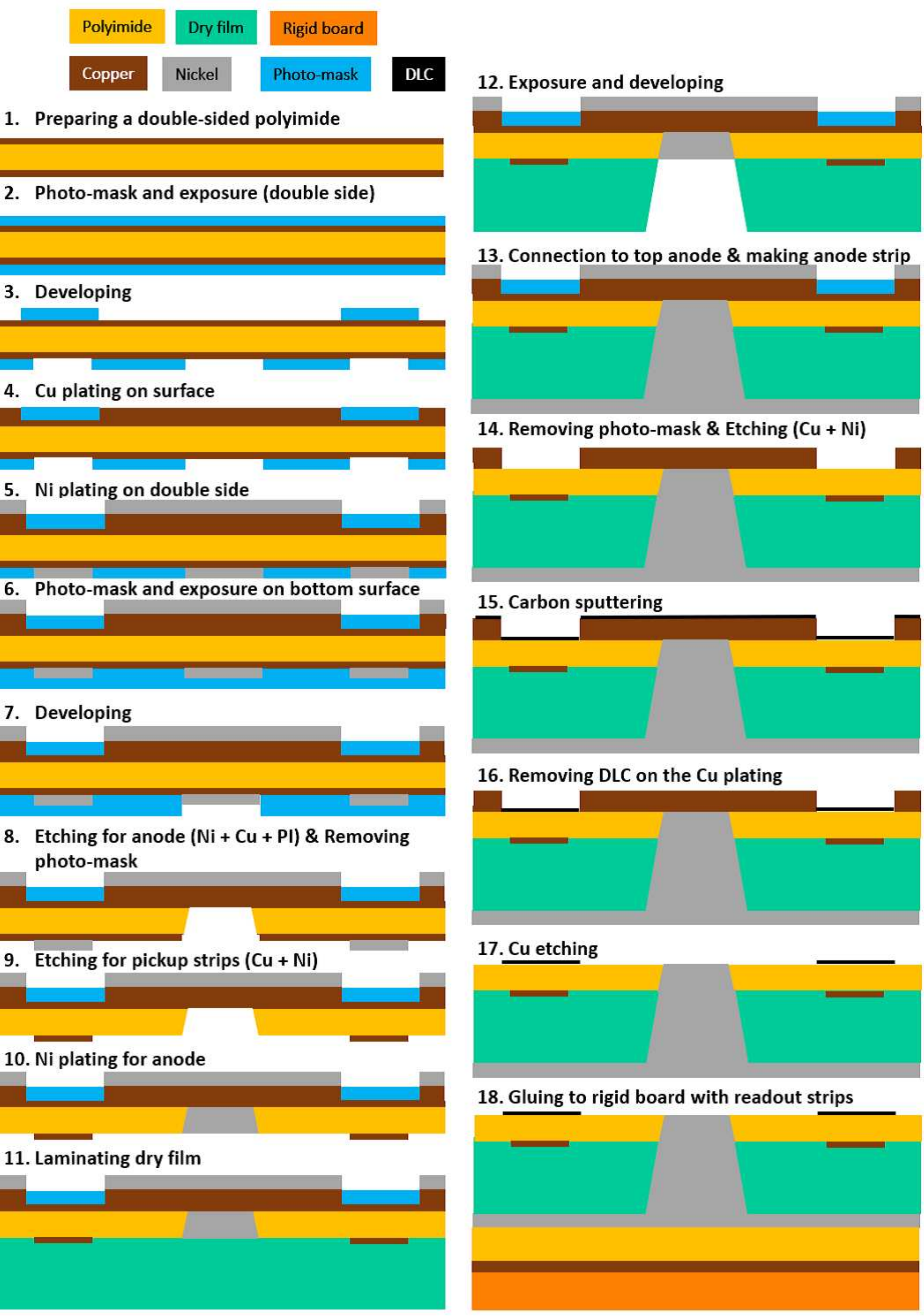}
\end{center}
\caption{Fabrication process of new $\mu$-PIC}
\label{fabrication}
\end{figure}

The resistive $\mu$-PIC was fabricated by Raytech Inc. Fig. \ref{fabrication} shows its fabrication-process. Mask patterns for the cathode and pickup strips were respectively formed on the top and the bottom, on the surface of the 25 $\mu$m-thick Flexible Print Circuit (FPC) (Steps 1-5 in the figure). Mask patterns for the DLC resistors for the HV bias are also formed in this process (not shown in this figure). The substrate at the center of the circles on the cathode is etched, and the anodes are formed by plating nickel in the etched hole (Steps 6-10). The dry film is laminated on the bottom surface (Step 11). Holes corresponding to the position of each anode are made by photolithography (Step 12). Anode strips are formed perpendicular to the cathode strips and connected to the top anodes (Step 13). Resistive cathodes are made by carbon sputtering (Steps 14-17). Finally, the detector is glued to the rigid PCB that has readout strips for the anodes (Step 18).

\subsection{Detector properties}
\begin{figure}[h]
\centering
\includegraphics[width=10cm]{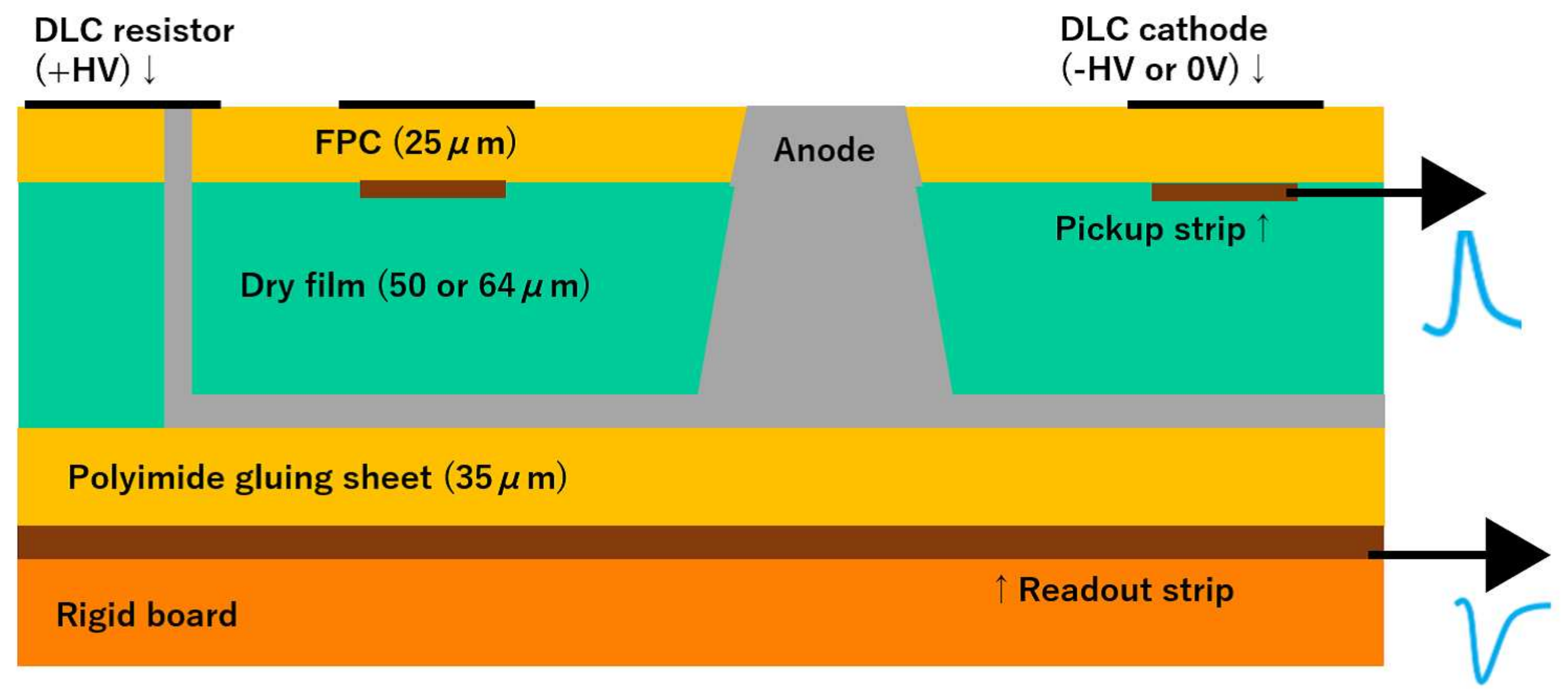}
\caption{Cross-section of improved $\mu$-PIC}
\label{upiccrosssection}
\end{figure}

\begin{figure}[htbp]
\begin{center}
\includegraphics[width=10cm,clip]{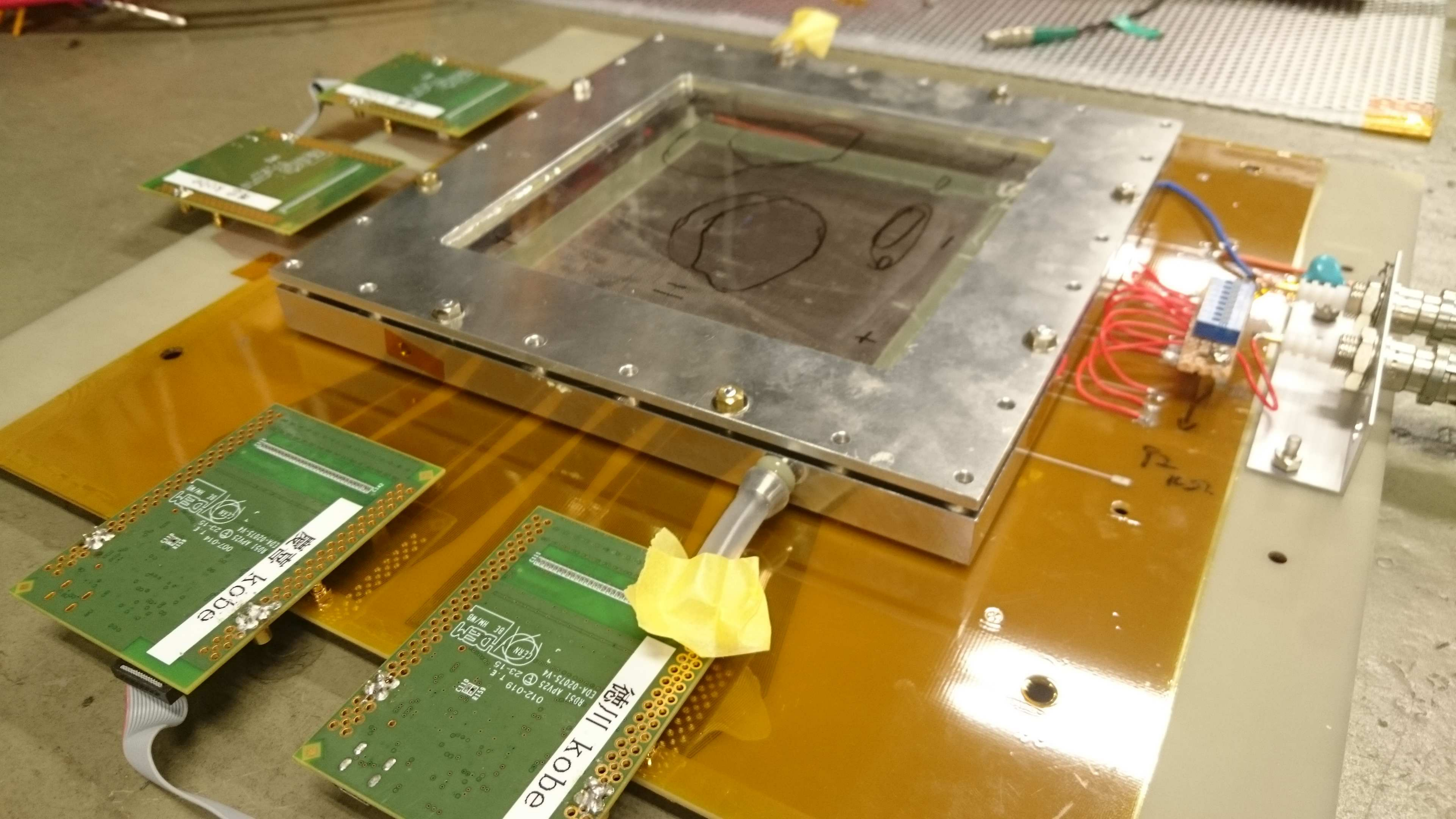}
\end{center}
\caption{New $\mu$-PIC in gas package: APV25 front-end chips are attached to connectors on $\mu$-PIC board.}
\label{upicpackage}
\end{figure}

Fig. \ref{upiccrosssection} shows a schematic view of the cross-section of the new $\mu$-PIC and describes the connection between a DLC resistor and an anode strip. Fig. \ref{upicpackage} shows the detector in a gas package. APV25 front-end chips \cite{apv} are attached to the detector by connectors. Table \ref{upicparameter} shows its parameters where four chambers are classified into two types. For RC37 and RC38, the anode diameter is 70-75 $\mu$m, the dry film is 50 $\mu$m thick, and the surface cathode resistivity is $\sim$180 k$\Omega$/sq. For RC41 and RC42, the anode diameter is 55-60 $\mu$m, the dry film is 64 $\mu$m thick, and the surface cathode resistivity is $\sim$600 k$\Omega$/sq. 

\begin{table*}[htbp]
\begin{center}
\begin{tabular}{ccccc}
\hline
Chamber name & \multicolumn{2}{c}{Diameter size} & Thickness of & Resistivity\\
\cline{2-3}
	& Anode[$\mu$m] & Cathode[$\mu$m] & dry film[$\mu$m] & [k$\Omega$/sq.] \\
\hline
RC37, RC38 & 70-75 & 240-250 & 50 & 180	\\
RC41, RC42 & 55-60 & 240-250 & 64 & 600 \\
\hline
\end{tabular}
\end{center}
\caption{Parameters of produced detectors.}
\label{upicparameter}
\end{table*}

\section{Basic performances of the new $\mu$-PIC}
\subsection{Gas gains}
\begin{figure}[h]
\centering
\includegraphics[width=10cm,clip]{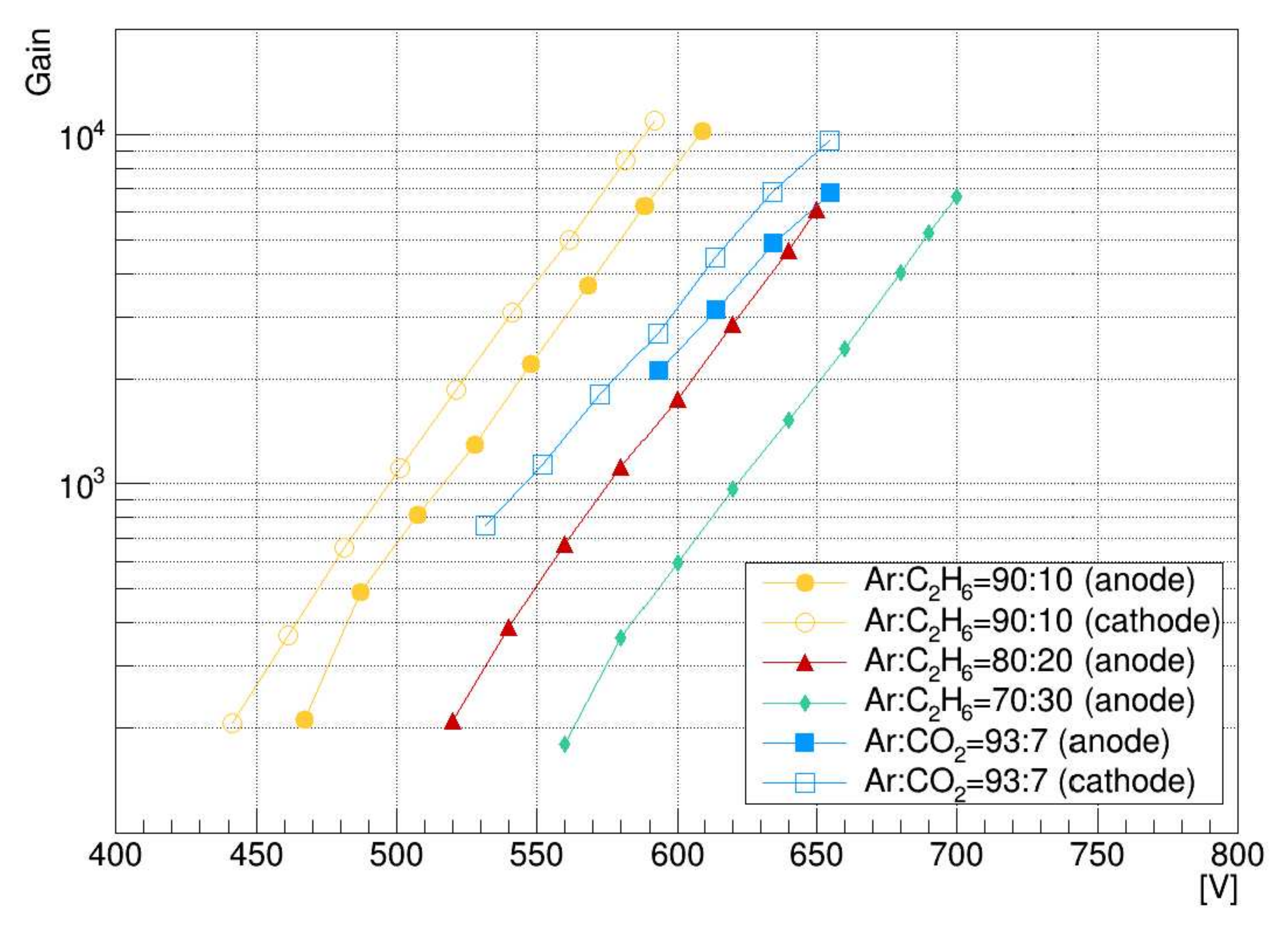}
\caption{Gas gains for various gas mixtures as a function of amplification voltage}
\label{gain}
\end{figure}

\begin{figure}[htbp]
\begin{tabular}{cc}
\begin{minipage}{0.5\hsize}
\begin{center}
\includegraphics[width=7cm]{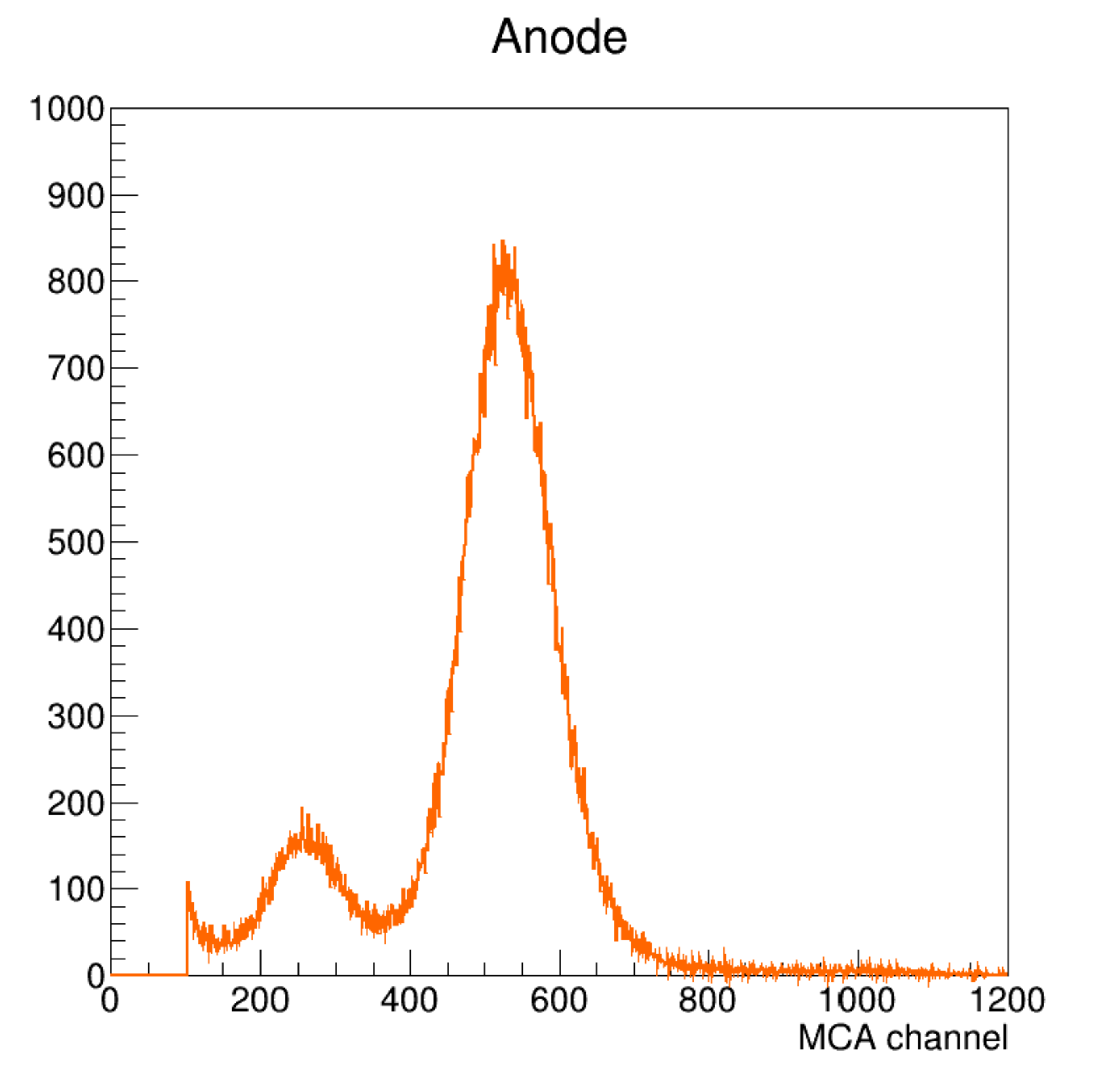}
\end{center}
\end{minipage}
\begin{minipage}{0.5\hsize}
\begin{center}
\includegraphics[width=7cm]{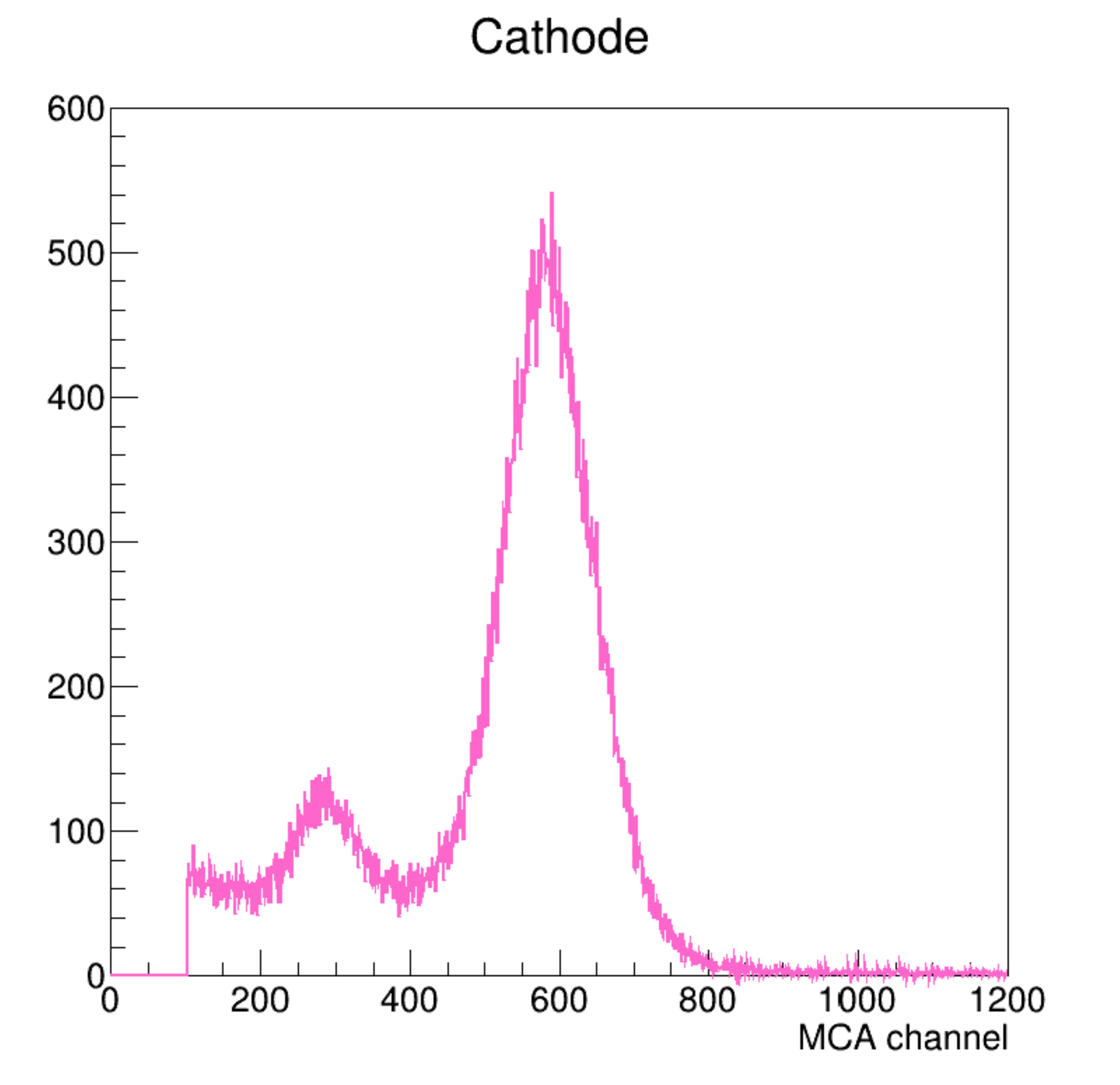}
\end{center}
\end{minipage}
\end{tabular}
\caption{Charge spectrum of $\mathrm{{}^{55}Fe}$ X-rays obtained from anodes (left) and cathodes (right) at identical 32 $\times$ 32 pixels irradiation area: Gas gain is $\sim$1000.}
\label{spe}
\end{figure}

Gas gains were measured using a $\mathrm{{}^{55}Fe}$ 5.9keV X-ray source with Ar/$\mathrm{C_2H_6}$ (9:1, 8:2, 7:3) and Ar/$\mathrm{CO_2}$ (93:7) gas mixtures. The drift field was set to 3 kV/cm for 3 mm drift gap. The cathode voltage was set to 0V. Gain curves were obtained by varying the anode voltages. ATLAS ASDs with an analogue output (charge gain: 0.8V/pC) \cite{asd}) were used for the preamplifier. The spectrum of the signals was measured by a Multi Channel Analyzer (MCA8000D). For the Ar/$\mathrm{C_2H_6}$ (9:1) and Ar/$\mathrm{CO_2}$ (93:7) gas mixtures, the signals from the cathodes were also measured. Fig. \ref{gain} shows the measured gas gains as a function of the amplification voltage. Gas gains exceeding 6000 were achieved with all the gas mixtures. The maximum achievable gain that exceeded $\mathrm{10^4}$ was obtained with the Ar/$\mathrm{C_2H_6}$ (9:1) and Ar/$\mathrm{CO_2}$ (93:7) gas mixtures. The signal charge of the cathodes was larger than those of the anodes. One reason for this result is that the charges decreased due to the anode's low capacitance. Fig. \ref{spe} shows the charge spectrum obtained from the anodes (left) and the cathodes (right), where 32 $\times$ 32 pixels were irradiated by collimated X-rays. The main peak corresponding to a photoelectric peak of 5.9keV and an escape peak (2.7keV) are clearly seen, and the resolution is $\sim$20\% (FWHM) for both readouts.

\subsection{Uniformity of gas gains}
\begin{figure}[htbp]
\begin{tabular}{cc}
\begin{minipage}{0.5\hsize}
\begin{center}
\includegraphics[width=7cm]{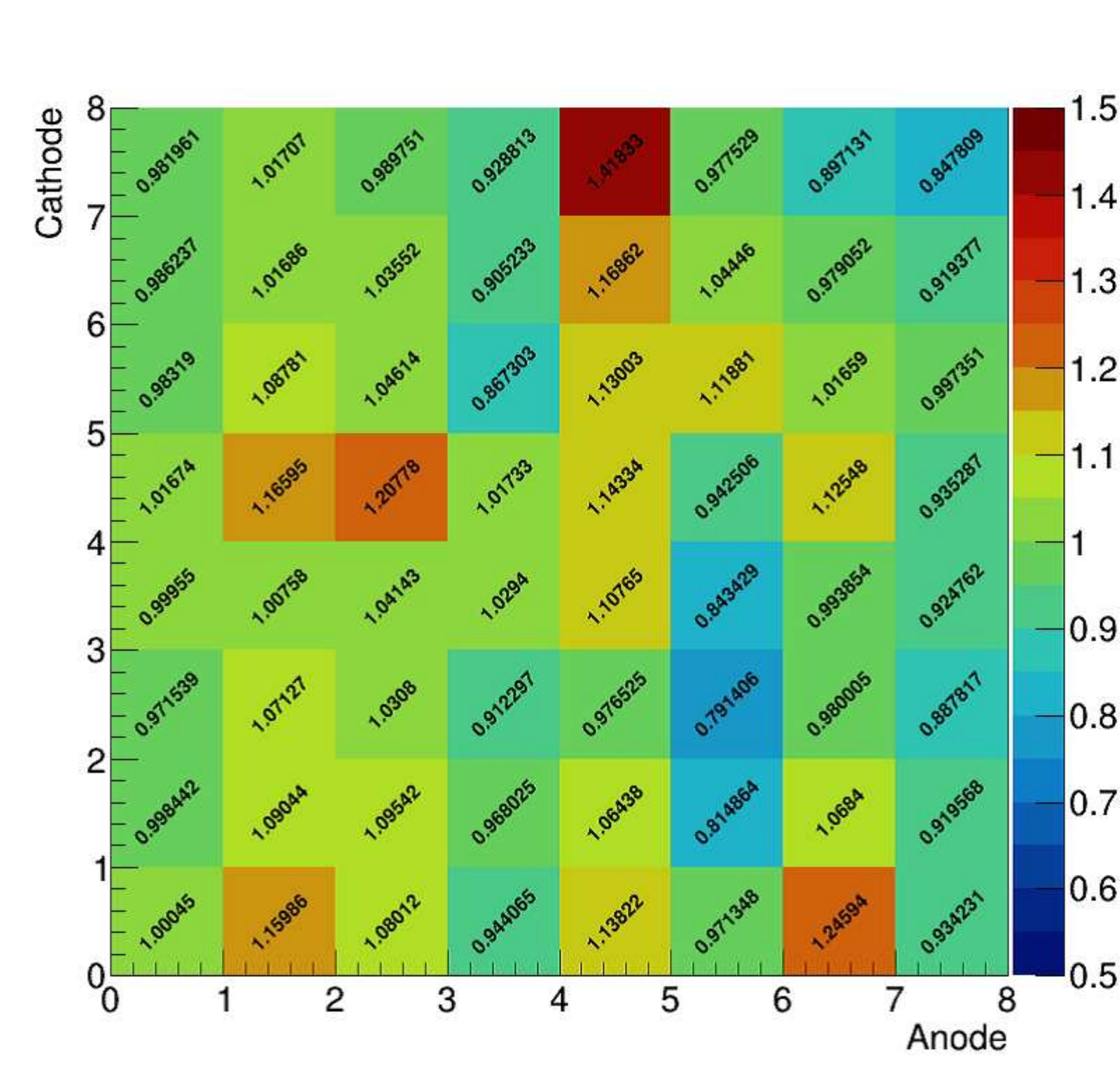}
\end{center}
\end{minipage}
\begin{minipage}{0.5\hsize}
\begin{center}
\includegraphics[width=7cm]{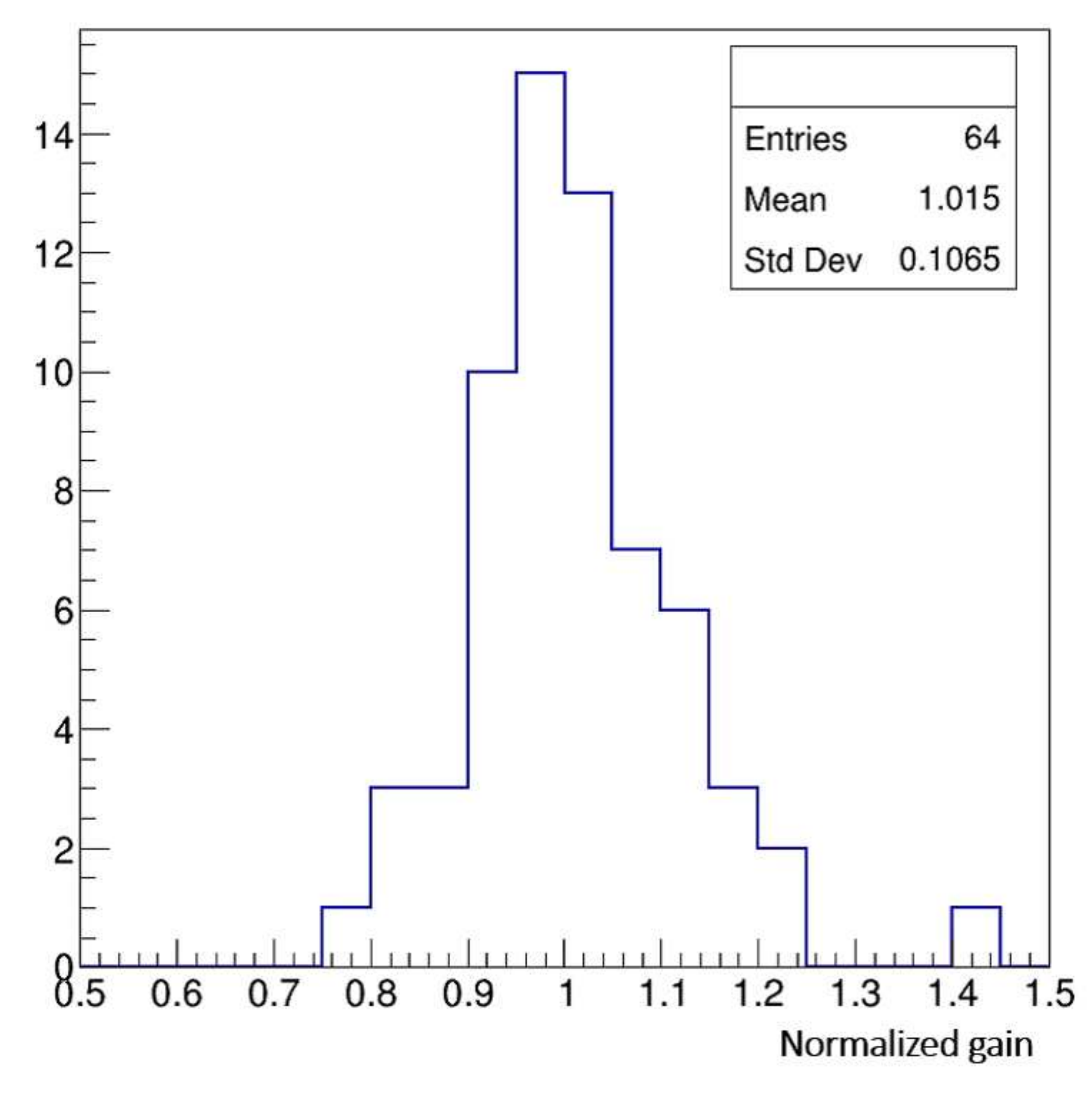}
\end{center}
\end{minipage}
\end{tabular}
\caption{Gain map (left) and its distribution (right) obtained from anodes}
\label{mapano}
\end{figure}

\begin{figure}[htbp]
\begin{tabular}{cc}
\begin{minipage}{0.5\hsize}
\begin{center}
\includegraphics[width=7cm]{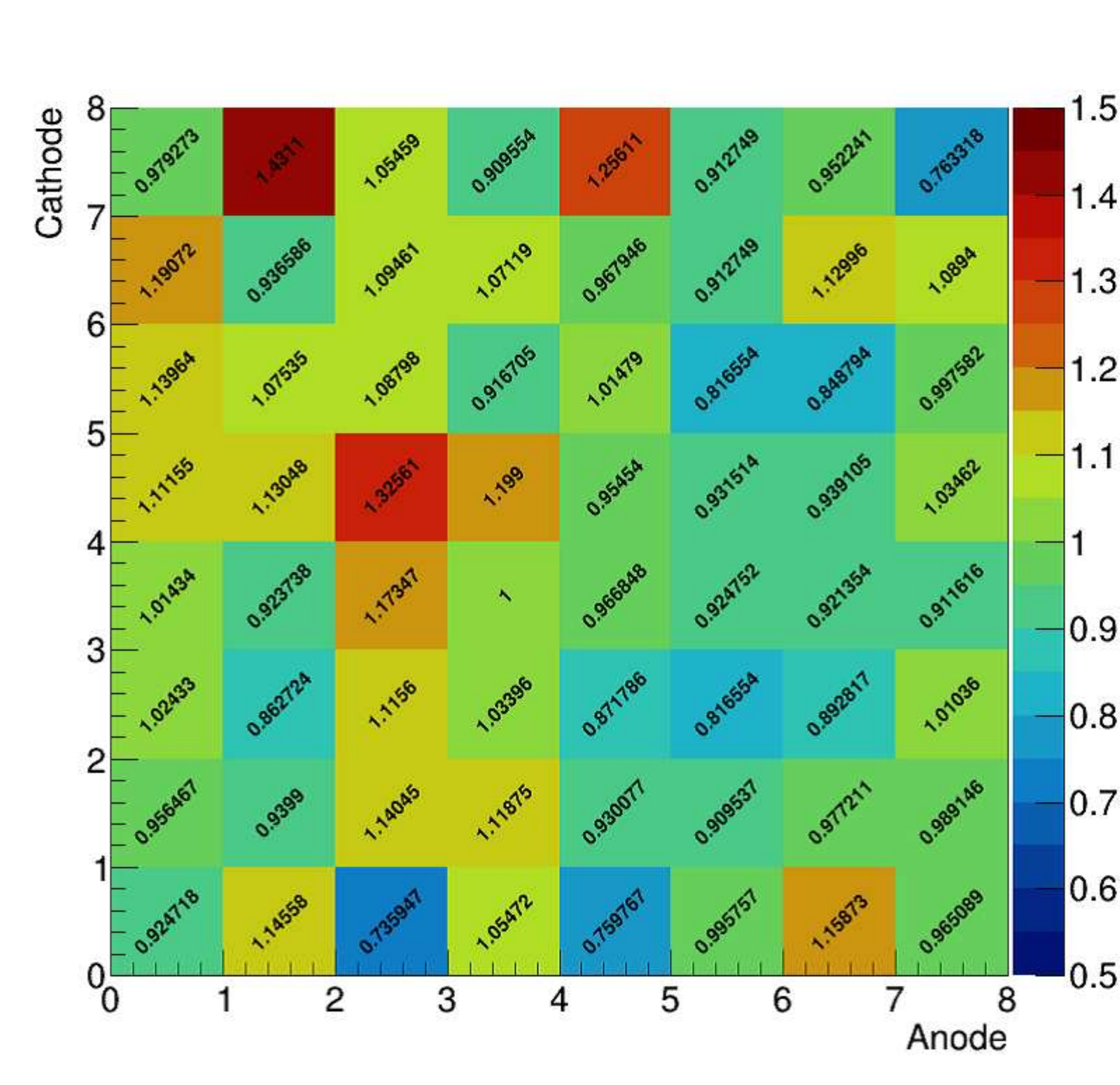}
\end{center}
\end{minipage}
\begin{minipage}{0.5\hsize}
\begin{center}
\includegraphics[width=7cm]{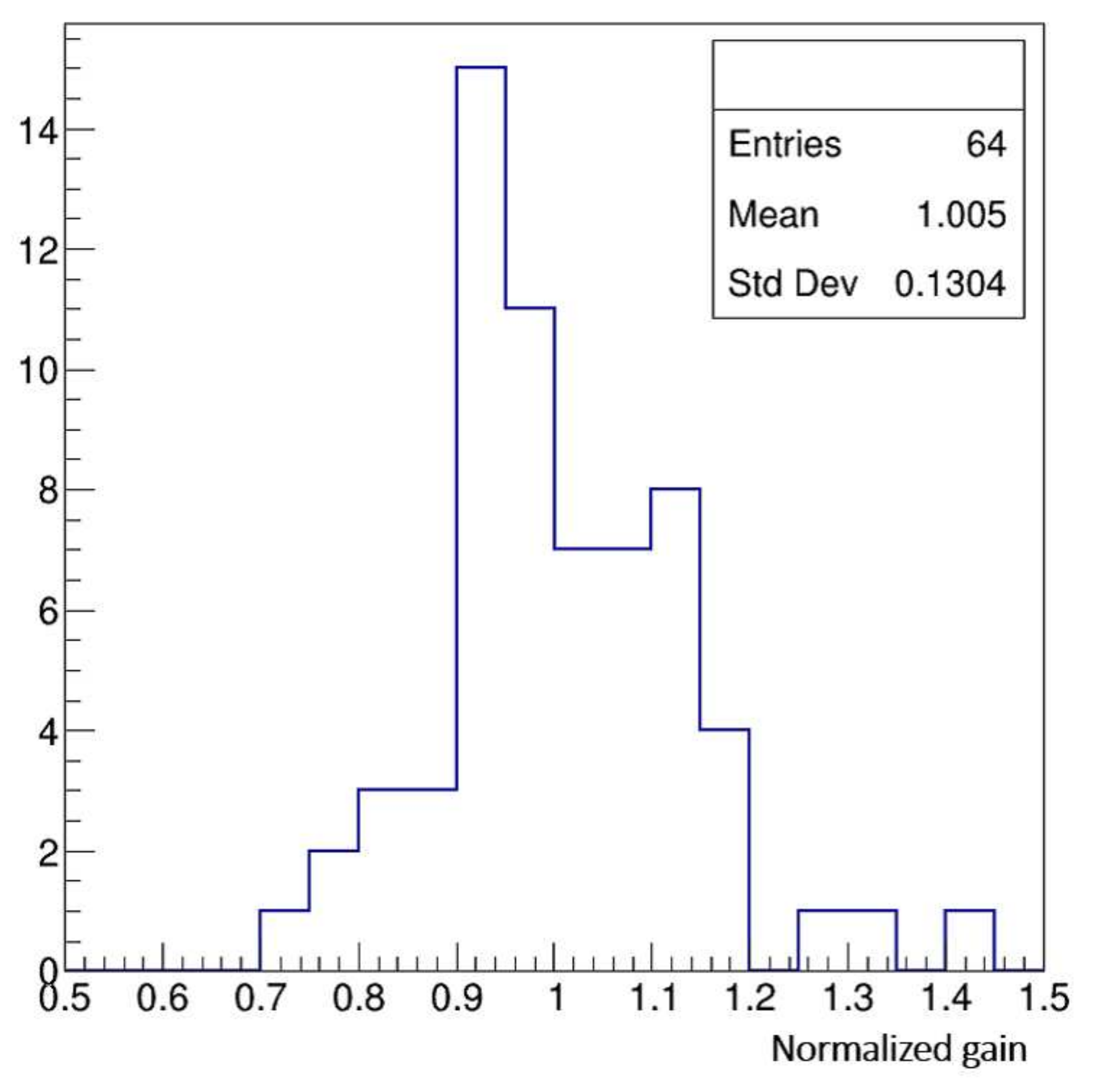}
\end{center}
\end{minipage}
\end{tabular}
\caption{Gain map (left) and its distribution (right) obtained from cathodes}
\label{mapcatho}
\end{figure}

The uniformity of the gas gains was tested over the entire detection area, which was divided into 8 $\times$ 8 areas that were each 32 $\times$ 32 pixels. The map and the distribution of the gains obtained from the anodes and the cathodes are respectively shown in Fig. \ref{mapano} and Fig. \ref{mapcatho}. Gas gains are normalized by the medium value of 1080 for the anode and 1090 for the cathode. Except for a few points, the gas gain variations do not exceed $\pm$30\%.

\subsection{X-ray image}
\begin{figure}[htbp]
\begin{tabular}{cc}
\begin{minipage}{0.5\hsize}
\begin{center}
\includegraphics[width=6cm]{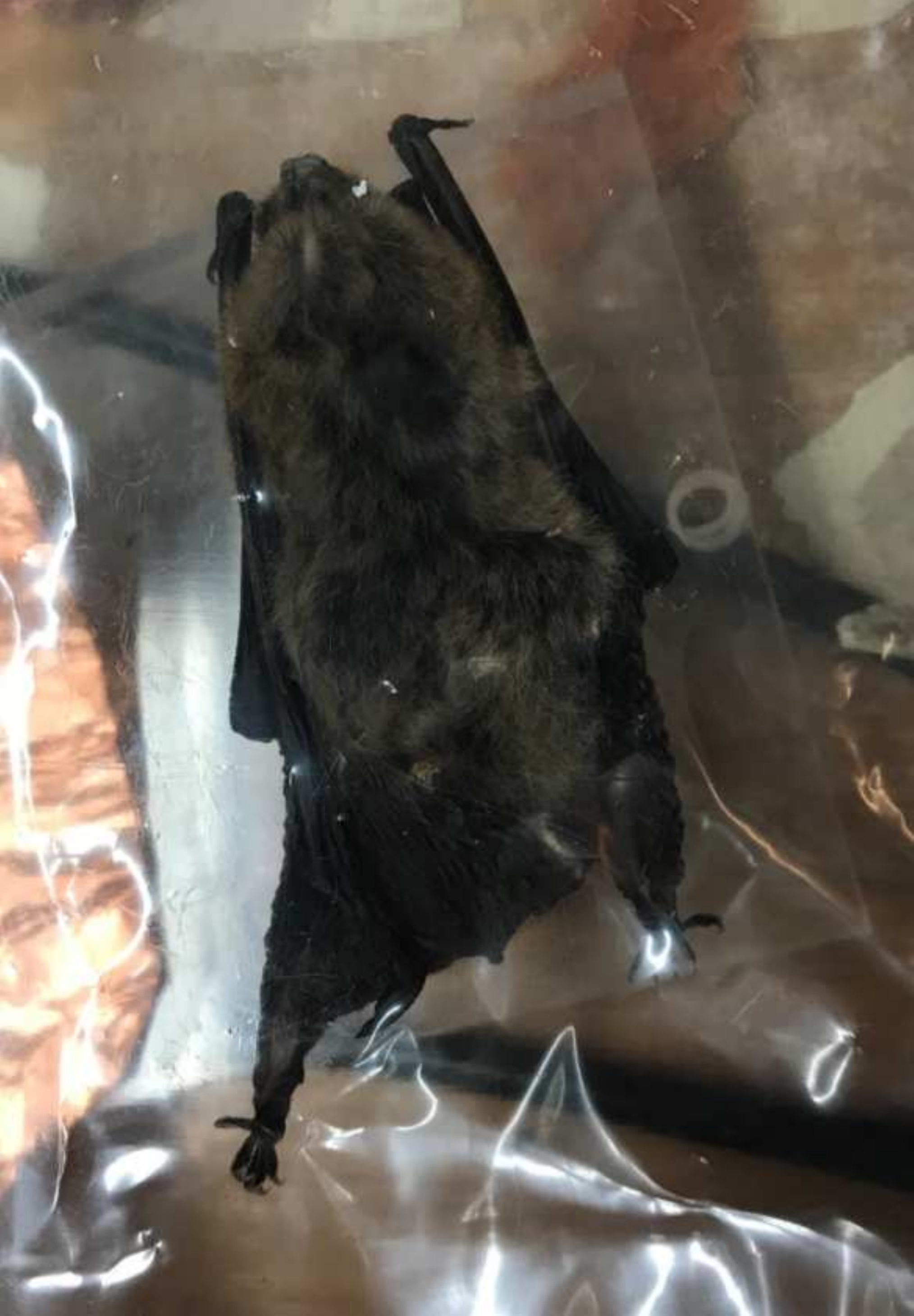}
\end{center}
\end{minipage}
\begin{minipage}{0.5\hsize}
\begin{center}
\includegraphics[width=6cm]{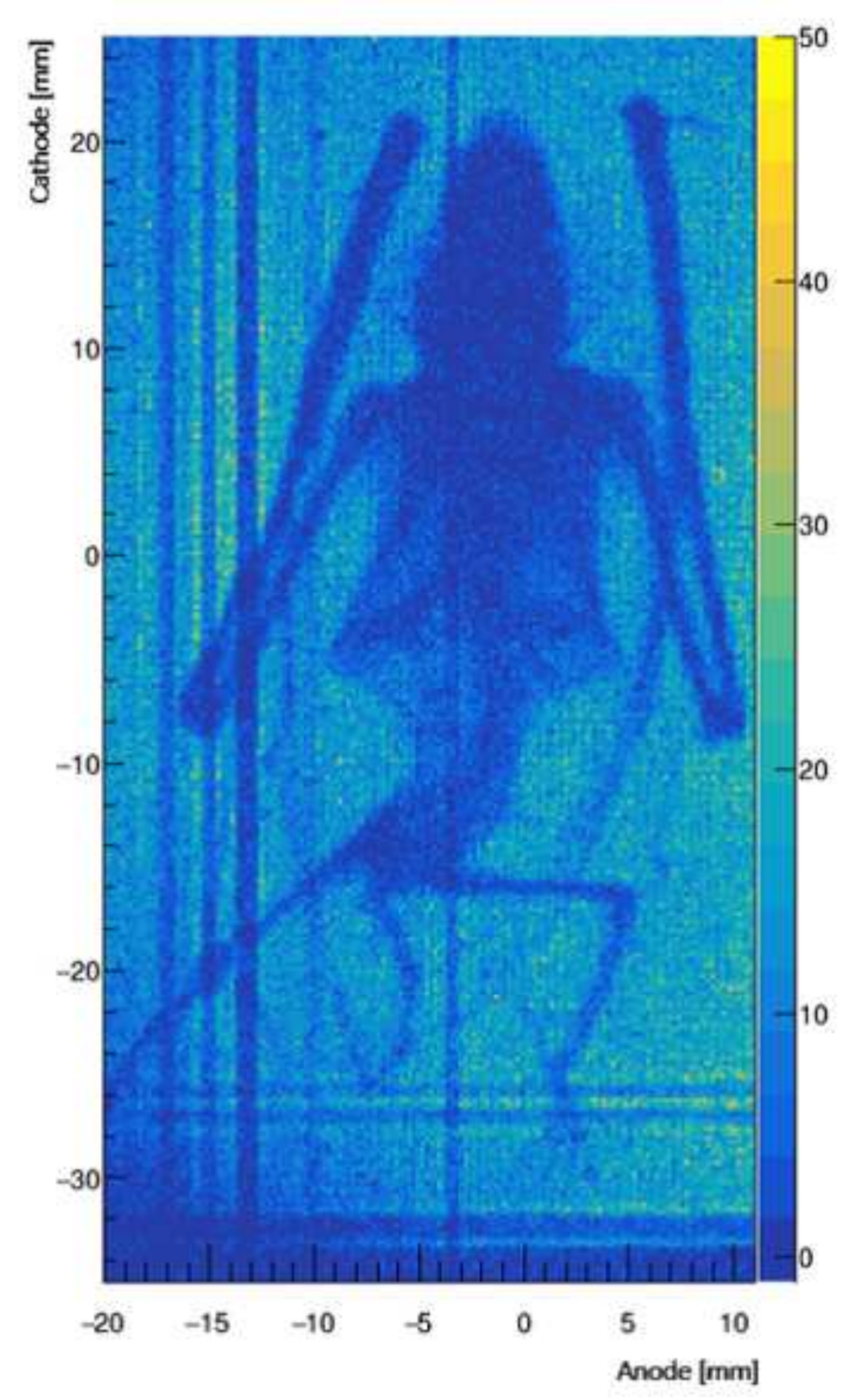}
\end{center}
\end{minipage}
\end{tabular}
\caption{Bat and its X-ray image}
\label{bat}
\end{figure}

An imaging test was performed using an 8keV X-ray source at CERN's RD51 Laboratory. The SRS/APV25 readout system was used for the data acquisition. Fig. \ref{bat} shows a bat and its X-ray image. Its skeleton, its body's contour, and its left thumb claw are clearly seen. Its left toe claws are less visible. Thus, fine X-ray images can be obtained using the new resistive $\mu$-PIC.

\section{Performance study for the charged particles at CERN SPS test beam}
\subsection{Experimental setup}
The tracking performances of $\mu$-PIC were studied using charged particles beam. The main objective of this test is to evaluate the two-dimensional position resolution of the new $\mu$-PIC. The time resolution and the detection efficiency were also measured. The test was performed in October 2017 at the SPS H4 beam line in CERN \cite{sps,h4}. The secondary beams (muons and pions) were delivered from the T2 target on which the 400 GeV/c primary proton beam is transported. The momentum of the secondary beams was set to 150GeV/c. Each beam spill lasts about four seconds. By closing the beam shutter, only the muons beam was derived, and its intensity of the muon beam was $\sim\mathrm{10^5}$/spill.

\begin{figure}[htbp]
\centering
\includegraphics[width=12cm,clip]{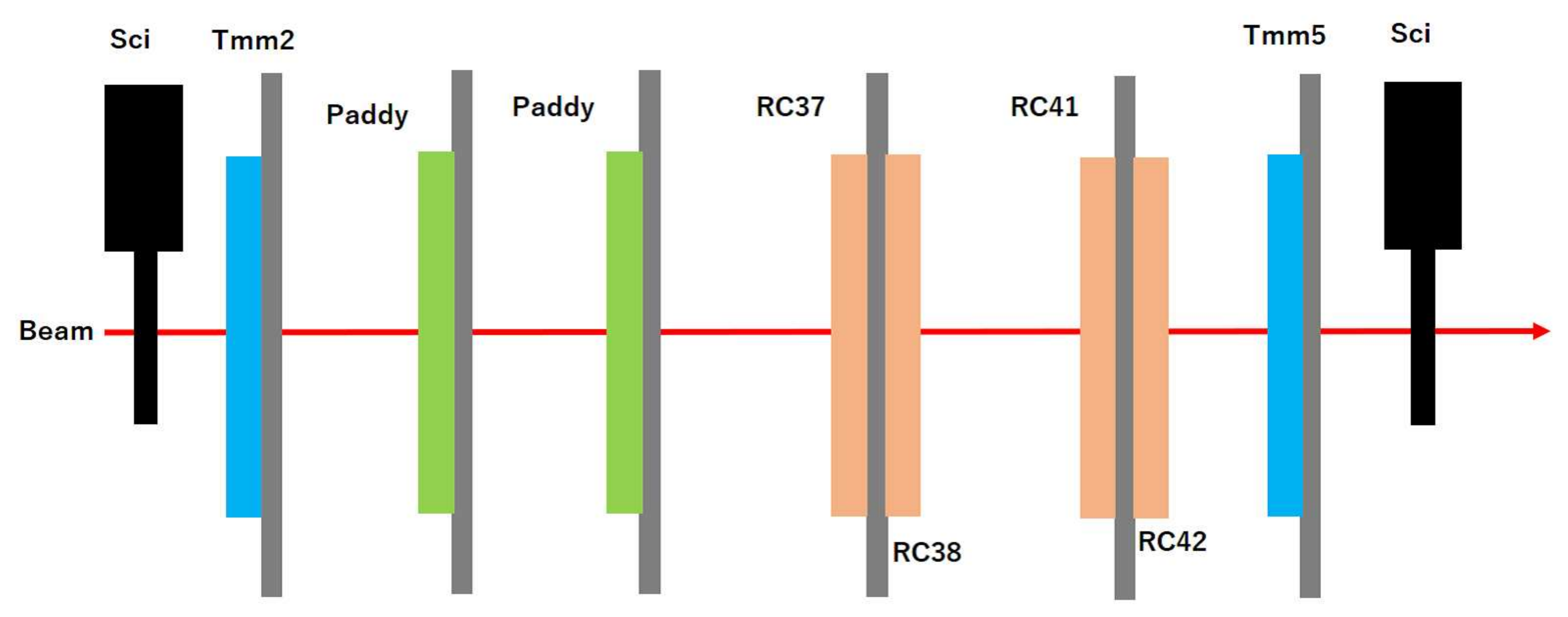}
\caption{Experimental setup of test: Four $\mu$-PICs (RC37, RC38, RC41, and RC42) were put between two Micromegas chambers (Tmm2, Tmm5) and trigger scintillators. Both two $\mu$-PICs are set in a back-to-back configuration. ''Paddy'' is another type of detector unrelated to this work.}
\label{h4setup}
\end{figure}

\begin{figure}[htbp]
\centering
\includegraphics[width=10cm,clip]{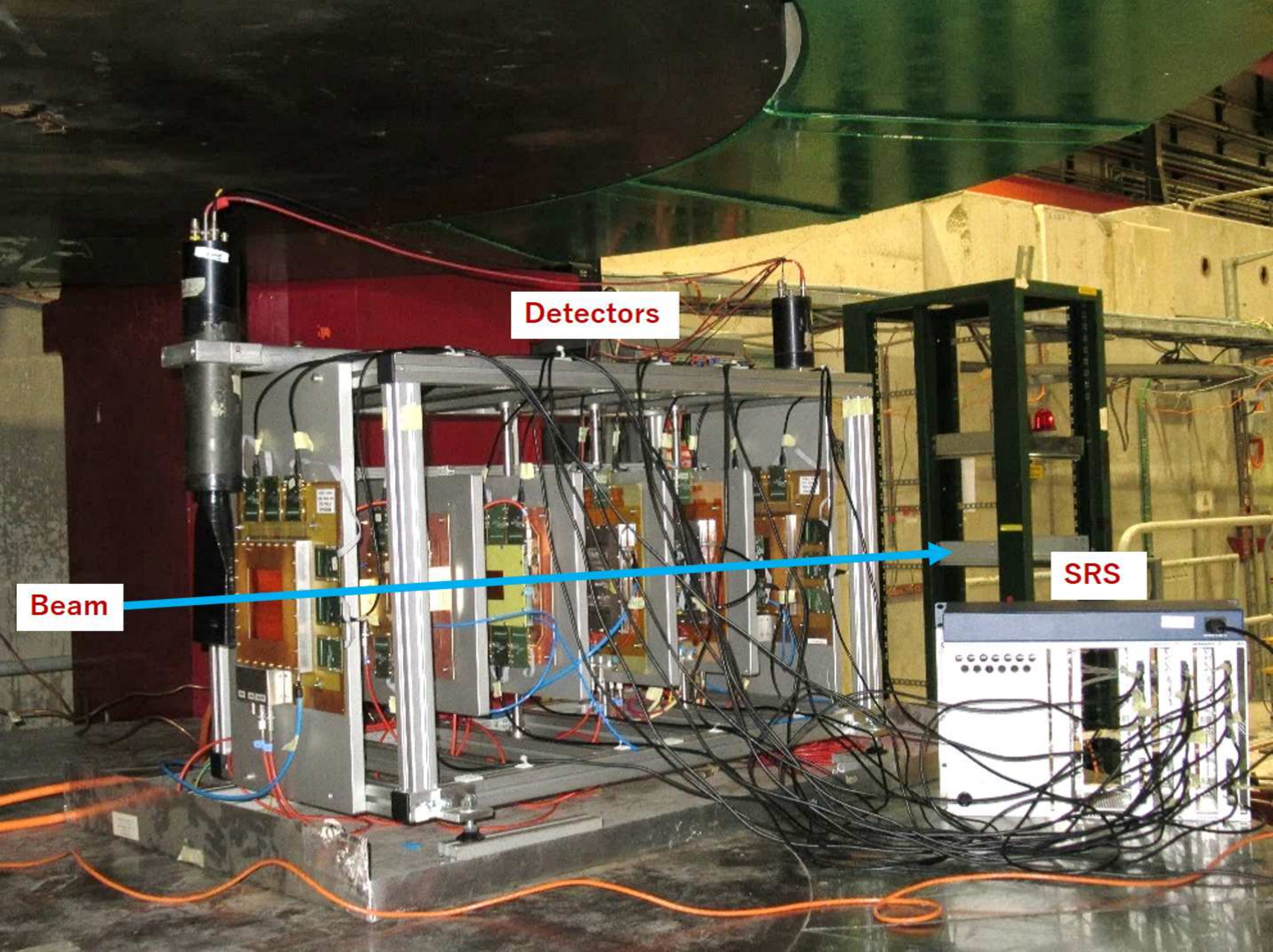}
\caption{Photograph of experimental setup}
\label{h4setupphoto}
\end{figure}

Four $\mu$-PICs were tested: RC37, RC38, RC41, and RC42. The drift fields of RC37 and RC38 were set to 3kV/cm with a 3 mm drift gap. The drift fields of RC41 and RC42 were set to 1 kV/cm with a 5mm drift gap. Figs. \ref{h4setup} and \ref{h4setupphoto} show the experimental setup of this test. Two $\mu$-PICs, which have identical properties, were put in a back-to-back configuration (Fig. \ref{h4setup}). Two Micromegas chambers with two-dimensional readouts (Tmm2, Tmm5) were used for beam telescopes \cite{tmm}. The Micromegas chambers have 360 strips with a 250 $\mu$m pitch for both readouts and a 9 $\times$ 9 $\mathrm{cm^2}$ active area. The drift field was set to 600 V/cm with a 5 mm drift gap. Two 10 $\times$ 10 $\mathrm{cm^2}$ plastic scintillators were used for the trigger.  $\mu$-PICs were operated with Ar/$\mathrm{CO_2}$ (93:7) or Ar/$\mathrm{C_2H_6}$ (7:3) gas mixtures. Micromegas chambers were operated with an Ar/$\mathrm{CO_2}$ (93:7) gas mixture.

\subsection{Data acquisition}
\begin{figure}[htbp]
\begin{center}
\includegraphics[width=10cm,clip]{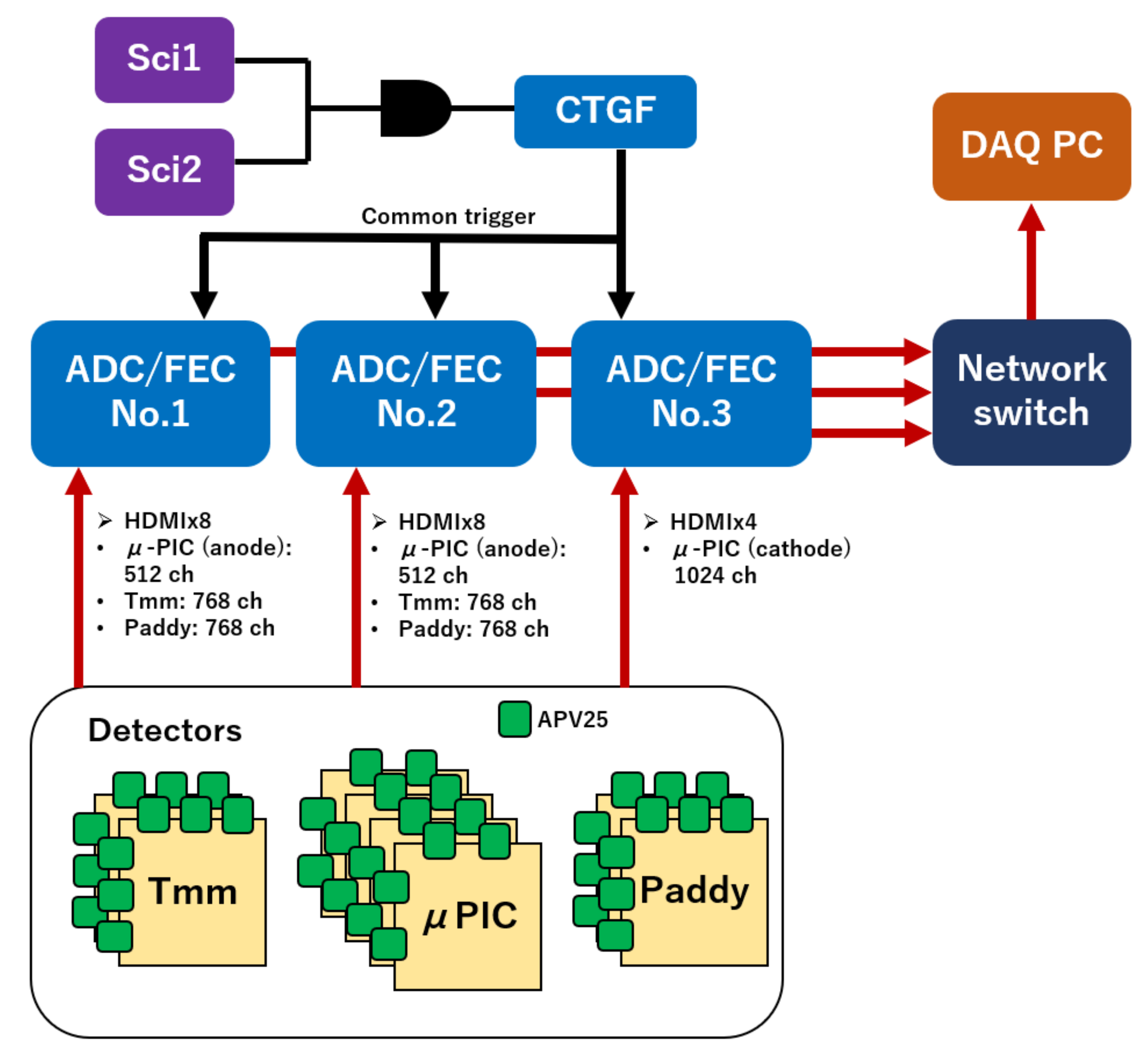}
\end{center}
\caption{Block diagram of readout process}
\label{h4blockdiagram}
\end{figure}

The data acquisition system was based on the Scalable Readout System (SRS) \cite{srs} with APV25 front-end electronics \cite{apv}. SRS, which was developed by the CERN RD51 collaboration, consists of an Analogue Digital Converter (ADC), Front-End Concentrator (FEC) cards, front-end electronics (APV25 in this test), and other parts, including a power supply. The APV25 chip with 128-ch readouts was originally developed for a silicon strip detector of the CMS tracker \cite{apv}. Detector signals are fed to charge amplifiers with a 50 ns CR-RC shaper and sampled by a 40 MHz clock. Analogue signals of 128 channels are sampled every 25 nsec, and multiplexed signals from two APVs (256 channels in total) are fed to the ADC card using a standard HDMI cable. Once a trigger is invoked, 15 cycles of multiplexed signals (corresponding to 375 nsec) are read and recorded as one frame. Signals are digitized by the ADC card and sent to the DAQ PC by the FEC via Gbit Ethernet. Fig. \ref{h4blockdiagram} shows a block diagram of the readout process in this setup. Two ADC/FEC combos were used for the readout of the negative signals of the $\mu$-PIC anodes and the Micromegas chambers. One ADC/FEC combo was used for the positive signals of the $\mu$-PIC cathodes. A trigger signal from two scintillators was sent once to an SRS Clock and Trigger Generator \& Fun-out (CTGF) card. Then a common trigger was distributed to three FEC cards by the CTGF. This scheme enabled data acquisition without any time lag among multiple FEC cards. Signals from the three FEC cards were merged by a network switch and sent to the DAQ PC by one Ethernet cable. The ''mmDAQ'' software developed by Muon ATLAS MicroMegas Activity (MAMMA) was used for data taking. The position, time, and charge informations are recorded with mmDAQ.

\subsection{Signal processing}
\begin{figure}[htbp]
\begin{center}
\includegraphics[width=10cm,clip]{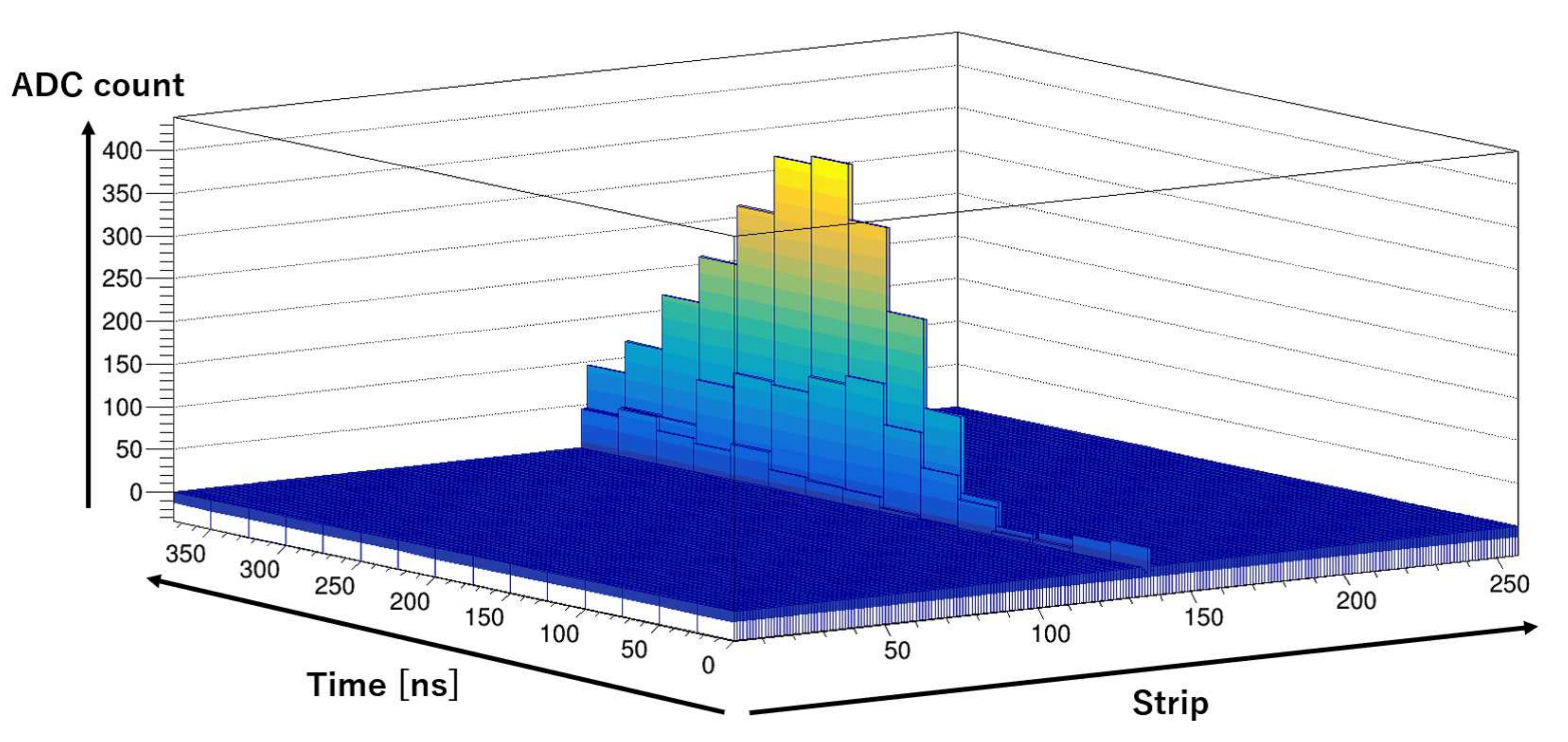}
\end{center}
\caption{Three-dimensional event display of a typical muon event on $\mu$-PIC}
\label{muonevent}
\end{figure}

\begin{figure}[htbp]
\begin{center}
\includegraphics[width=10cm,clip]{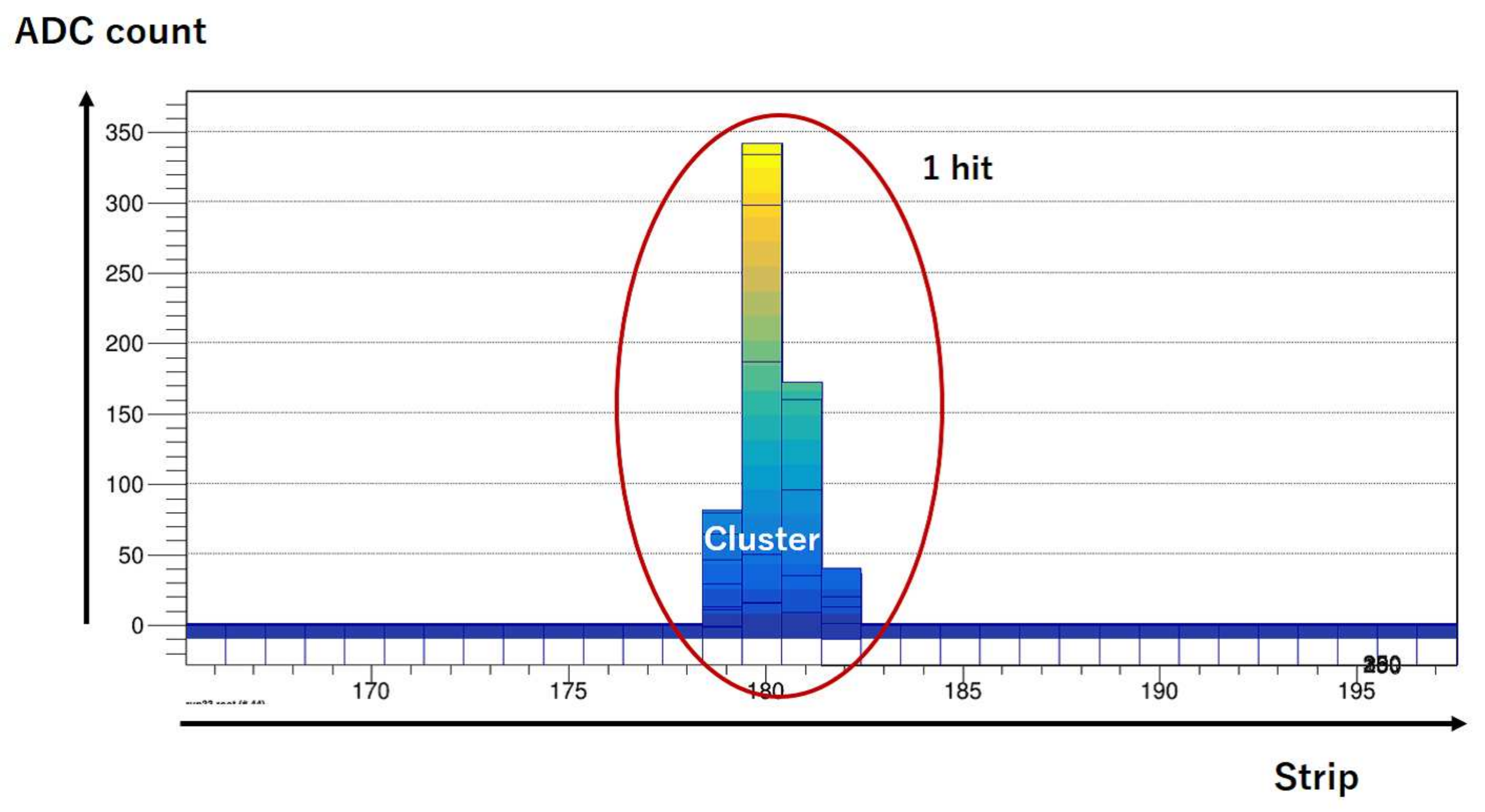}
\end{center}
\caption{Two-dimensional event display: The horizontal axis is ''Strip'' and vertical axis is ''ADC count''.}
\label{stripq}
\end{figure}

Fig. \ref{muonevent} shows a three-dimensional event display of a typical muon event on the $\mu$-PIC with three axes, ''Strip'', ''Time [ns]'' and ''ADC count''. ''Strip'' denotes the number of strips with a range of 1-256 for each readout. The shaper output signal is sampled every 25 ns during 15 cycles for each strip in each event. Therefore, each event has these three informations. Following parameters were defined for the analysis.
\begin{itemize}
\item strip: Hit strip number (position).
\item time: Bin number on the Time axis with a range of 1$\leq$time$\leq$15 in one event frame. Each bin has a 25 ns time window.
\item Q: ADC count per strip per time.
\item $\mathrm{Q_{max}}$: Maximum Q among 15 time samples in each strip, defined as the strip charge.
\end{itemize}
Fig. \ref{stripq} shows a two-dimensional event display for a muon track, where the horizontal axis is ''Strip'' and the vertical axis is ''ADC count''. Signals are clearly seen on four consecutive strips. When one particle passes the detector and generates primary electrons, these primary electrons fall on the multiple strips depending on the incident angle and the drift gap. For example, 2-5 strips will have hits by a normal incident track into a 5 mm drift gap. Therefore, adjacent hit strips are merged into one cluster. The number of clusters corresponds to the number of incident particles. The following are the cluster's parameters.
\begin{itemize}
\item cluster charge: Sum of $\mathrm{Q_{max}}$ in one cluster ($\mathrm{\sum{Q_{max}}}$).
\item cluster position: Defined as the charge-weighted centroid by the following formula;
\begin{equation}
\mbox{Cluster position} = \frac{\sum^{}_{}((strip)\times Q_{max})}{\sum^{}_{}{Q_{max}}}
\end{equation}
This matches the particle's position if it perpendicularly enters the detector.
\end{itemize}

\subsection{Detection efficiency}
\begin{figure}[htbp]
\begin{center}
\includegraphics[width=13cm,clip]{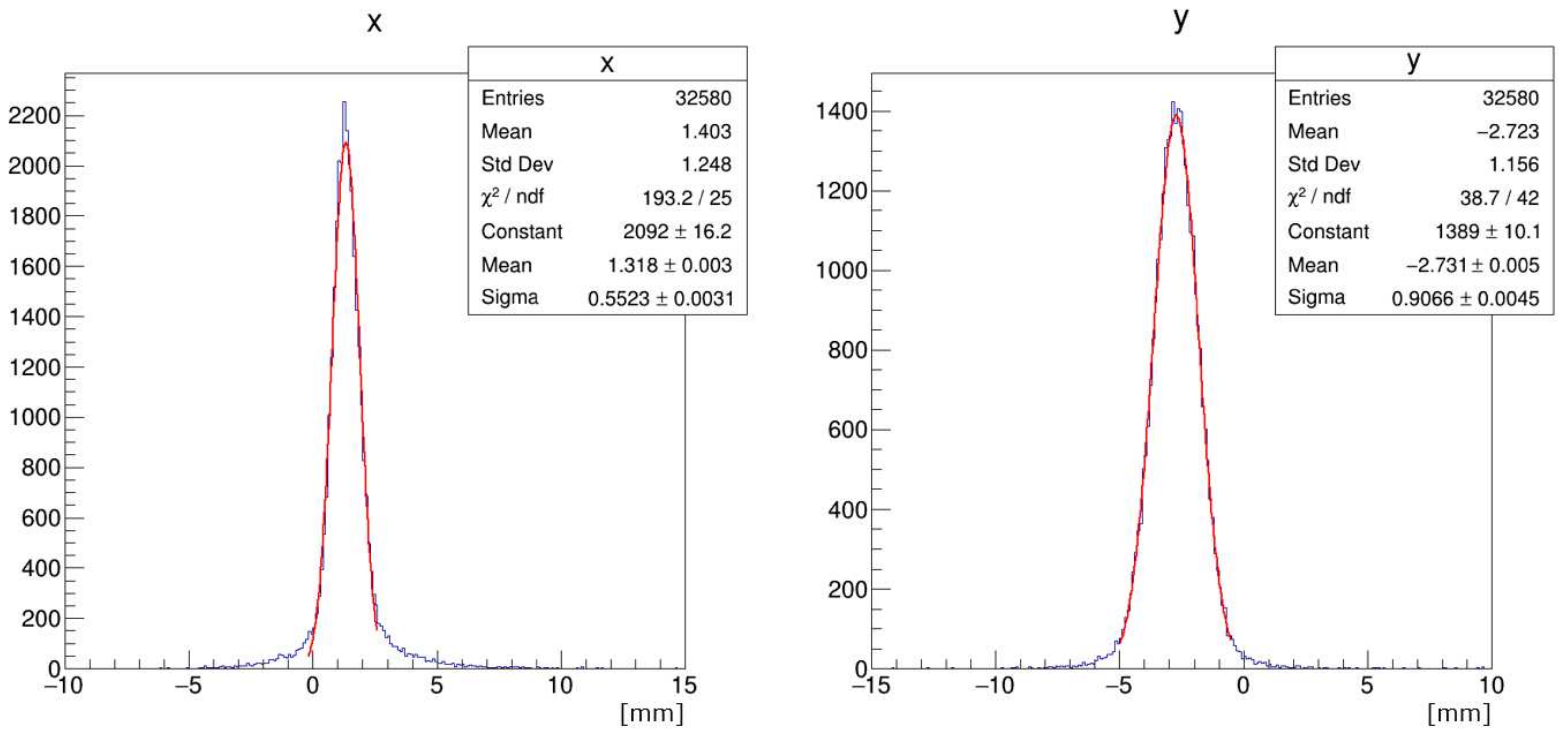}
\end{center}
\caption{Residual distributions of hit positions of two Tmm chambers: X coordinate is parallel to resistive strips of Tmm and Y coordinate is perpendicular to them.}
\label{tmmresi}
\end{figure}

Next, the detection efficiencies of the $\mu$-PIC were measured for RC42. Detection efficiency is defined by the following equation.

\begin{equation}
\mbox{Detection efficiency} = \frac{N_{both}}{N_{Tmm}}
\end{equation} 
Here,
\begin{itemize}
\item $\mathrm{N_{Tmm}}$: Number of events in which there is one cluster on each of two Tmm chambers.
\item $\mathrm{N_{both}}$: Number of events in which there is at least one cluster within 5 mm from the interpolated hit position by two Tmm chambers.
\end{itemize}
If false tracks were included in $\mathrm{N_{Tmm}}$, they were removed. Fig. \ref{tmmresi} shows the residual distributions of the hit position of two Tmm chambers. Here the X coordinate is parallel to the resistive strips of the Tmm and the Y coordinate is perpendicular to them. When the residual was within 5 mm from the mean value obtained from the gaussian fit, that event was selected. 

\begin{figure}[htbp]
\begin{tabular}{cc}
\begin{minipage}{0.5\hsize}
\begin{center}
\includegraphics[width=7cm]{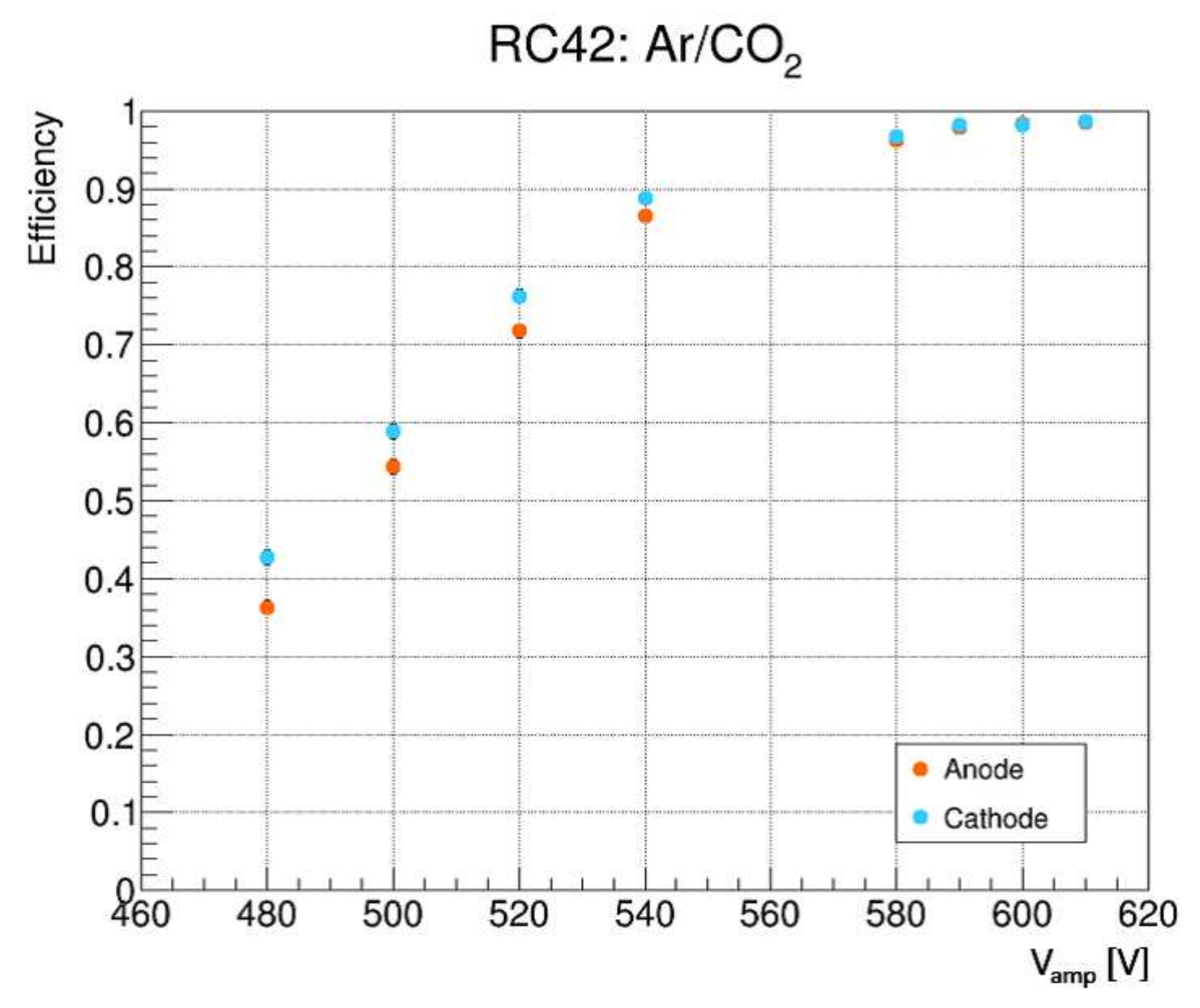}
\end{center}
\end{minipage}
\begin{minipage}{0.5\hsize}
\begin{center}
\includegraphics[width=7cm]{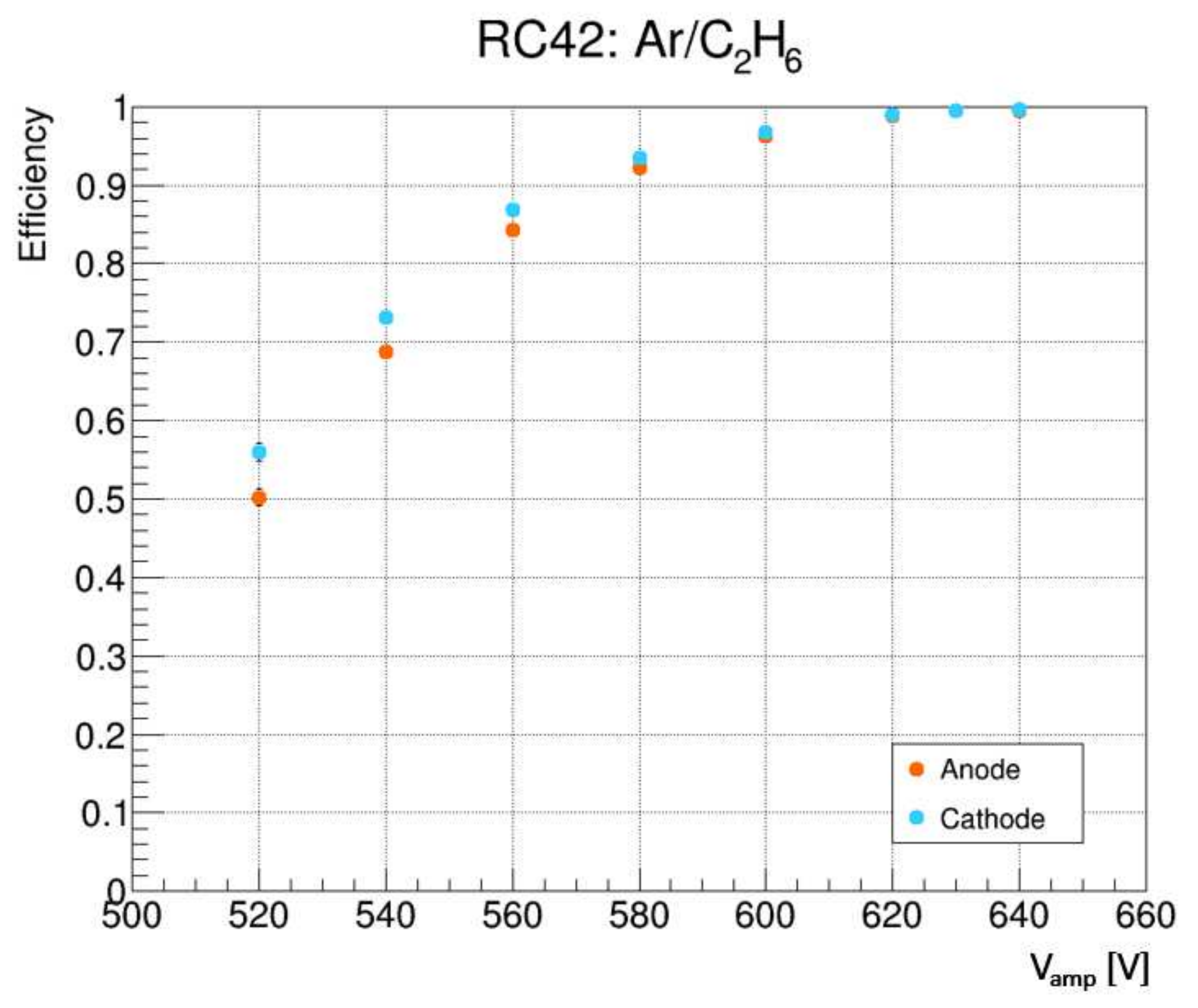}
\end{center}
\end{minipage}
\end{tabular}
\caption{Detection efficiency of RC42 as a function of amplification voltage for Ar/$\mathrm{CO_2}$ and Ar/$\mathrm{C_2H_6}$ (7:3) gas mixtures}
\label{eff_RC42}
\end{figure}

Fig. \ref{eff_RC42} shows the detection efficiency of RC42 as a function of the amplification voltage for the Ar/$\mathrm{CO_2}$ (93:7) and Ar/$\mathrm{C_2H_6}$ (7:3) gas mixtures. These results were obtained using a detection area of about 8 cm $\times$ 8 cm, which is equal to the cross-section of the muons beam. At sufficient amplification voltage, the detection efficiencies of both coordinates exceeded 98\% for Ar/$\mathrm{CO_2}$ (93:7) and 99\% for Ar/$\mathrm{C_2H_6}$ (7:3). These results show that $\mu$-PIC can detect charged particles in a large detection area.

\subsection{Time resolution}
\begin{figure}[htbp]
\begin{center}
\includegraphics[width=10cm,clip]{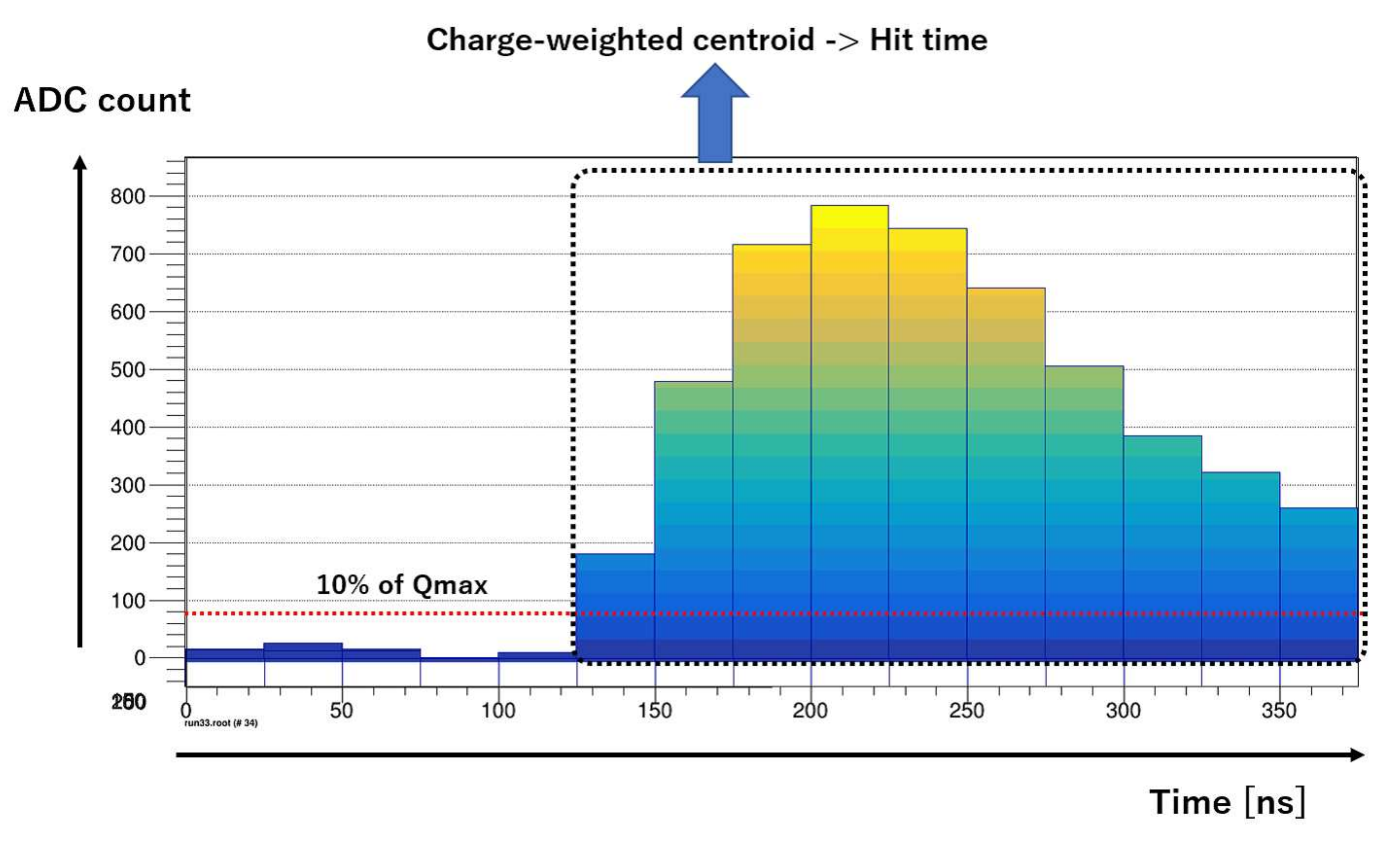}
\end{center}
\caption{Two-dimensional event display for a muon track, where horizontal axis is ''Time [ns]'' and vertical axis is ''ADC count''}
\label{timeq}
\end{figure}

Since signals are sampled at 40 MHz, there is a time jitter of $\pm$12.5ns. To remove its effect, the time resolution was measured using the difference of the hit time between two chambers, which were set back-to-back. Fig. \ref{timeq} shows a two-dimensional event display of RC42 for a muon track, where the horizontal axis is ''Time [ns]'' and the vertical axis is ''ADC count''. To determine the cluster's hit time, the hit time of each strip in the cluster was calculated by a charge-centroid for the time bins that exceeded 10\% of $\mathrm{Q_{max}}$. The following is the equation.
\begin{equation}
\mbox{Hit time} = \frac{\sum^{}_{}((time)\times Q)}{\sum^{}_{}Q}
\end{equation}
where {\it time} is the time bin number and {\it Q} denotes ''Q'' for a corresponding time bin. The cluster's hit time was defined as the earliest hit time in the cluster. When the strip signal is too small, the signal shape is deformed, which worsens the time resolution. When $\mathrm{Q_{max}}$ is lower than a threshold, the strip was removed. The threshold was set to 200 when the amplification voltage was over 580 V for Ar/$\mathrm{CO_2}$ and over 600 V for Ar/$\mathrm{C_2H_6}$. The threshold was set to 100 for lower voltages.

\begin{figure}[htbp]
\begin{center}
\includegraphics[width=13cm,clip]{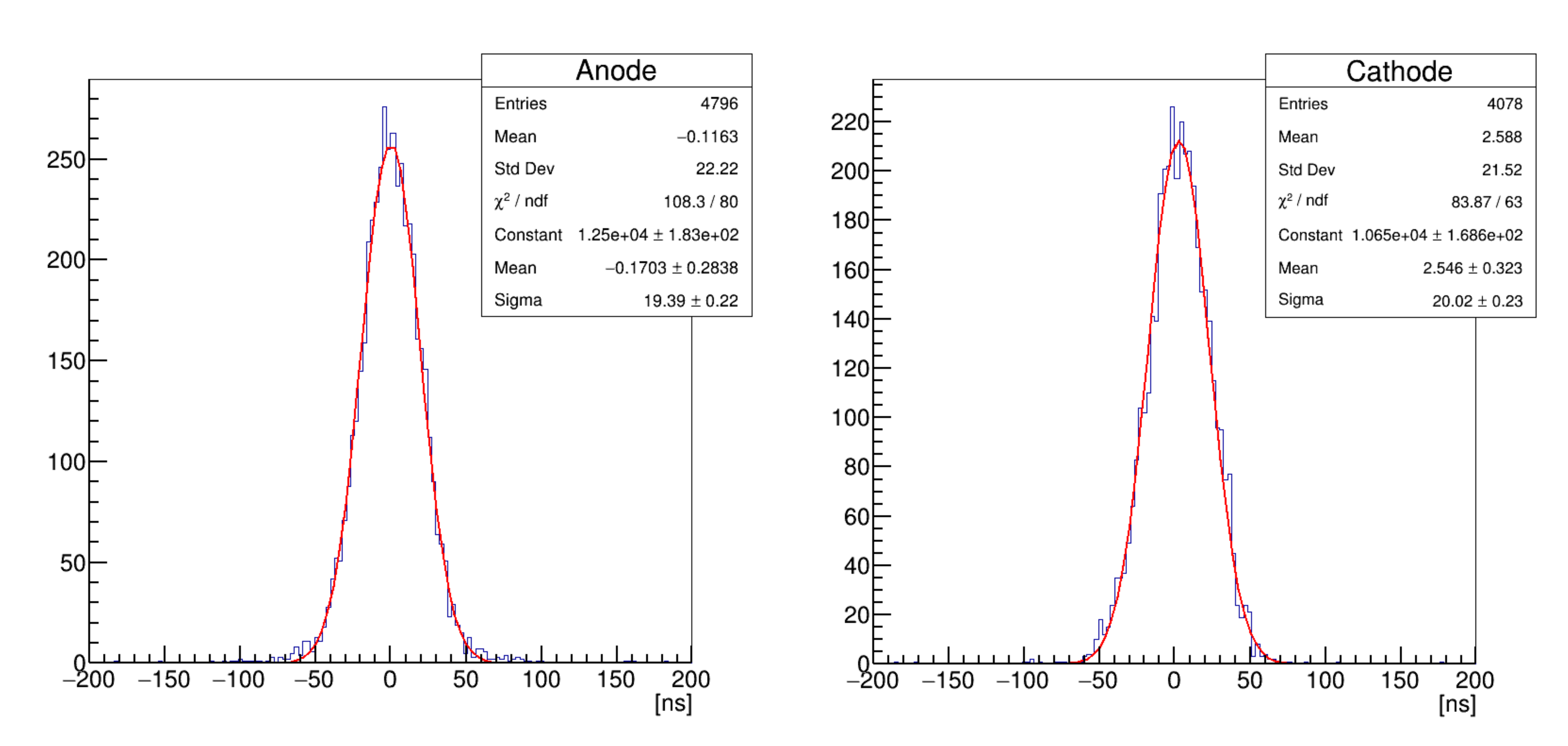}
\end{center}
\caption{Distribution for time differences between RC41 and RC42}
\label{timedist}
\end{figure}

\begin{figure}[htbp]
\begin{center}
\includegraphics[width=9cm,clip]{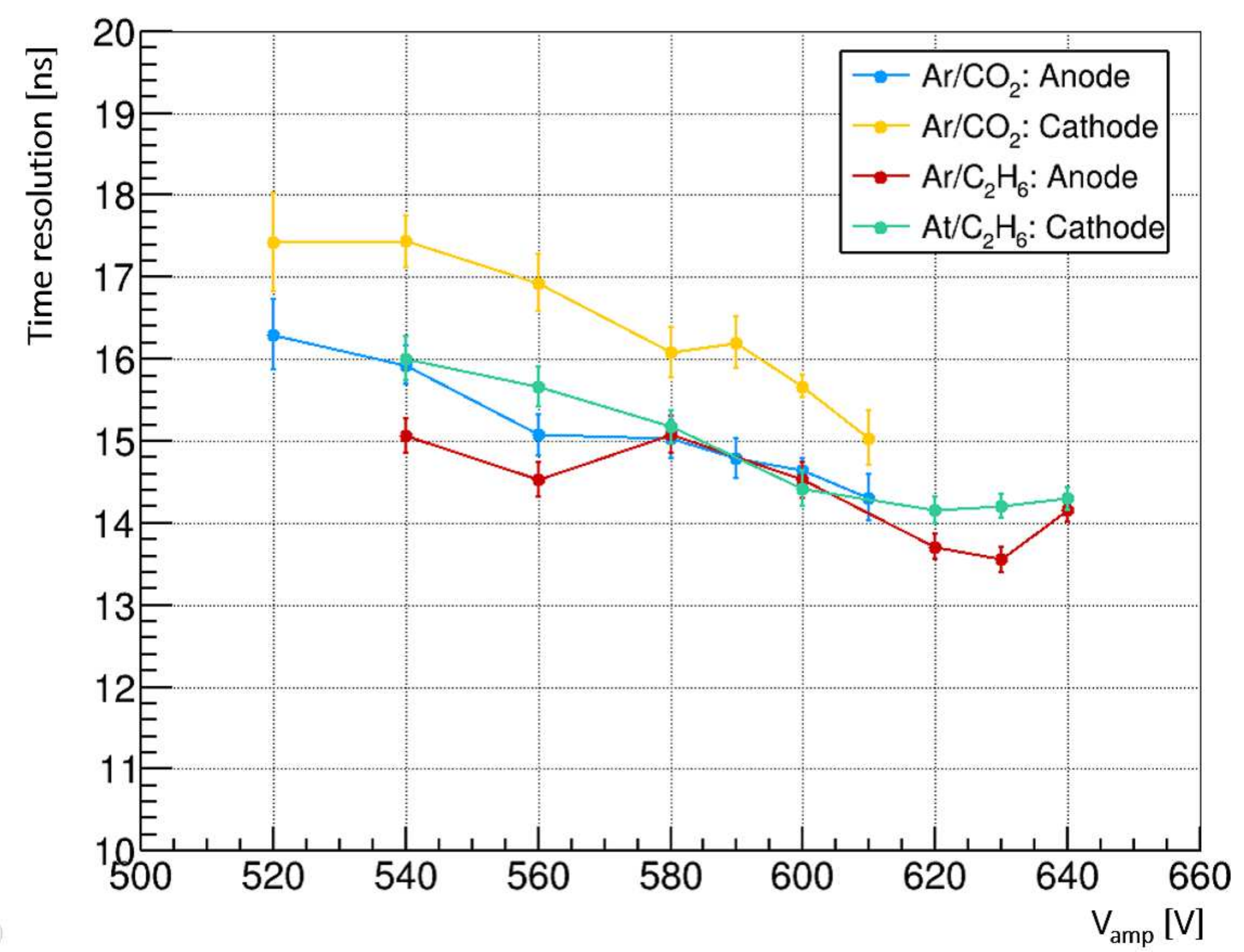}
\end{center}
\caption{Plots of time resolution as a function of amplification voltage}
\label{timereso}
\end{figure}

Fig. \ref{timedist} shows the distribution for the time difference between two chambers with gaussian fit. Assuming that the two chambers have identical resolution, the time resolution is given by dividing $\sigma$ by $\sqrt2$. Fig. \ref{timereso} shows the plots of the time resolution as a function of the amplification voltage. At adequate amplification voltage, the time resolutions were between 13.5-16 ns. There is no significant difference between the readout coordinates and the gases. Note that the measured time resolution depends on the configuration of the electronics and a method that determines the hit time. The expected time resolution in this setup is 10-15 ns, which is consistent with our results.

\subsection{Position resolution}
\begin{figure}[htbp]
\begin{center}
\includegraphics[width=13cm,clip]{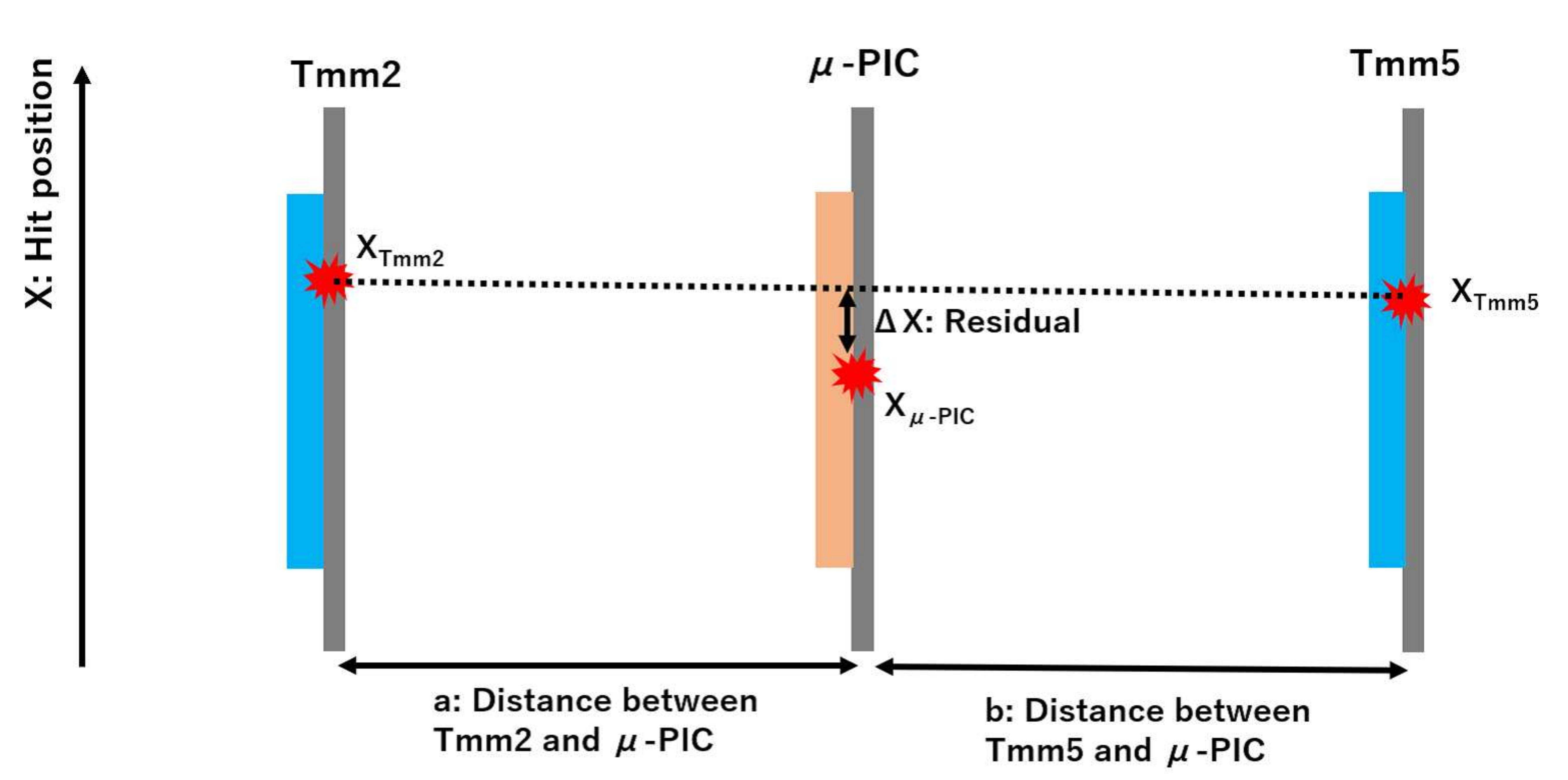}
\end{center}
\caption{Schematic view of one $\mu$-PIC and two Tmm chambers} 
\label{residualposition}
\end{figure}

Two Tmm chambers were used, as a reference, to measure the position resolution of RC37 and RC42. Here, the measurement was taken from the residual between the hit position on the $\mu$-PIC and the interpolated position using the Tmm chambers. The residual is defined by the following equation.

\begin{equation}
\Delta X = X_{\mu PIC} - \frac{bX_{Tmm2}+aX_{Tmm5}}{a+b}
\end{equation}
Here,

\begin{itemize}
\item $\mathrm{X_{\mu PIC}, X_{Tmm2}, X_{Tmm5}}$: Hit position of each chamber (Fig. \ref{residualposition}).
\item {\it a}: Distance between Tmm2 and $\mu$-PIC.
\item {\it b}: Distance between Tmm5 and $\mu$-PIC.
\item $\Delta$X: Residual.
\end{itemize}

\begin{figure}[htbp]
\begin{tabular}{cc}
\begin{minipage}{0.5\hsize}
\begin{center}
\includegraphics[width=7cm]{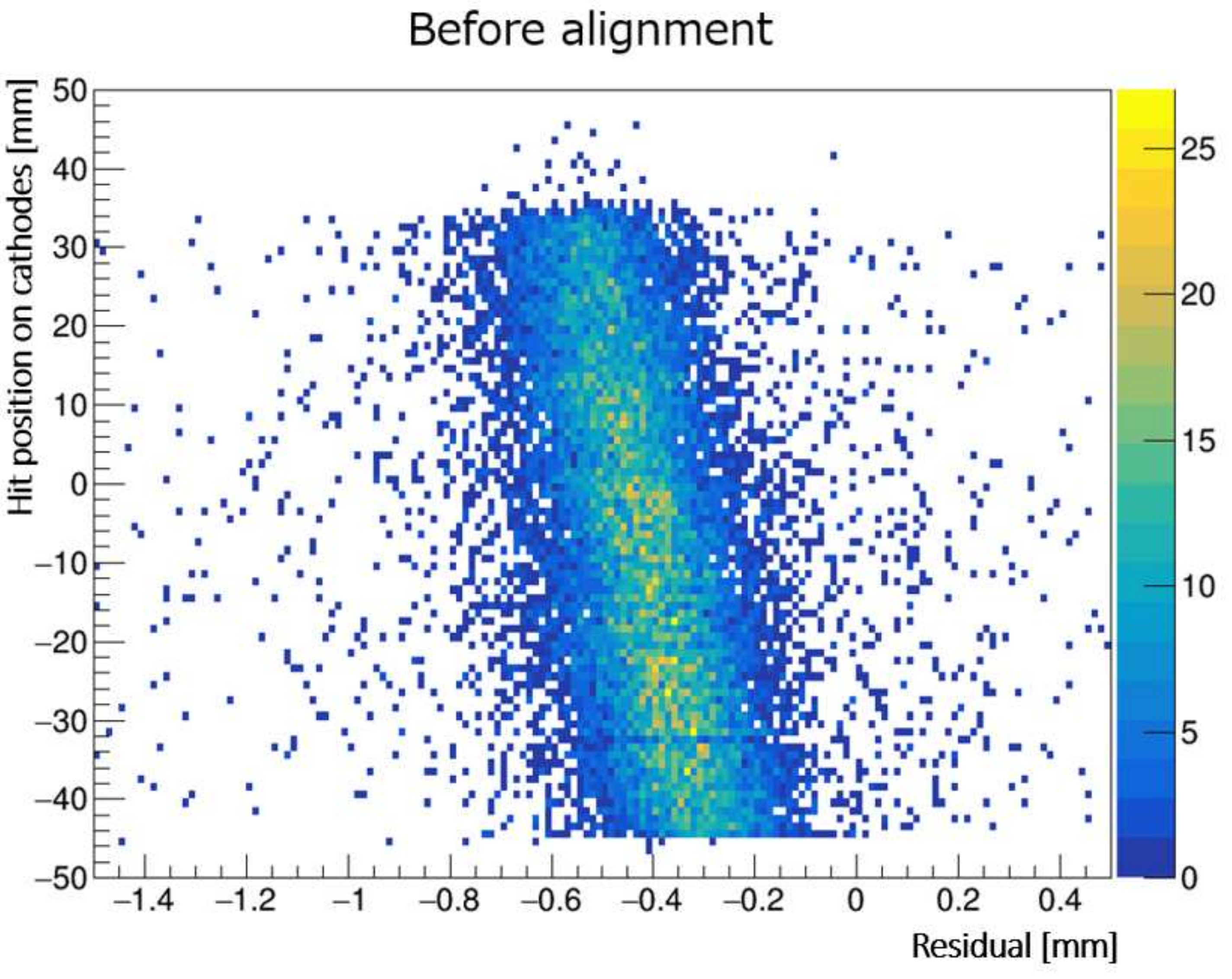}
\end{center}
\end{minipage}
\begin{minipage}{0.5\hsize}
\begin{center}
\includegraphics[width=7cm]{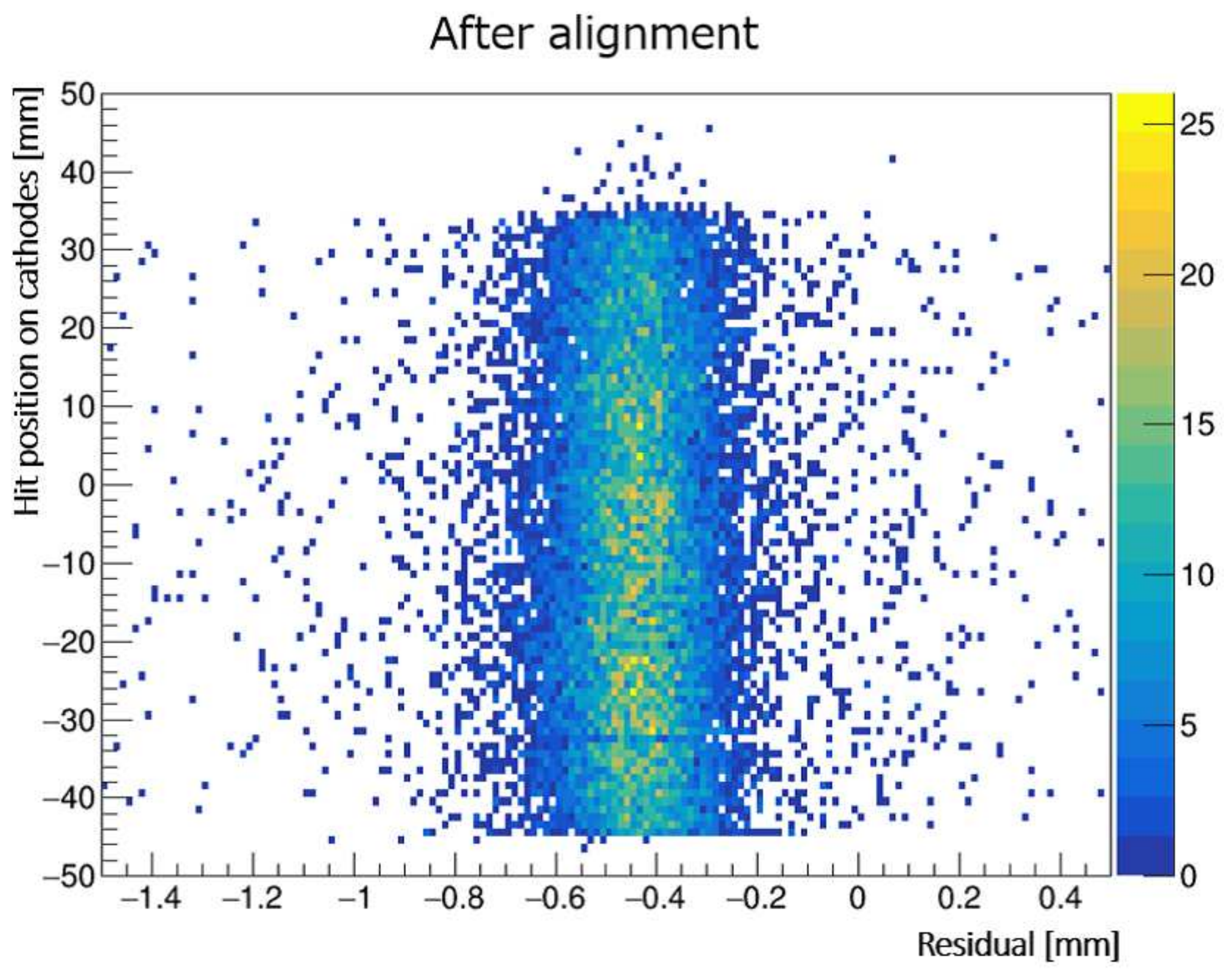}
\end{center}
\end{minipage}
\end{tabular}
\caption{Residual distribution obtained from anode coordinate against hit position for cathode coordinate of RC42: Left: before alignment correction. Right: after alignment correction.}
\label{rotate}
\end{figure}

The directions of the anode and cathode coordinates of the $\mu$-PIC are respectively parallel to X and Y of the Tmm. However, since the directions of the three chambers (two Tmms and one $\mu$-PIC) were not aligned perfectly, the alignments were corrected. Fig. \ref{rotate} shows the residual distribution obtained from the anode coordinate against the hit position for the cathode coordinate of RC42. The left side of the figure shows before the alignment correction. The residual distributions differ depending on the cathode positions. The right side shows after the alignment correction, which measured the precise resolution.

\begin{figure}[htbp]
\begin{center}
\includegraphics[width=15cm,clip]{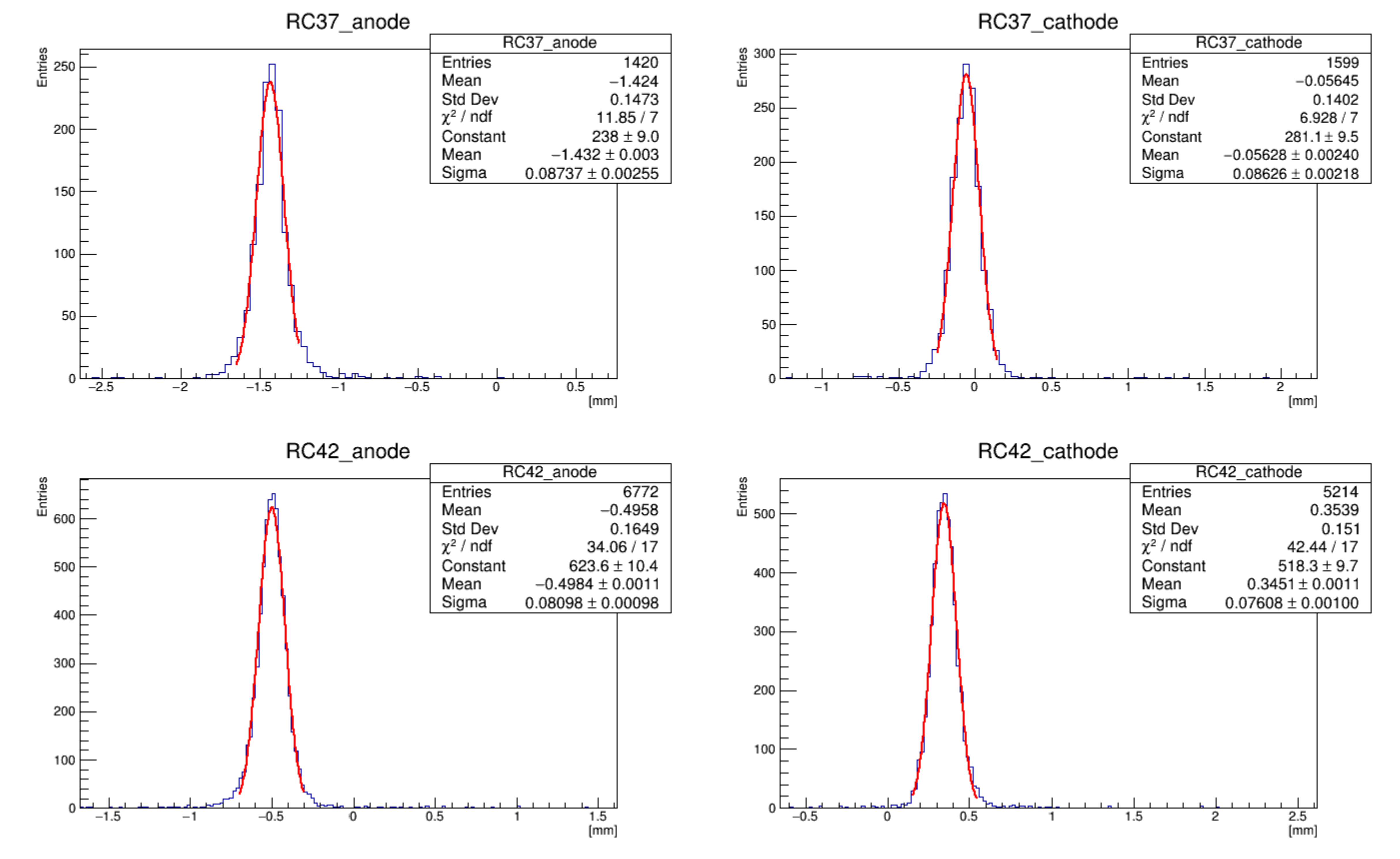}
\end{center}
\caption{Residual distribution of RC37 and RC42 for both coordinates}
\label{residual}
\end{figure}

Fig. \ref{residual} shows a residual distribution with the gaussian fit. Assuming that the two Tmm chambers have identical resolution ($\mathrm{\sigma_{Tmm}}$), the position resolution of $\mu$-PIC ($\mathrm{\sigma_{\mu PIC}}$) is given by the following equation.

\begin{equation}
\sigma_{\mu PIC}^2 = \sigma_{residual}^2 - \frac{a^2+b^2}{(a+b)^2}\sigma_{Tmm}^2
\end{equation}
A previous study of two-dimensional Micromegas reported that the position resolutions of Tmm are 56 $\mu$m for the X coordinate and 55 $\mu$m for the Y coordinate \cite{tmmreso}.

\begin{figure}[htbp]
\begin{tabular}{cc}
\begin{minipage}{0.5\hsize}
\begin{center}
\includegraphics[width=7cm]{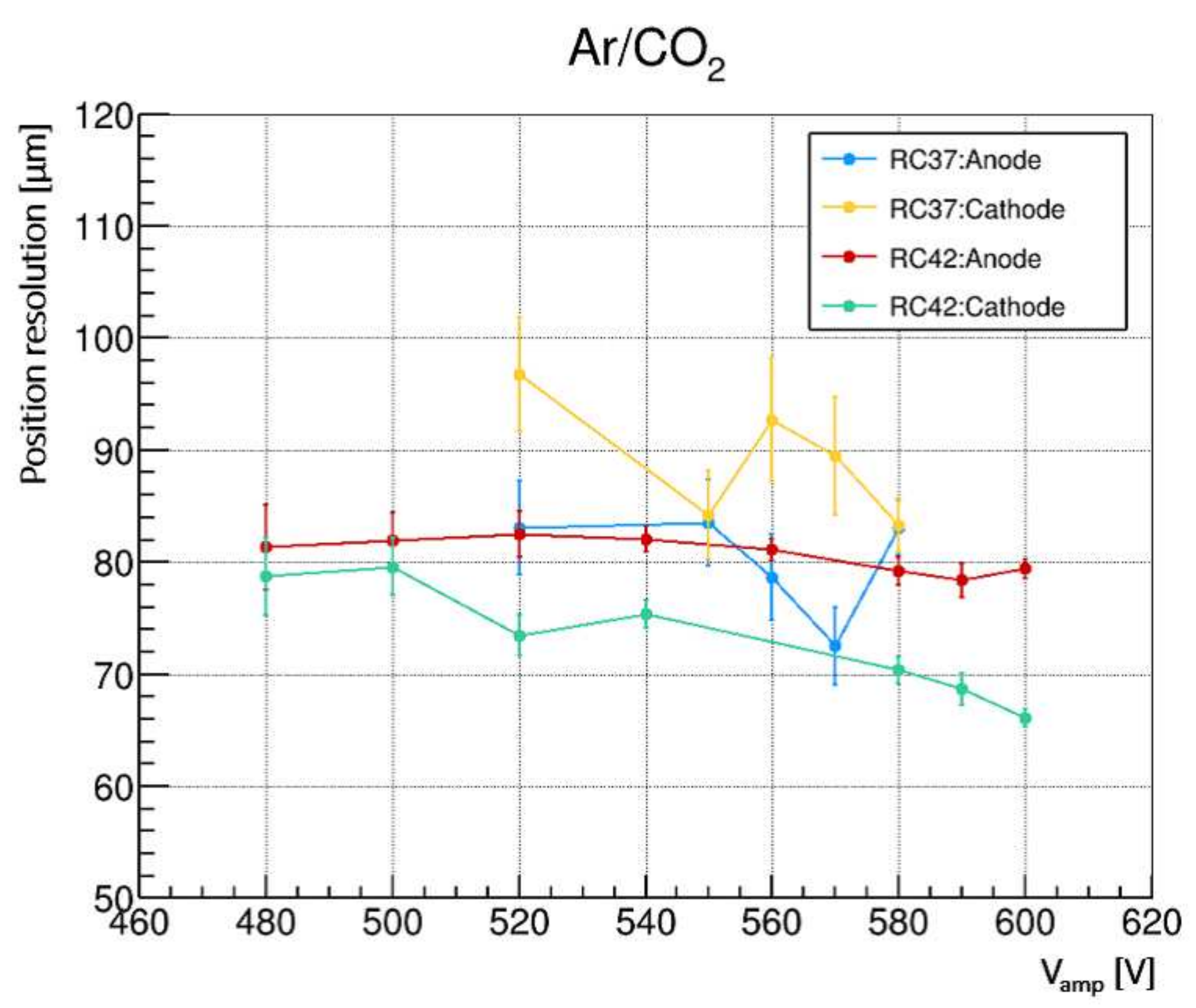}
\end{center}
\end{minipage}
\begin{minipage}{0.5\hsize}
\begin{center}
\includegraphics[width=7cm]{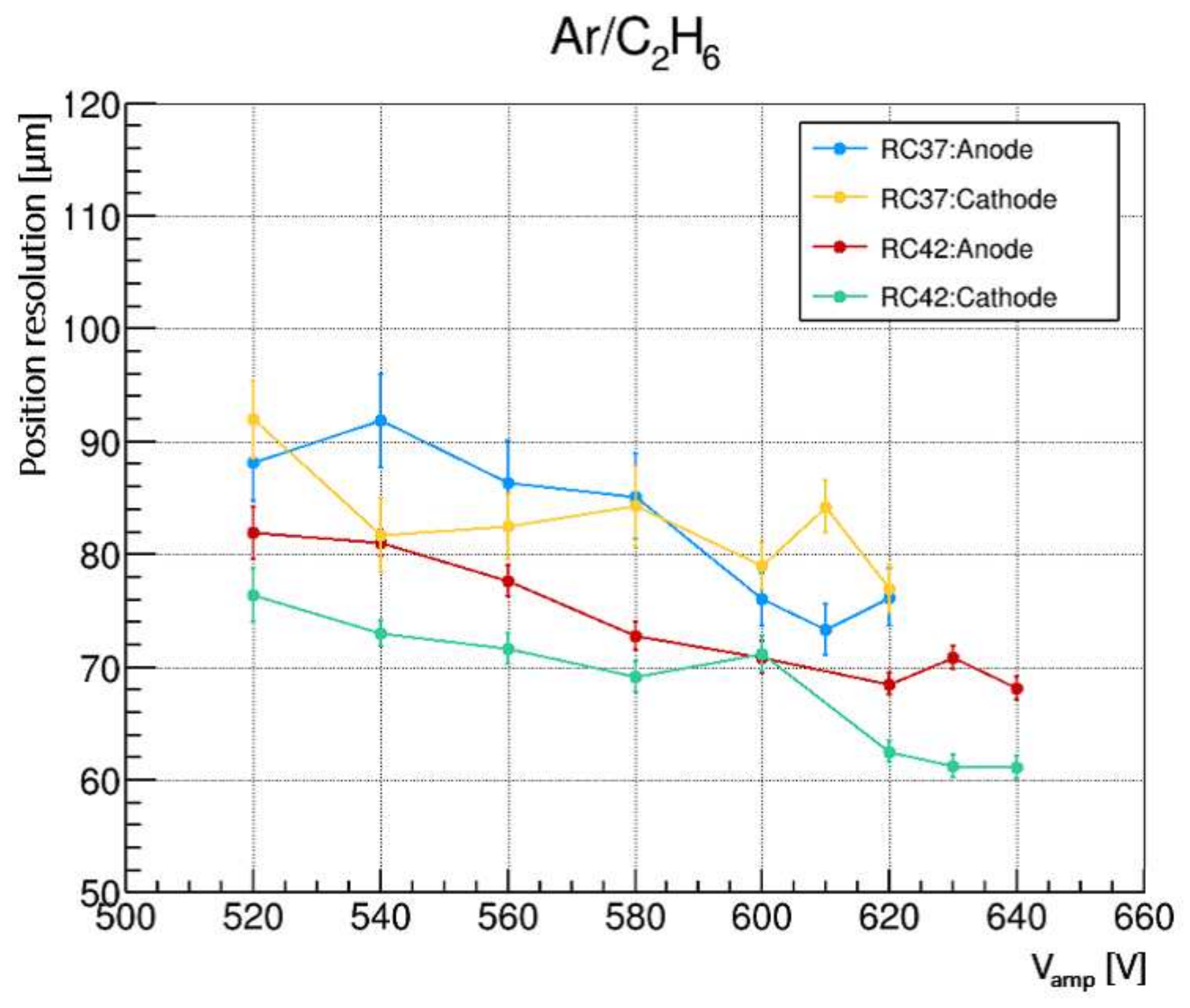}
\end{center}
\end{minipage}
\end{tabular}
\caption{Position resolution measured using interpolated position from two Tmms: Left is of Ar/$\mathrm{CO_2}$, and right is of Ar/$\mathrm{C_2H_6}$.}
\label{residualreso}
\end{figure}

Fig. \ref{residualreso} shows the position resolutions of RC37 and RC42 as a function of the amplification voltage. For RC37, the position resolutions have large variation due to less statistics. For RC42, they improve as the amplification voltage increases. By using a charge-weighted centroid, position resolutions better than 100 $\mu$m were obtained for both coordinates.

\section{Performance study under fast-neutrons irradiation}
\subsection{Experimental setup}
The resistive $\mu$-PIC was designed for a charged particle detector in a harsh background of heavily ionizing particles. As described in Section 1, in the Hadron collision experiment, the atomic nuclei of the chamber materials and the gas atoms are recoiled by fast neutrons. Those nuclei heavily ionize gas atoms, and huge charges of $\mathrm{\sim10^{4-5}}$ electrons are deposited in the gas volume. When the detector is operated with a gas gain of several thousands, the electrons density exceeds the Raether limit ($\mathrm{\sim10^{6-7}}$), and large sparks easily occur. To evaluate the detector performance, $\mu$-PIC was operated in an intense fast neutrons environment.

A test was performed in July 2017 at the tandem electrostatic accelerator facility at the Faculty of Maritime Science, Kobe University. In this facility, fast neutrons with a few MeV can be produced by a $\mathrm{{}^9Be(d,n)^{10}B}$ reaction. The 1 mm thick Be target was set at the end of the deuteron beam, isolated from the ground, and connected to a current monitor to measure the beam current. A bias voltage of several tens of volts was applied to it for suppressing the secondary electrons emitted from the target to accurately measure the beam current. The total amount of accumulated charge was calculated using the beam current and the irradiation time. This value, which determines the neutron yield, is estimated to be $\mathrm{2\times10^9 n/\mu C}$ with a deuteron energy of 3MeV \cite{be}. This is an experimental value, and the test's condition is not as same as previous studies. Therefore, it was assumed that the neutron yield had a range of $\pm$50\%.

The $\mu$-PIC was placed in front of the Be target. The distance between the detector surface and the target was set to 4, 10, 30, or 60 cm, depending on the required neutron flux, which was also controlled by varying the deuteron beam intensity from 20 to 1000 nA. The total amount of irradiated neutrons was calculated as follows:

\begin{equation}
N = 2 \times 10^9 {\rm [n/\mu C]} \times I {\rm [\mu A]} \times T {\rm [s]} \times \frac{S}{4\pi d^2}
\end{equation}
where {\it I} is the beam current on the Be target, {\it T} is the irradiation time, {\it S} is the operated area on the chamber, and {\it d} is the distance between the Be target and the chamber. Since $\mu$-PIC's detection area is 10 $\times$10$ \mathrm{cm^2}$, {\it d} depends on the position in the chamber. When the chamber is put near the Be target, the difference of {\it d} increases. Therefore, it was treated as systematic error in this test.

\begin{figure}
\begin{center}
\includegraphics[width=13cm,clip]{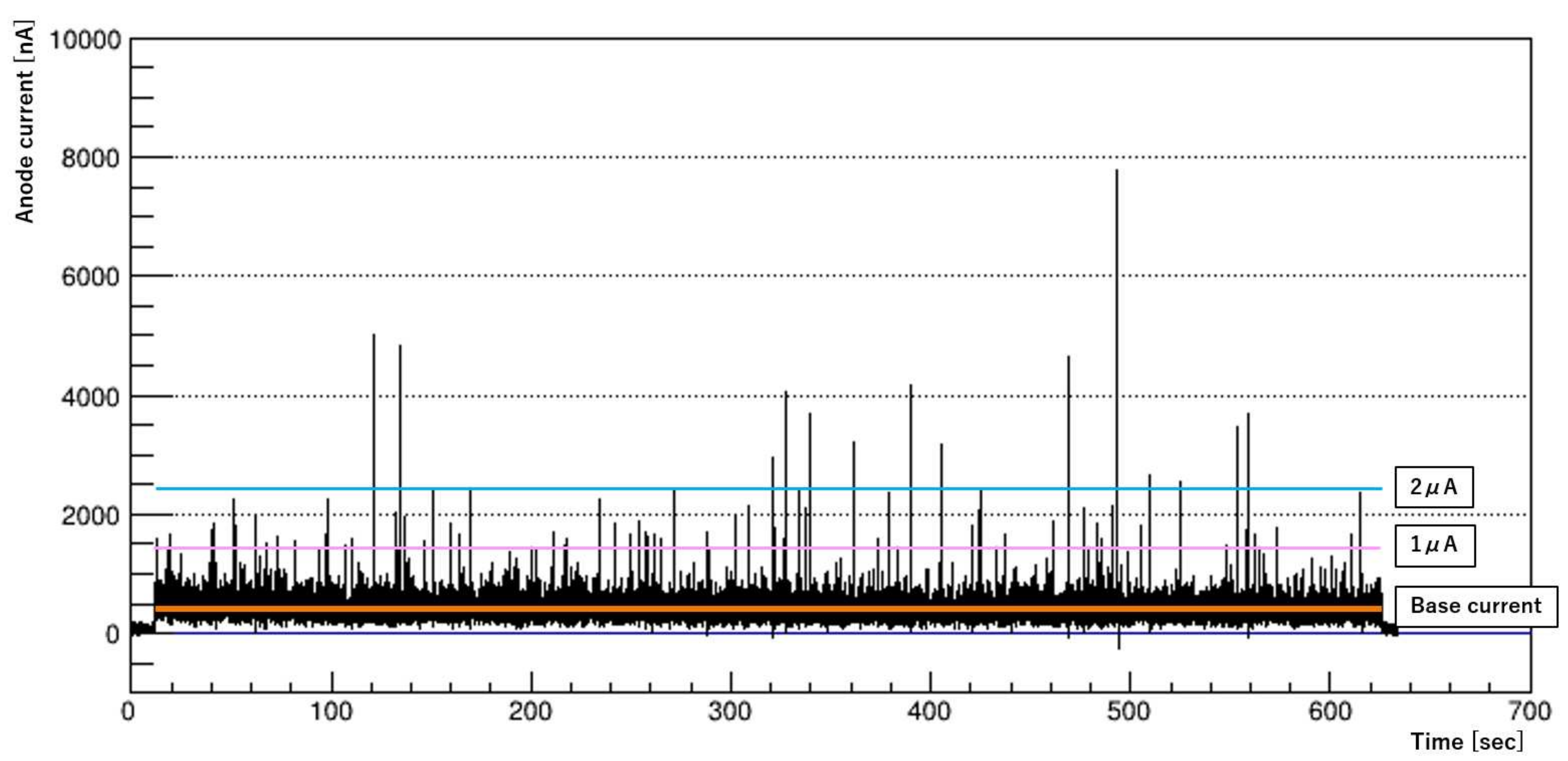}
\end{center}
\caption{Current monitor on anode of RC37 under neutron irradiation recorded for ten minutes: Gas gain is $\sim$2000.}
\label{currentmonitor}
\end{figure}

RC37 and RC38 (Table\ref{upicparameter}) were operated in this condition. The drift field was set to 3 kV with 3 mm drift gap. An Ar/$\mathrm{C_2H_6}$ (9:1) gas mixture was used. The anode current was recorded using an USB digital oscilloscope (UDS-1G02S-HR) to observe and record sparks. Fig. \ref{currentmonitor} shows a recorded current monitor on the anode. Applying fast-neutron irradiation (orange line in Fig. \ref{currentmonitor}) constantly yielded a base current on the electrode. When a spark occurred, a large current over 1$ \mu$A momentarily flowed. When the anode current exceeded the threshold, it was counted as a spark. The current threshold was set to 1 $\mu$A and 2 $\mu$A against the base current.

\begin{figure}[h]
\begin{center}
\includegraphics[width=10cm,clip]{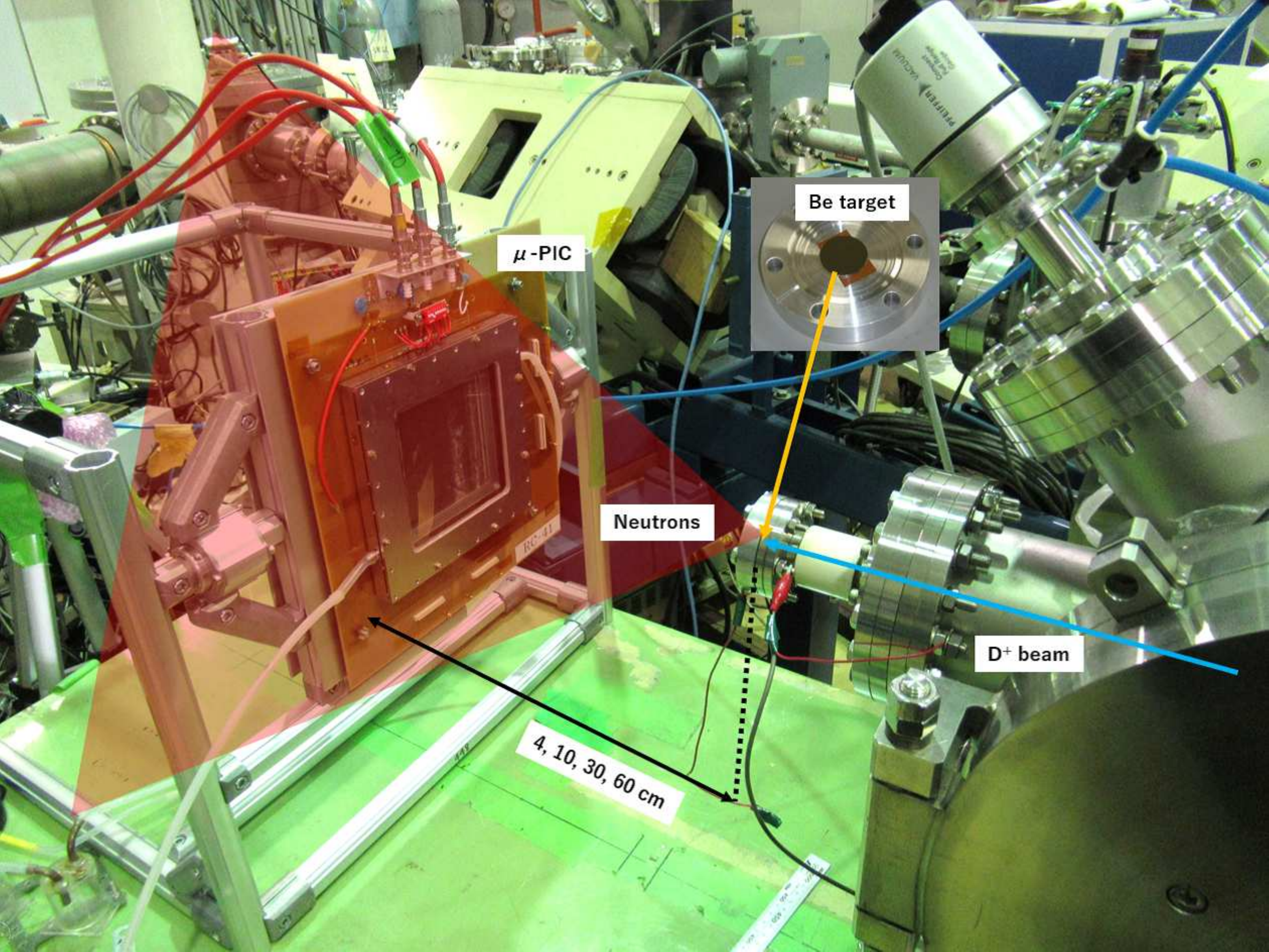}
\end{center}
\caption{Experimental setup}
\label{tandemsetup}
\end{figure}

\subsection{Spark rate}
\begin{figure}
\begin{center}
\includegraphics[width=10cm,clip]{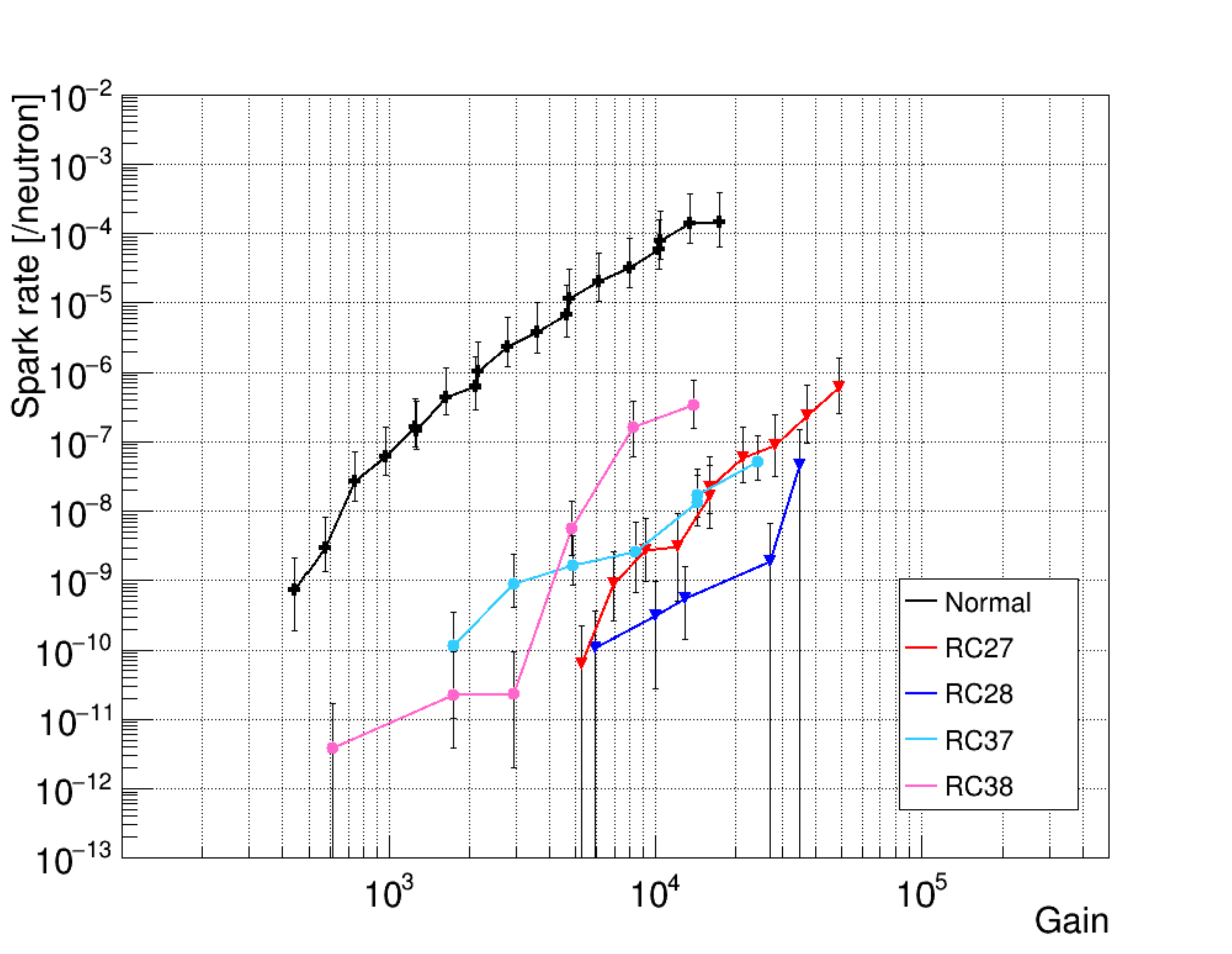}
\end{center}
\caption{Spark rate of $\mu$-PICs: RC37 and RC38 are new $\mu$-PICs. RC27 and RC28 are former resistive $\mu$-PICs. ''Normal'' denotes original $\mu$-PIC without resistive cathodes.}
\label{sparkrate}
\end{figure}

\begin{figure}
\begin{center}
\includegraphics[width=10cm,clip]{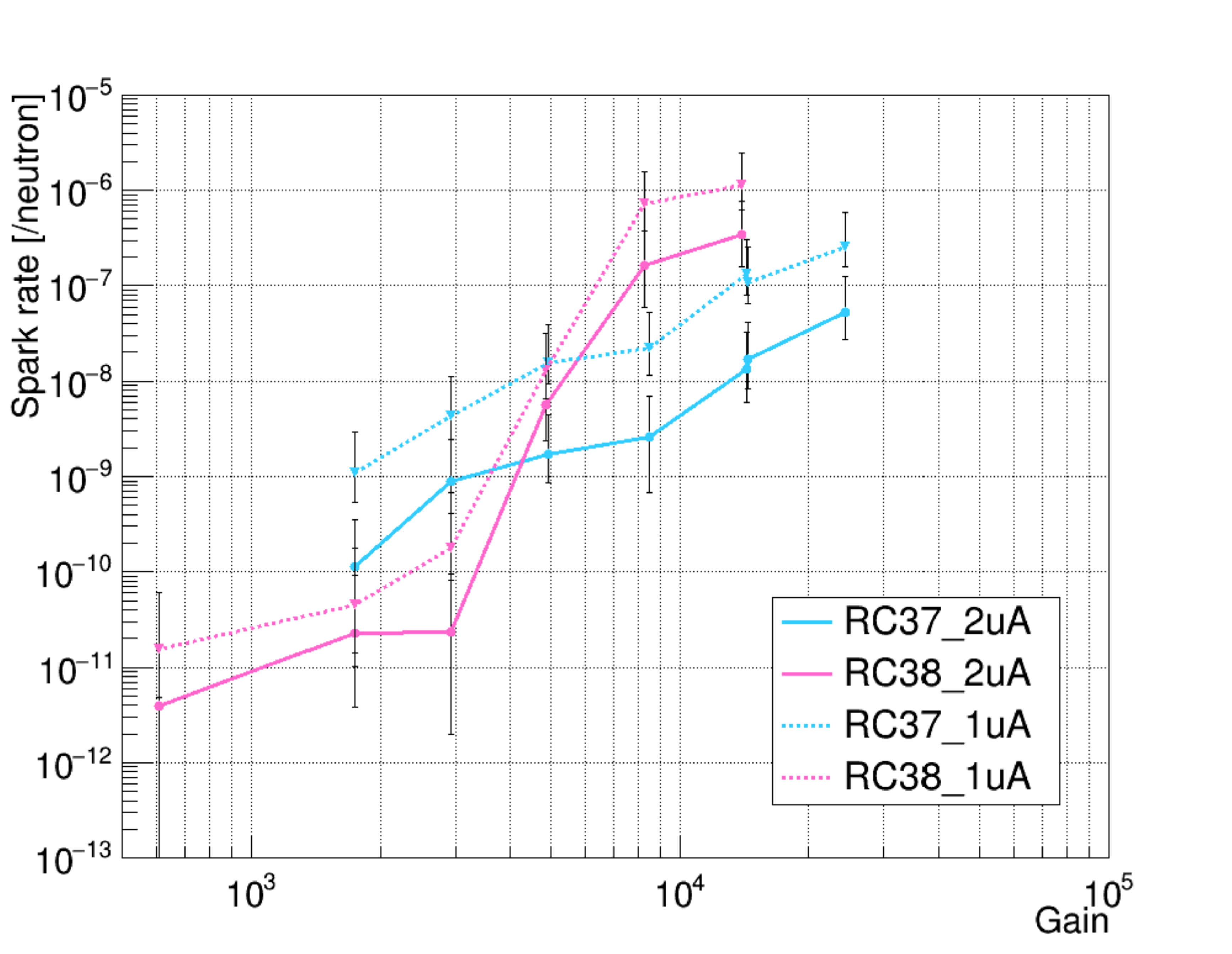}
\end{center}
\caption{Spark rate of new $\mu$-PICs with current thresholds of 1 $\mu$A and 2 $\mu$A}
\label{sparkrate_1uA}
\end{figure}

The spark rate was defined as the ratio of the number of sparks to the total number of irradiated neutrons in the chamber. The spark rates of $\mu$-PICs are plotted in Fig. \ref{sparkrate} with a current threshold of 2$\mu$A. Here, RC37 and RC38 are the new $\mu$-PICs. RC27 and RC28 are the former resistive $\mu$-PICs, where carbon polyimide paste was used for the resistive cathodes (Section 3.2). ''Normal'' denotes the original $\mu$-PIC without resistive cathodes. The spark rate of RC37 is consistent with that of the previous result above a gas gain of 8000, and it is $\mathrm{10^{3-5}}$ times lower than that of the conventional $\mu$-PIC. The spark rate of RC38 is $\mathrm{10^{1-2}}$ times higher than that of RC37 above a gas gain of 8000. Various reasons might explain the differences of the results. One is the operated area of the chambers. In a previous study (\cite{reupic}), the area was only 0.64 $\times$ 10.24 $\mathrm{cm^2}$. RC37's operated area was 4-6 times larger and that of RC38 was 6-8 times larger than in the previous study. The others are the individual differences of the chambers and the conditions in the operation. In this test, these effects could not been evaluated. Fig. \ref{sparkrate_1uA} shows the spark rate of the new chambers using different current thresholds of 1$\mu$A and 2$\mu$A. The spark rate with a threshold of 1$\mu$A is about ten times higher than that of 2$\mu$A, meaning that almost all of the sparks are suppressed below 1$\mu$A. These result show that large sparks are strongly suppressed by the $\mu$-PIC with the new structure and the DLC electrode.

\subsection{Base current on the electrode}
\begin{figure}
\begin{center}
\includegraphics[width=8cm,clip]{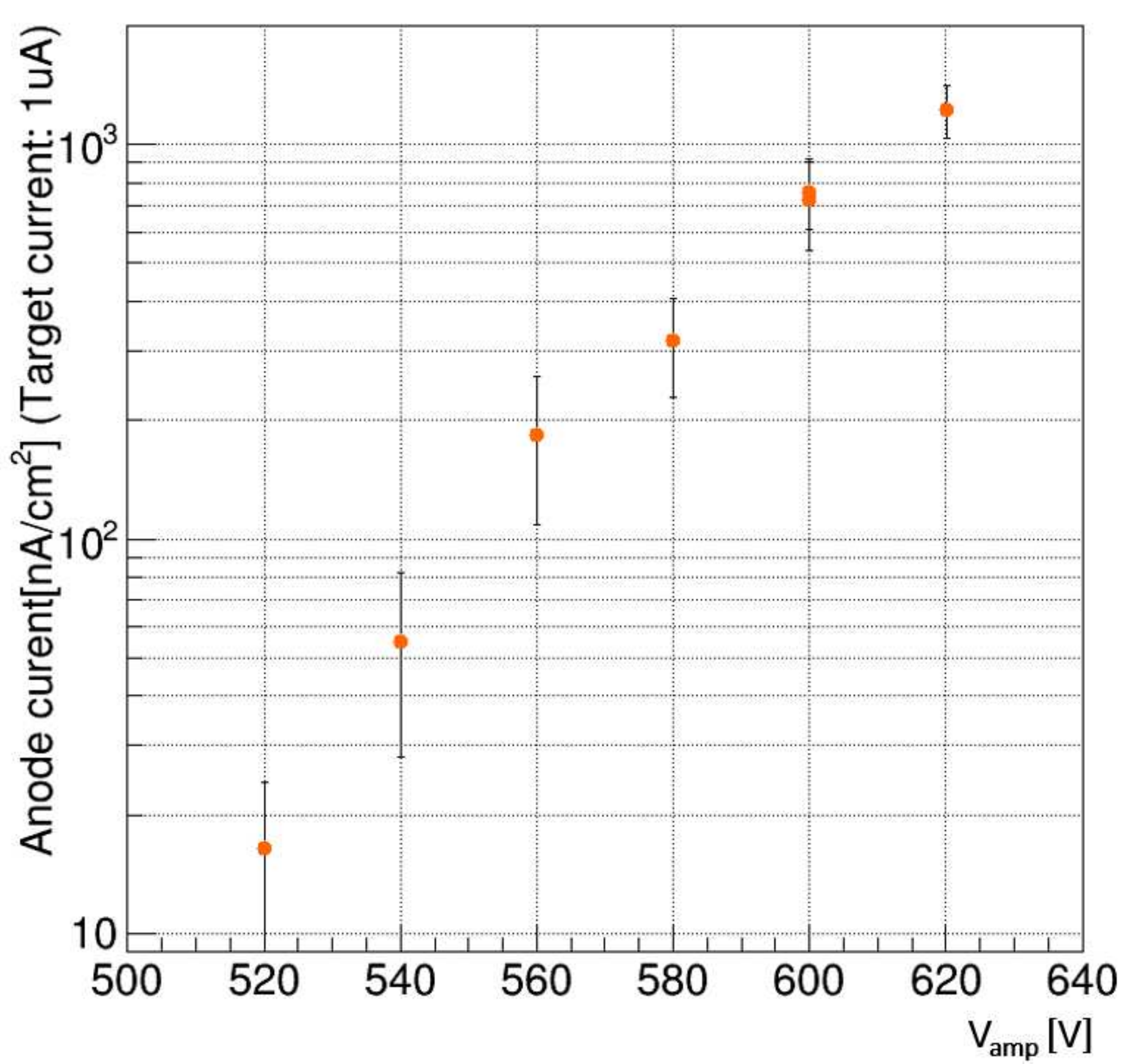}
\end{center}
\caption{Plots of base current [nA/$\mathrm{cm^2}$] as a function of amplification voltage: Current of Be target was normalized with 1 $\mu$A. Neutrons flux was estimated to be $\mathrm{60-250 kHz/cm^2}$. Gain drops were not observed.}
\label{anodecurrent10cm}
\end{figure}

\begin{figure}
\begin{center}
\includegraphics[width=8cm,clip]{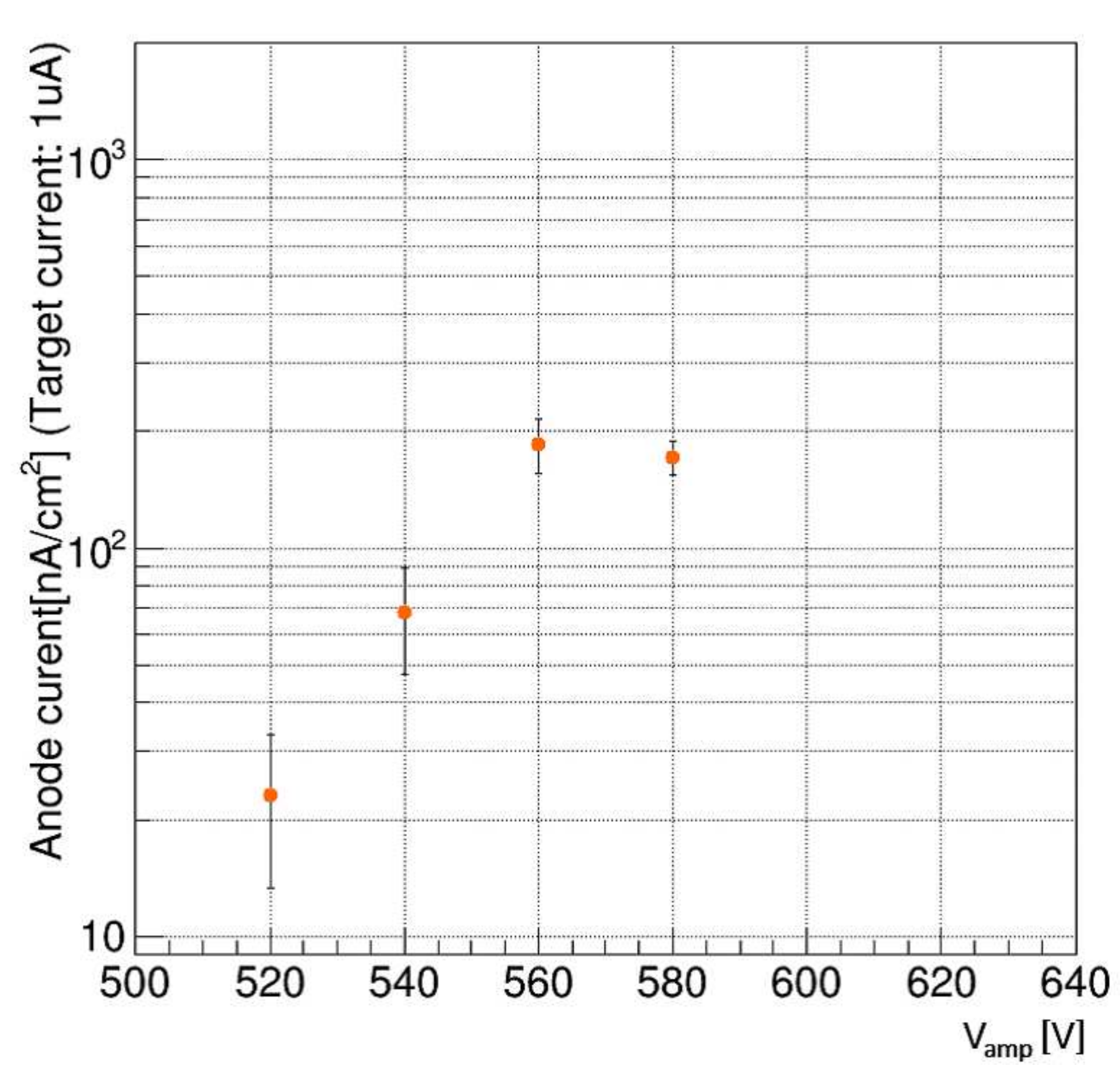}
\end{center}
\caption{Plots of base current [nA/$\mathrm{cm^2}$] as a function of amplification voltage: Current of Be target was normalized with 1 $\mu$A. Neutrons flux was estimated to be $\mathrm{1-4 MHz/cm^2}$. Base current increases linearly up to 560V in which gas gain is around 5000.}
\label{anodecurrent7cm}
\end{figure}

Other than the spark rate, another important thing to consider is the amount of voltage drop in the resistive electrodes under the high radiation condition. Voltage drop denotes the gain drop that may worsen the detection efficiency. This was evaluated using the base current on the electrodes under neutron irradiation (Fig. \ref{currentmonitor}). The surface resistivity of the cathodes is $\sim$180 k$\Omega$/sq. The operated area was 2.56 $\times$ 10.24 $\mathrm{cm^2}$. Fig. \ref{anodecurrent10cm} shows the plots of the base current [nA/$\mathrm{cm^2}$] as a function of the amplification voltage under the neutrons flux of $\mathrm{60-250 kHz/cm^2}$. The current of the Be target was normalized with 1 $\mu$A. The base current increases linearly up to the gas gain of $\sim$20000. This means that the gain did not drop, and $\mu$-PIC can be operated under a fast-neutrons flux around $\mathrm{100 kHz/cm^2}$. Fig. \ref{anodecurrent7cm} shows the results where the neutrons flux is $\mathrm{1-4 MHz/cm^2}$. In this case, the base current decreases at 580 V. However, it increases linearly up to 560 V in which the gas gain is around 5000. If fast-neutrons flux becomes higher, it might be difficult to keep enough gain. In that case, the cathode surface resistivity should be reduced to less than 180 k$\Omega$/sq.

\subsection{Duration and occupied region of the spark}
\begin{figure}[h]
\begin{center}
\includegraphics[width=8cm,clip]{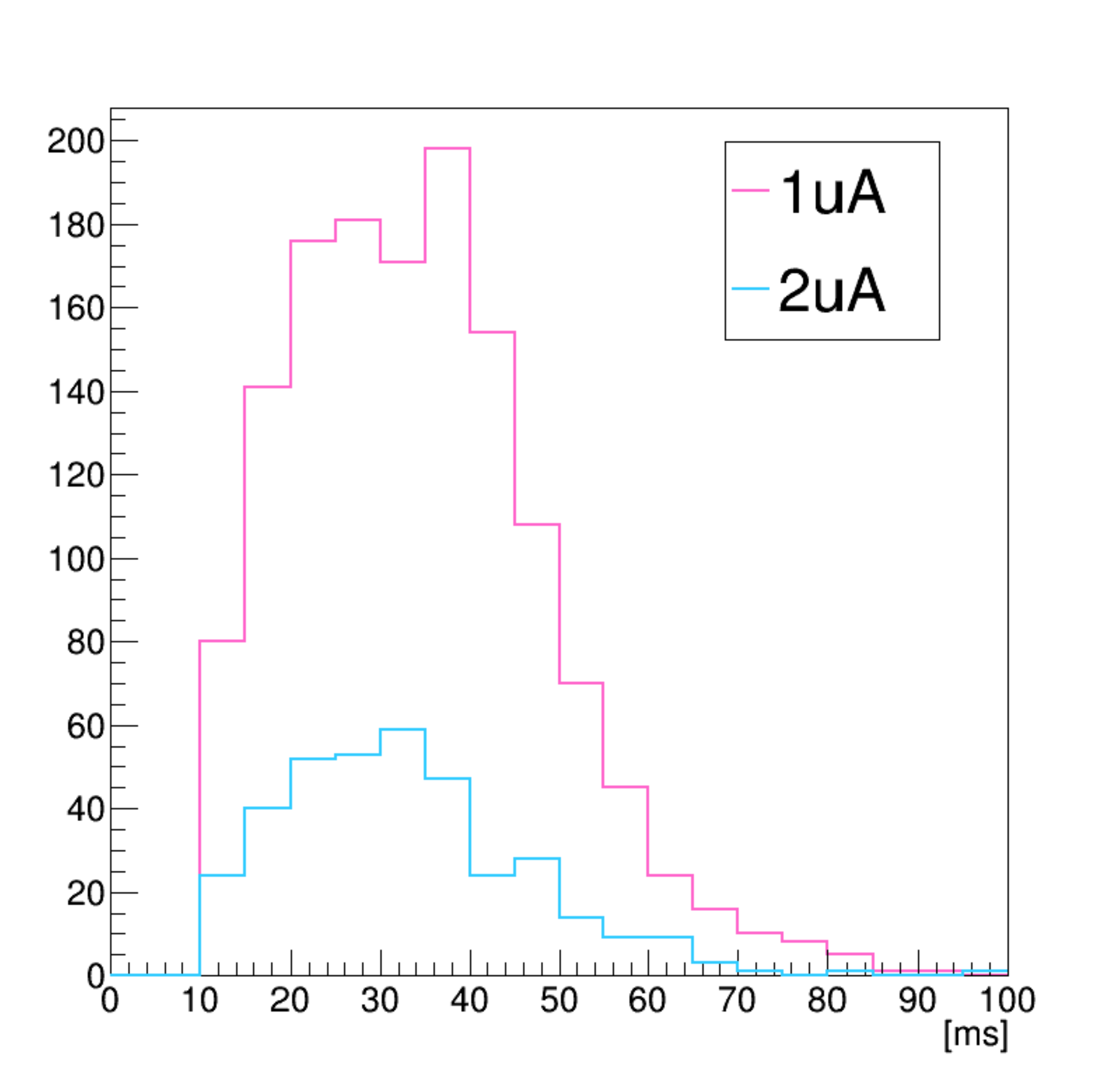}
\end{center}
\caption{Duration time of spark with current thresholds of 1 $\mu$A and 2 $\mu$A}
\label{deadtime}
\end{figure}

The spark's duration time was measured using the recorded current data of high voltage on the anodes. The current data were recorded with a sampling rate of 5 ms. The time during which the current exceeded the threshold (1 $\mu$A or 2 $\mu$A) was measured for each spark. Fig. \ref{deadtime} shows the histogram of the spark's duration time. For both thresholds, most sparks have a duration time less than 50 ms, and no sparks exceed 100 ms. This duration time corresponds to the detector's dead time. With the SRS-APV readout system, it was found that when a spark occurred, it affected all the strips that belongs to the same readout card. From those results, it can be estimated that when a spark occurred, a dead time below 100 ms is caused at the 128 strips, assuming that 128-ch readout cards are used for the same front-end.

\section{Conclusion}
A novel design of the resistive $\mu$-PIC has been proposed and tested. By using photolithography instead of laser drilling, anode and cathode electrodes were well aligned in the entire 10 $\times$ 10 $\mathrm{cm^2}$ detection area. DLC thin film made by a carbon-sputtering technique was used for resistive electrodes instead of carbon-loaded paste. This novel technique enabled flexible configuration of the resistivity at high uniformity. The resistors for HV bias and capacitors for AC coupling were completely removed by applying PCB and carbon sputtering techniques. These ideas made $\mu$-PIC a robust and very compact detector in which fine two-dimensional position measurement can be performed.

The performance of the proposed resistive $\mu$-PIC has been measured in various ways. With a $\mathrm{{}^{55}Fe}$ 5.9keV X-rays source, gas gain over $\mathrm{10^4}$ was attained. The gain's uniformity was within 30\% in the entire detection area. $\mu$-PIC is expected to have the capability to detect MIPs with high detection efficiency in the entire 10$\times$10$\mathrm{cm^2}$ detection area. Two-dimensional X-ray images were taken, and fine 1 mm structures were clearly seen. The performances for charged particles have been measured using a 150 GeV/c muons beam at the SPS/H4 beam line in CERN. The detection efficiency was more than 98\%, and the time resolution was 13-16 ns. The position resolution was 60-90 $\mu$m. To evaluate the tolerance to sparks, a fast-neutrons irradiation test has been performed at the tandem electrostatic accelerator at Kobe University Faculty of Maritime Science. The spark rate closely resembled the previous result of our former resistive $\mu$-PICs: $\mathrm{\sim10^{-8}}$/neutron for a gas gain of 5000. Furthermore, it has been found that when the surface resistivity of the cathode was $\sim$180 k$\Omega$/sq., the $\mu$-PIC could be operated with a gas gain of 5000 under fast-neutrons flux of up to $\mathrm{1-4 MHz/cm^2}$ without any gain reduction. 

Thus a novel design of a resistive $\mu$-PIC has been established, and performance studies have showed that it is a promising detector for future high-rate applications.

\section*{Acknowledgements}
The authors are grateful to Hideo Uehara at Raytech Inc. for the production of our detectors. We also thank RD51 collaboration for supporting our work at CERN, especially Eraldo Oliveri, Givi Sekhniaidze and Rui De Oliveira. We also thank the staff of the tandem electrostatic accelerator facility at the Faculty of Maritime Science, Kobe University. This work was supported by JSPS KAKENHI Grant Numbers JP26610069, and JP26104707.

\end{document}